\documentclass[floatfix,prd,aps,preprint,showpacs,amsmath,amssymb,nofootinbib]{revtex4}
\usepackage{graphicx}
\usepackage{dcolumn}
\usepackage{bm}
\usepackage{hyperref}
\newcommand{\bt}{\beta}

\def\chpt{$\chi$PT}

\def\efdsofd{f_{D_s}/f_D}

\def\frootm{$f_{Qq}\sqrt{M_{Qq}}$}
\def\efrootm{f_{Qq}\sqrt{M_{Qq}}}
\def\frootmp{$f'_{Qq}\sqrt{M_{Qq}}$}
\def\efrootmp{f'_{Qq}\sqrt{M_{Qq}}}
\def\bbbar{$B$-$\bar B$}
\def\bsbsbar{$B_s$-$\bar B_s$}
\def\fb{$f_B$}
\def\fbs{$f_{B_s}$}
\def\fd{$f_D$}
\def\fds{$f_{D_s}$}
\def\fbofd{$f_{B}/f_D$}
\def\fbsofb{$f_{B_s}/f_B$}
\def\fdsofd{$f_{D_s}/f_D$}
\def\fbofds{$f_B/f_{D_s}$}
\def\fbsofds{$f_{B_s}/f_{D_s}$}

\def\leftvec{\raise1.5ex\hbox{$\leftarrow$}\kern-.85em}
\def\half{{\scriptstyle \raise.15ex\hbox{$\frac{1}{2}$}}}
\def\threehalves{{\scriptstyle \raise.15ex\hbox{$\frac{3}{2}$}}}
\def\third{{\scriptstyle \raise.15ex\hbox{$\frac{1}{3}$}}}
\def\twothirds{{\scriptstyle \raise.15ex\hbox{$\frac{2}{3}$}}}
\def\fourthirds{{\scriptstyle \raise.15ex\hbox{$\frac{4}{3}$}}}
\def\fourth{{\scriptstyle \raise.15ex\hbox{$\frac{1}{4}$}}}
\def\gtwid{\raise.3ex\hbox{$>$\kern-.75em\lower1ex\hbox{$\sim$}}}
\def\ltwid{\raise.3ex\hbox{$<$\kern-.75em\lower1ex\hbox{$\sim$}}}

\def\ie{{\it i.e.},\ }
\def\eg{{\it e.g.},\ }

\def\et{{\it et al.}}

\def\vs{{\it vs.}\ }

\def\cO{{\cal O}}

\def\eq#1{eq.~(\ref{eq:#1})}
\def\eqs#1#2{eqs.~(\ref{eq:#1}) and (\ref{eq:#2})}
\def\eqsthree#1#2#3{eqs.~(\ref{eq:#1}), (\ref{eq:#2}) and (\ref{eq:#3})}
\def\eqsfour#1#2#3#4{eqs.~(\ref{eq:#1}), (\ref{eq:#2}), (\ref{eq:#3}) and (\ref{eq:#4})}

\def\prl#1{Phys.\ Rev.\ Lett.\ {\bf #1}}
\def\prd#1{Phys.\ Rev.\ {\bf D#1}}
\def\plb#1{Phys.\ Lett.\ {\bf #1B}}
\def\npb#1{Nucl.\ Phys.\ {\bf B#1}}

\def\dallas#1{Nucl.\ Phys.\ {\bf B} (Proc.\ Suppl.) {\bf 34}, #1 (1994)}

\def\pisa#1{Nucl.\ Phys.\ {\bf B} (Proc.\ Suppl.) {\bf 83-84}, #1(2000)}
\def\bangalore#1{Nucl.\ Phys.\ {\bf B} (Proc.\ Suppl.) {\bf 94}, #1 (2001)}
\def\berlin#1{Nucl.\ Phys.\ {\bf B} (Proc.\ Suppl.) {\bf 106-107}, #1 (2002)}
\def\boston{presented at the International Symposium,
{\it Lattice 2002}, Boston, June 24--29, 2002, to be published
in Nucl.\ Phys.\ {\bf B} (Proc.\ Suppl.)}

\def\MeV{{\rm Me\!V}}
\def\GeV{{\rm Ge\!V}}

\begin{document}
\title{
Lattice Calculation of Heavy-Light Decay Constants with Two Flavors of Dynamical Quarks
}
\author{
C.~Bernard$^1$, 
S.\ Datta$^{2,3}$,
T.\ DeGrand$^4$,
C.\ DeTar$^5$,
Steven Gottlieb$^2$, Urs M.~Heller$^6$,
C.\ McNeile$^7$,
K.\  Orginos$^8$,
R.~Sugar$^9$, D.~Toussaint$^{10}$\\
(The MILC Collaboration)
}
\affiliation{
\vspace{5mm}   
$^1$Washington University, St.~Louis, Missouri 63130, USA\\
$^2$Indiana University, Bloomington, Indiana 47405, USA\\
$^3$Fakult\"at f\"ur Physik, Universit\"at Bielefeld, D 33615
Bielefeld, Germany\\
$^4$University of Colorado, Boulder, Colorado 80309, USA \\
$^5$University of Utah, Salt Lake City, Utah 84112, USA\\
$^6$CSIT, The
Florida State University, Tallahassee,
   Florida 32306-4120, USA\\
$^7$Dept.\  of Math.\  Sci., University of Liverpool, Liverpool, L69 3BX, UK\\
$^8$RIKEN BNL Research Center, Upton, New York 11973, USA\\
$^9$University of California, Santa Barbara, California 93106, USA\\
$^{10}$University of Arizona, Tucson, Arizona 85721, USA\\
}
\date{\today}

\vspace{2cm}

\begin{abstract}\noindent
We present results for  \fb, \fbs, \fd, 
\fds\  and their ratios in the presence of two flavors
of light sea quarks ($N_f=2$).
We use Wilson light valence quarks 
and Wilson and static heavy valence quarks; the sea quarks
are simulated with staggered fermions.
Additional quenched simulations with nonperturbatively
improved clover fermions allow us to improve our control of
the continuum extrapolation.
For our central values the masses of the sea quarks are not
extrapolated to the physical $u$, $d$ masses; that is, the central
values are  ``partially quenched.'' 
A calculation using ``fat-link clover'' valence fermions is also
discussed but is not included in our final results.
We find, for example,
$f_B  =  190 (7) ({}^{+24}_{ -17}) ({}^{+11}_{ -2}) ({}^{+8}_{ -0})\;\MeV$,
$f_{B_s}/f_B  =  1.16 (1) (2) (2)({}^{+4}_{ -0}) $,
$f_{D_s}  =  241 (5)  ({}^{+27}_{ -26}) ({}^{+9}_{ -4})  ({}^{+5}_{ -0}) \;\MeV$, and
$f_{B}/f_{D_s} = 0.79 (2) ({}^{+5}_{-4})  (3) ({}^{+5}_{ -0})  $,
where in each case the first error is statistical and the 
remaining three are systematic: the error within the partially quenched
$N_f=2$ approximation, the error due to the missing
strange sea quark and to partial quenching, and an estimate
of the effects of chiral logarithms at small quark mass.
The last error, though quite significant in decay constant ratios,
appears to be smaller than has been recently suggested by Kronfeld and Ryan,
and Yamada.
We emphasize, however, that as in other lattice computations to date,
the lattice $u,d$ quark  masses are not very light  and
chiral log effects may not be fully under control.
\end{abstract}

\maketitle

\section{INTRODUCTION}
\label{sec:intro}
Accurate values for the leptonic decay constants of $B$ and $B_s$ mesons
are crucial for interpreting experimental measurements of 
\bbbar\ mixing and bounds on, or future
measurements of, \bsbsbar\ mixing.
Knowledge of the decay constants (coupled with knowledge of the corresponding
B-parameters) makes possible a determination of
the CKM elements $V_{td}$ and $V_{ts}$ from these experiments.

In the $D$-meson sector, CLEO-c will measure leptonic decay rates 
at the 3--4\% level \cite{CLEOc}. Assuming
3-generation unitarity, this translates into 
determinations  of \fds\ and \fd\ with roughly 2\% accuracy. Coupled with accurate 
theoretical
computations of ratios such as $f_B/f_{D_s}$, this will provide crucial
information about the $B$-sector. In addition, if computations
of the $D$ and $D_s$ decay constants themselves can be performed at the few 
percent level, the experiments
will directly determine $V_{cs}$ and $V_{cd}$ with similar precision.

At least in principle, lattice QCD offers a means
to compute quantities such as $f_B$ or $f_B/f_{D_s}$ with control over all sources
of systematic error.
Here, we present a 
computation by the MILC collaboration of the decay constants
\fb, \fbs, \fd, \fds, and their ratios. We take into account
the effects of virtual quark loops from
two light flavors of sea quarks; \ie we have
two ``dynamical quarks.''  
Additional discussion and preliminary results for the dynamical calculation
can be 
found in Refs.~\cite{LAT00,CBreview}.  Our earlier work, which
focused on the quenched approximation and used dynamical configurations
only for an estimate of the quenching errors, appeared in
\cite{MILC-PRL}, with further details in  \cite{PRELIM-MILC}.

This paper is organized as follows:
Our lattice formalism is presented in Sec.~\ref{sec:formalism}.
We discuss the Fermilab approach to heavy
quarks on the lattices \cite{EKM}, and explain how we adapt it
to Wilson and 
nonperturbatively improved \cite{ALPHA-CSW-CA-ZA} clover
quarks.  We also explain our use of perturbative
renormalization and the choice of scale (``$q^*$'').

Section \ref{sec:details} gives the lattice computational details.
We discuss the generation of configurations, the evaluation
of quark propagators for Wilson, clover, and static quarks,
and various aspects of the analysis, including fitting and
extrapolation.  
The most significant open issue here involves the effect
of chiral logarithms on the light quark mass extrapolations.
In important recent work,
Kronfeld and Ryan \cite{KRONFELD} and Yamada \cite{YAMADA} (building on
work of the JLQCD collaboration \cite{JLQCD-LAT01}) have argued that
standard linear or quadratic extrapolations from typical lattice
light quark masses miss the logarithms at low mass and drastically
underestimate such quantities as \fbsofb.  Since the quark masses
available to the present calculation are of this typical size,
it is imperative that we estimate the  chiral logarithm effects as best
we can.  We devise a method to estimate, at least crudely,
such effects.  The method is based on the extrapolation of the ratio of the
light-light to the heavy-light decay constant. 

In Sec.~\ref{sec:quenched} we revisit
the quenched approximation.  The dominant source of systematic
error in our previous quenched computation \cite{MILC-PRL} 
was the continuum extrapolation. Two new features of the current
analysis have significantly reduced that error:  (1) new
running with both Wilson and clover quarks and (2) a new central value for the
scale $q^*$ for the
heavy-light axial current \cite{CB-TD}. 

We then turn to the dynamical quark data in Sec.~\ref{sec:dynamical}.
The improved control over
discretization errors in the quenched approximation gives us more
confidence in the central value and errors deduced
from the continuum extrapolation of our dynamical quark data.
Other sources of systematic error, including the chiral extrapolation,
higher order perturbation theory,  and the effects of partial quenching
are also discussed in detail.
Finally we clarify the effects of ``fat-link'' fermions using some 
new test runs in the quenched case.  These shed light on why the 
preliminary values for heavy-light decay constants 
with fat-link fermions on dynamical configurations
were anomalously low \cite{MILC-LAT99}.

Our conclusions and
the outlook for reducing the main systematic uncertainties 
are discussed
in Sec.~\ref{sec:conclusions}.  We describe work in progress that
addresses the outstanding issues in the chiral and continuum
extrapolations.

The computation presented in this paper is rather complicated:
we use several different actions, operators,
renormalizations, fitting techniques, and extrapolations. 
Part of the reason for this is that the simulations with dynamical quarks
are extremely demanding computationally and therefore have taken
years to complete.  During that time, as our understanding of the physics
and analysis issues grew, our methods evolved.
The variety of methods used does have one important virtue: it allows us to estimate
many of the systematic effects in a direct way.

\section{FORMALISM}
\label{sec:formalism}

In a groundbreaking paper \cite{EKM}, El-Khadra, Kronfeld and Mackenzie (EKM)
show that one can make sense of heavy 
Wilson-like\footnote{``Wilson-like'' means that the fermion action includes the
naive discretization of the Dirac equation plus a Wilson term to
remove doublers. There may be further additional correction terms
to reduce lattice artifacts.  Standard Wilson fermions as well
as ``clover'' fermions \cite{SW} fall into this class.}
fermions on the
lattice even when $am_Q\gtwid 1$, where $a$ 
and $m_Q$ are the lattice spacing and heavy quark mass, respectively.
Indeed, in the nonrelativistic limit $m_Q\gg \Lambda_{QCD}$, they show that 
the effective Hamiltonian has the form (after Foldy-Wouthuysen-Tani
transformation)
\begin{equation}
\label{eq:hamiltonian}
H = \bar Q\Big(M_1 + \gamma_0 A_0 -\frac{\vec D^2}{2M_2}
-\frac{i\vec \Sigma\cdot\vec B}{2M_3}\Big)Q + \cO(1/m_Q^2)\ ,
\end{equation}
where $Q$ is the effective heavy quark field, $\vec D$ is the 
spatial covariant derivative, $\vec B$ is the
chromomagnetic field, $\vec \Sigma$ are the Pauli matrices,
and $m_Q$ is a generic heavy quark mass. The
masses $M_1$, $M_2$,  and $M_3$ are particular functions of 
the bare heavy quark mass $am_0$
that depend on the quark action. Here
$m_0$ is given by
\begin{equation}
am_0 = \frac{1}{2\kappa_Q} - \frac{1}{2\kappa_c}\ ,
\label{eq:mzero}
\end{equation}
where $\kappa_Q$ is the heavy quark hopping parameter and
$\kappa_c$ is its critical value.\footnote{We assume throughout this
paper that the
spatial and temporal hopping parameters are chosen equal,
that the Wilson parameter $r=1$, and that the spatial and
temporal parts of the clover term, if present, have equal
coefficients.  This is not the complete
generality of Ref.~\cite{EKM}, but is sufficient for our purposes.}
The ``pole mass,'' $M_1$, controls the exponential decay of the
zero-momentum propagator in Euclidean time, but is just an additive
constant in bound state energies.  The nontrivial physics of a heavy quark
in a heavy-light bound state is controlled at this order by the ``kinetic
mass,'' $M_2$, which fixes the heavy-quark energy-momentum dispersion relation,
and the ``magnetic mass,'' $M_3$, which governs chromomagnetic effects,
such as hyperfine splittings.

For computations of heavy-light decay constants, one also needs to know how
the lattice axial current, $\bar q\gamma_0\gamma_5 Q$, renormalizes.  
At tree level but through order ($1/m_Q$), EKM show that the renormalization 
is given simply by the replacement, $Q\to Q_I$, where
the tree-level improved field is
\begin{equation}
\label{eq:Q-I}
Q_I(x) = \sqrt{2\kappa_Qe^{aM_1}}\;\; [1 + ad_1\vec\gamma\!\cdot\!\! \vec D]\,Q(x)\ ,
\end{equation}
with $d_1$ another function of $am_0$. 
We have included the standard $\sqrt{2\kappa_Q}$ factor needed
to go from lattice-normalized to continuum-normalized fields.  

At tree level, one has
\begin{eqnarray}
\label{eq:treemasses}
aM_1 &=& \ln (1 + am_0)\ ,\cr
\noalign{\smallskip}
aM_2 &=& \frac{am_0 (1+am_0)(2+am_0)}{2+4am_0 +(am_0)^2}\cr
\noalign{\smallskip}
&=& \frac{e^{aM_1}\sinh(aM_1)}{1+\sinh(aM_1)}\ ,\cr
\noalign{\smallskip}
aM_3 &=& \frac{am_0 (1+am_0)(2+am_0)}{2(1+am_0)+c_{SW}am_0(2+am_0)}\ ,\cr
\noalign{\smallskip}
d_1 &=& \frac{am_0}{2(1+am_0)(2+am_0)}\ ,
\end{eqnarray}
where $c_{SW}$ is the coefficient of the clover term. 

A large fraction of the one-loop corrections to \eq{treemasses}
can be included with tadpole renormalization \cite{LandM}.  We use
$\kappa_c$ to define
the mean field value, $u_0$, of the gauge link: $u_0\equiv 1/(8\kappa_c)$.
Absorbing $u_0$ into $\kappa$ gives a tadpole-improved
hopping parameter, $\tilde\kappa$, and bare mass, $\tilde m_0$:
\begin{eqnarray}
\label{eq:tildekappa}
\tilde\kappa &\equiv& u_0\kappa= \frac{\kappa}{8\kappa_c}\;,\quad \tilde\kappa_c=1/8\cr
a\tilde m_0 &\equiv&a m_0/u_0= 
\frac{1}{2\tilde\kappa} - \frac{1}{2\tilde\kappa_c}  = 4\frac{\kappa_c}{\kappa}-4\ .
\end{eqnarray}

We denote the tadpole-improved versions of the quantities in 
\eq{treemasses} by
$\tilde M_1$, $\tilde M_2$, $\tilde M_3$, and $\tilde d_1$.
They are found simply by replacing $m_0\to \tilde m_0$.
Similarly, the tadpole-improved version
of \eq{Q-I} is
\begin{eqnarray}
\tilde Q_I(x) &=& \sqrt{2\kappa_Qu_0e^{a\tilde M_1}}\;\;
[1 + a\tilde d_1\vec\gamma\!\cdot\!\! \vec D]\,Q(x)\cr
&=& \sqrt{1-\frac{3\kappa_Q}{4\kappa_c}}\;\;
[1 + a\tilde d_1\vec\gamma\!\cdot\!\! \vec D]\,Q(x)\ .
\label{eq:tildeQ-I}
\end{eqnarray}
For some applications, the $\tilde d_1$ term here may be neglected.
It is therefore convenient also to define 
\begin{equation}
\label{eq:tildeQ-I0}
\tilde Q_I^0(x) \equiv
\sqrt{1-\frac{3\kappa_Q}{4\kappa_c}}\; Q(x) \ .
\end{equation}

We take the physical mass of our lattice heavy-light mesons to be
the meson kinetic mass, $M_{Qq,2}$.  Although $M_{Qq,2}$ could be determined
directly from the meson dispersion relation, that would require
the computation of meson propagators with nonzero momenta, which in
any case are rather noisy.  Instead we define $M_{Qq,2}$ by \cite{BLS}
\begin{equation}
\label{eq:mhl2}
M_{Qq,2} = M_{Qq,1} + \tilde M_2 - \tilde M_1\ ,
\end{equation}
where $M_{Qq,1}$ is the pole mass of the meson determined on the lattice,
and $\tilde M_2$ and $\tilde M_1$ refer to the heavy quark, $Q$.
The UKQCD collaboration, in Fig.~8 of Ref.~\cite{UKQCD-fB00}, compare the
kinetic meson mass determined by the dispersion relation with that given by
\eq{mhl2} (but without tadpole improvement).  The agreement is good, and would
in fact be still better if the tadpole improved version were used.  

\subsection{Wilson fermions}
\label{sec:wilson-formalism}

For Wilson fermions ($c_{SW}=0$), the magnetic mass $M_3$ is not equal
to the kinetic mass $M_2$, even at tree level.  As discussed in
Refs.~\cite{JLQCD98} and \cite{MILC-PRL}, this produces an
error at fixed $a$
of $\cO((c_{mag}-1)\Lambda_{QCD}/M_{Qq})$,
where $c_{mag}\equiv M_2/M_3$.  Hence there
is little point in keeping the 
$\tilde d_1$ term in \eq{tildeQ-I}, which is also of
$\cO(\Lambda_{QCD}/M_{Qq})$.  (Indeed, keeping such terms without including
at least the $\cO(g^2)$ perturbative corrections to them is likely to
increase the systematic error \cite{CBreview}.) We thus set $\tilde d_1=0$
in the Wilson case and use \eq{tildeQ-I0}.

The 1-loop mass-dependent perturbative matching for the heavy-light axial current
has been calculated by Kuramashi \cite{KURAMASHI}. Since we include
tadpole renormalization
for both the heavy and light
quarks through \eq{tildeQ-I0}
(with $\tilde d_1=0$), we adjust the 
result to reflect our  choice of $\kappa_c$ to
define the mean link.  (Ref.~\cite{KURAMASHI} uses the Feynman-gauge link.)
The continuum contribution to the matching generates a logarithm
of the heavy quark mass.  In \cite{KURAMASHI}, this is taken as
$\log{M_1}$.  Since we take $M_2$  as the physical mass, we also
adjust the result of \cite{KURAMASHI} to replace $M_1$ with $M_2$
in the logarithm.

Additional issues for perturbative matching are the definition of
the coupling constant and the scale at which it is evaluated.
We use the coupling $\alpha_V$, whose value at scale $3.4018/a$ is
defined in terms of the plaquette \cite{LandM,ALPHAV}.  It has become
standard to evaluate the coupling at the 
Lepage-Mackenzie scale $q^*$ \cite{LandM}, defined by
\begin{equation}
\log((q^*)^2) = \frac{\int d^4q\; I(q) \log(q^2)}{\int d^4q\; I(q)}\ , 
\label{eq:qstar} 
\end{equation} 
where $I(q)$ is the complete integrand for the quantity of interest.
In other words, the 1-loop 
axial current renormalization constant, $Z_A$, is given by
$Z_A = 1 + \alpha_V(q^*) C_F z_A/(4\pi)$, where $C_F$ is the quadratic
Casimir and $z_A\equiv \int d^4q\; I(q)$.  Here, we need
$Z_A$ in three cases (light-light, static-light, and heavy-light),
which we denote by $Z_A^{\rm qq}$, $Z_A^{\rm Statq}$, and $Z_A^{\rm KUR}$,
respectively, where ${\scriptstyle{\rm KUR}}$ emphasizes that we are talking about
the heavy-light renormalization constant computed by Kuramashi \cite{KURAMASHI}.
We adopt corresponding notation for $z_A$ and $I$.

Unfortunately, at the time the analysis described here was performed,  $q^*$
for $Z_A^{\rm KUR}$  
had not been determined \cite{HARADA}.
For $Z_A^{\rm qq}$
with $\kappa_c$-tadpole improvement,  $q^*=2.32/a$ \cite{BGM},
which we use here when fixing the scale through $f_\pi$. For $Z_A^{\rm Statq}$
with plaquette tadpole improvement, Hernandez and 
Hill \cite{HERNANDEZandHILL}
found $q^*=2.18/a$.  Since the light-light and static-light values of
$q^*$ were so close, it was argued in \cite{MILC-PRL} that either could be
used in the heavy-light case, and in fact the light-light value $q^*=2.32/a$
was chosen for the standard computation (central value).

Recently, Bernard and DeGrand \cite{CB-TD} have repeated the Hernandez and Hill
computation. They find a significantly different value of 
$q^*$ for the $\kappa_c$-tadpole-improved $Z_A^{\rm Statq}$.  
Their result depends on 
the heavy-light mass, which enters through the continuum part of the matching.  
However the mass
dependence is rather weak over the range of masses used in the current numerical
work, and it is therefore adequate to use an average value $q^*\approx 1.43/a$.
Since Ref.~\cite{CB-TD} has not yet appeared, it may be helpful to summarize
here the reasons for the disagreement with Ref.~\cite{HERNANDEZandHILL}. 

First of all, Ref.~\cite{HERNANDEZandHILL} sets to zero certain parts of the
lattice integrand whose contributions to the matching vanish by 
contour integration.  This is a standard
procedure \cite{EICHTENandHILL} for evaluating integrals involving a 
static quark propagator.  However, such integrals do not vanish when the 
integrand is first multiplied by $\log(q^2)$, as in \eq{qstar}.
Bernard and DeGrand argue that it is incorrect to
discard parts of the integrand unless their contributions to both
the numerator and denominator of \eq{qstar} vanish.

Secondly,  there are ``constant'' terms in the matching coming from
the dimensionally regularized continuum integrals. Hernandez and Hill
treat these as constant over the 4-dimensional Brillouin zone.
Similarly, the $\log(am_Q$) term in $Z_A^{\rm Statq}$, which comes
from both continuum and lattice integrals, 
is set to zero in \cite{HERNANDEZandHILL}
by the choice $a=1/m_Q$.  In contrast, Ref.~\cite{CB-TD} keeps the
full continuum integrands as well as the full lattice integrands.
This does introduce a small amount of arbitrariness: the dimensionally
regulated continuum integrals must be replaced by finite, 
subtracted 4-dimensional
integrals, and there is some freedom in how the subtraction is done.
However, as long as the subtraction is ``reasonable,'' the arbitrariness
in $q^*$ is small.  
If we accept the results of Ref.~\cite{CB-TD}, then $q^*$ for $Z_A^{\rm Statq}$ 
is no longer very close to $q^*$ for $Z_A^{\rm qq}$.  
Instead, we take the static-light $q^*\approx 1.43/a$ for
$Z_A^{\rm KUR}$.  Since $am_B$ is quite large on our lattices
($\approx 1.2$ to $4$), we believe this is a reasonable choice.
Of course, it is always necessary to consider a range of $q^*$ to estimate
perturbative errors, and the range we pick (see Sec.~\ref{sec:quenched})
includes the light-light $q^*$, as well as the values in Ref.~\cite{HARADA}.

Summarizing the results of this section, we may express the
$0^{\rm th}$ component of our renormalized
heavy-light current as
\begin{equation}
\label{eq:A-KUR}
A^{\rm KUR}_0 = Z_A^{\rm KUR}(q^*) 
\bar{\tilde q}^0_I \gamma_0 \gamma_5  \tilde Q_I^0 \ ,
\end{equation}
with $q^*=1.43/a$, with $\tilde Q_I^0$ given by \eq{tildeQ-I0}, 
and with a corresponding expression
for the tadpole improved light quark field, ${\tilde q^0_I}$.
For convenience, we also use \eq{A-KUR} for the light-light
pseudoscalars (``pions''), even though the mass dependence
of Ref.~\cite{KURAMASHI} is negligible in that case.  We take 
$q^*=2.32/a$ \cite{BGM}
for light-light renormalization.

The errors in our heavy-light Wilson calculation are formally 
$\cO(a\Lambda_{QCD})$ and $\cO(\alpha_V^2)$.  Note that,
in the Fermilab formalism,
one should think of these errors as multiplied by an arbitrary
(but presumably $\cO(1)$) function of $aM_Q$, since we are working
to all orders in  $aM_Q$.  Thus the dependence on $a$ is in general 
complicated. An example of such complicated behavior
is the difference between lattice chromomagnetic effects,
which go like $a\Lambda_{QCD}/(aM_3)$, and the  desired behavior
$a\Lambda_{QCD}/(aM_2)$ (see \eq{treemasses}). Of course, in the
truly asymptotic regime where $aM_Q\ll 1$, the leading errors are indeed
linear in $a$, but this region is not currently accessible in practical
calculations.

In the static-light case, we have
\begin{equation}
\label{eq:A-STAT}
A^{\rm STAT}_0 = Z_A^{\rm STAT}(q^*)
\bar{\tilde q}^0_I \gamma_0 \gamma_5  h \ ,
\end{equation}
where $h$ is the static quark field,
$Z_A^{\rm STAT}$ is the one-loop renormalization
constant for the static-light current \cite{EICHTENandHILL,BOUCAUD}
with tadpole improvement,
and $q^*=1.43/a$.

\subsection{Nonperturbative Clover fermions}
\label{sec:clover-formalism}

For our computations with clover fermions, we take the clover
coefficient $c_{SW}$ calculated nonperturbatively by the ALPHA
collaboration \cite{ALPHA-CSW-CA-ZA}.  The $0^{\rm th}$ component
of the renormalized, improved (through
$\cO(a)$) 
light-light axial current
(which is needed here to set the
scale with $f_\pi$) is then  
\begin{eqnarray}
A_0^{\rm NP} &=& Z^{\rm NP}_A\sqrt{4\kappa_{q_1}\kappa_{q_2}}
(1+ b_Aa\bar{m}_0)
[A_0 + c_Aa\partial_0P_5
]\ ,\cr
A_0 &=& \bar q_1\gamma_0 \gamma_5 {q_2}\ ; \qquad
P_5 = \bar q_1 \gamma_5 {q_2} \ ,
\label{eq:A-NP}
\end{eqnarray}
where $\kappa_{q_1}$, $\kappa_{q_2}$ are the hopping
parameters of the light quarks, and $\bar{m}_0= 
(m_{{q_1},0} +m_{{q_2},0})/2$ 
is their average bare mass.
$Z_A^{\rm NP}$ and $c_A$  are the nonperturbative values given in
\cite{ALPHA-CSW-CA-ZA}.  The coefficient $b_A$ has not been
determined nonperturbatively by the  ALPHA collaboration, although
the difference $b_A-b_P$ has \cite{ALPHA-bm}. 
Bhattacharya \et\ \cite{LANL-NP} have determined $b_A$ at $\beta=6.0$
and $6.2$, but not at $\beta=6.15$, which is one of the couplings used here.
Our $b_A$ is therefore taken from perturbation theory \cite{SINTandWEISZ}, but
with coupling $\alpha_V(q^*)$, with $q^*$ chosen as the value ($\cong\!1/a$)
that produces the nonperturbative result \cite{ALPHA-CSW-CA-ZA} for the
similar quantity $b_V$.  This gives $b_A=1.47$ at $\beta=6.15$
and $1.42$ at $\beta=6.0$.  In the systematic error analysis,
we allow $b_A$ to vary over a range of values\footnote{At $\beta=6.0$,
reference~\cite{LANL-NP} gets $b_A=1.28(3)(4)$, which is within our range.
Note however, that Refs.~\cite{LANL-NP,ALPHA-CSW-CA-ZA} get quite different
values of $c_A$ at $\beta=6.0$, indicating that effects of $\cO(a^2)$
and higher play a significant role at this coupling.}
(see Sec.~\ref{sec:quenched}).

For chiral extrapolations, our canonical procedure (see Sec.~\ref{sec:chiral}) for
both Wilson and clover quarks is to
use the kinetic quark mass $aM_2$ (\eq{treemasses}) as the independent variable.
Through $\cO(a)$, this is equivalent to the clover 
``improved quark mass,'' $\tilde{am_q}=am_0(1+b_mam_0)$, with the choice $b_m=-0.5$.
A nonperturbative determination by the ALPHA collaboration \cite{ALPHA-bm} gives instead
$b_m\approx-0.709$ at $\beta=6.0$ and  
$b_m\approx-0.695$ at $\beta=6.15$; while boosted perturbation theory with
the result of Ref.~\cite{SINTandWEISZ} 
gives $b_m=-0.662$ at $\beta=6.0$ and $b_m\approx-0.655$ at $\beta=6.15$.
In the clover case, we have tried both these sets of $b_m$ values instead 
of our canonical procedure,
but found only negligible 
changes in the central values, errors,
and goodness of fits.  For example, $f_B$ changes by less than $0.4\%$ at $\beta=6.0$
and $0.1\%$ at  $\beta=6.15$.
We do not, therefore,
consider the standard improved quark mass further.

Since \eq{A-NP} is valid only through $\cO(a)$, it is likely to
produce large scaling errors if
applied to the renormalization of the heavy-light
axial current for heavy quarks with $am_{Q,0}\gtwid 1$.  Instead,
the straightforward approach is to use the 1-loop, $\cO(a)$, 
perturbative matching for clover fermions 
as calculated by Ishikawa, Onogi and Yamada \cite{IOY,IOY-COM}. We call
this approach ``NP-IOY,'' where IOY refers to the authors,
and NP indicates nonperturbative, because $c_{SW}$ has the value given
in \cite{ALPHA-CSW-CA-ZA}, and \eq{A-NP} is used in computing
$f_\pi$.  

Since Ref.~\cite{IOY} uses tadpole improvement defined through $\kappa_c$
\cite{IOY-COM},
just as we did above, we may summarize our application
of their result as
\begin{equation}
A_0^{{\rm NP\!-\!IOY}} = Z_A^{\rm IOY}\bar{\tilde q}_I \gamma_0 \gamma_5 \tilde Q_I 
 +Z^{\rm IOY}_{12} \bar{\tilde q}^0_I \gamma_0 \gamma_5  a\vec\gamma\!\cdot\!\! \vec D\,\tilde Q_I^0 \ ,
\label{eq:A-IOY}
\end{equation}
where $\tilde Q_I$ and $\tilde Q_I^0$ are given by \eqs{tildeQ-I}{tildeQ-I0} (and
similarly for $\tilde q_I$ and $\tilde q_I^0$), and where
\begin{eqnarray}
Z_A^{\rm IOY} &=& 1 + \alpha_V(q^*)\rho^{(0)}_A \cr
Z^{\rm IOY}_{12} &=& -\alpha_V(q^*)(\rho^{(1)}_A + \rho^{(2)}_A)/(2a\tilde M_2)\ ,
\label{eq:Z-IOY}
\end{eqnarray}
with $\rho^{(0)}_A$, $\rho^{(1)}_A $, and $\rho^{(2)}_A$ defined 
in \cite{IOY}. We have used the fact that the operators 
$\bar{\tilde q}_I \gamma_0 \gamma_5  \vec\gamma\!\cdot\! \vec D\,\tilde Q_I$
and $-\bar{\tilde q}_I \vec\gamma\!\cdot\!\! \leftvec D\,  
\gamma_0 \gamma_5  \tilde Q_I$
have equal matrix elements between zero-momentum states,
as is the case for our evaluation of decay constants, to combine
the coefficients $\rho^{(1)}_A$ and $\rho^{(2)}_A$.
For a central value of $q^*$, we take the result from the static-light
calculation of Ref.~\cite{CB-TD}, using the appropriate value of
$c_{SW}$ and taking the heavy quark mass to be the mass of the $B$.
This gives $q^*=3.34/a$ at $\beta=6.0$ and $q^*=2.85/a$ at $\beta=6.15$.
Unlike \cite{IOY}, we include the $\tilde d_1$ factor for the light 
quark in the first term of \eq{A-IOY}.
This is just for convenience, since 
$\tilde d_1$ is negligible for our light quarks.
We have neglected $d_1$ for heavy and light quarks in
the correction term (proportional
to $Z^{\rm IOY}_{12}$) in \eq{A-IOY}.  We remark here that with our current data
we use NP-IOY for \fb, \fbs\ but not for \fd, \fds\ because
the approximations  in \cite{IOY} are not applicable near the $D$ mass.

The errors in the NP-IOY calculation are formally 
$\cO(a^2\Lambda^2_{QCD})$ and $\cO(\alpha_V^2)$.  Again,
one should think of these errors as multiplied by an arbitrary
$\cO(1)$ function of $aM_Q$, making the $a$-dependence complicated
in general.  
For example, if $M_Q$ is held fixed and $a$ is varied in the region
where $aM_Q\sim1$,  linear $a$ dependence ($\sim\!a\Lambda^2_{QCD}/M_Q$)
is possible.

Although NP-IOY is a well-defined approach 
to heavy-light physics with nonperturbative clover fermions, it 
does not take advantage for heavy quarks of the nonperturbative information in 
\eq{A-NP}.
An alternative, which has been used by the APE \cite{APE-fB-BB00}
and UKQCD \cite{UKQCD-fB00} collaborations, it to apply \eq{A-NP}
for moderate $am_{Q,0}$, where it is still approximately justified,
and then extrapolate up to the $B$ mass.  That approach has two main
systematic errors: 
First, one has
no guidance from HQET about the order of the polynomial 
in $1/m_Q$ with which to extrapolate to the $B$ mass, and second  
$\cO(a^2)$ errors, while relatively small for the moderate masses
studied, can grow rather larger when extrapolated over a wide mass range.
We prefer instead a different method, which we call
``NP-tad'' for reasons discussed below.  The idea here is to replace
\eq{A-NP} by an equivalent expression through $\cO(a)$ but
which has the advantage of having a reasonable limit for large
$am_{Q,0}$.  The modified \eq{A-NP} is then used directly
at or near the $B$ mass.  

Equation~(\ref{eq:A-NP}) applied to a heavy-light current gives
\begin{eqnarray}
A_{0}^{\rm NP} &=& Z^{\rm NP}_A\sqrt{4\kappa_{Q}\kappa_{q}}
\Big(1+ b_A\frac{am_{Q,0}+ am_{q,0}}{2}\Big)
[A_{0} + c_Aa\partial_0P_{5}
]\ ,\cr
A_{0} &=& \bar q\gamma_0 \gamma_5 Q\ ; \qquad
P_{5} = \bar q \gamma_5 Q \ .
\label{eq:A-NPQq}
\end{eqnarray}
Note that
$A_{0}^{\rm NP}$ does 
not approach a static limit for $am_{Q,0}\to\infty$.
Instead, it goes to $-\infty$ because $c_A<0$ and $\partial_0P_5
\sim \sinh (M_{Qq,1}) \sim am_{Q,0}$. ($M_{Qq,1}$ is the meson pole mass.) 
Even if $c_A$ were zero, $A_{0}^{\rm NP}$
would still blow up because of the term $b_Aam_{Q,0}$.
For this reason, small discretization errors for moderate
$am_{Q,0}$ may be magnified significantly if \eq{A-NPQq} is used
to extrapolate to the $B$.

To define the NP-tad alternative, we first let
\begin{equation}
\label{eq:Rdef}
R(M_{Qq})\equiv \frac {\langle 0|\partial_0P_{5}|Qq \rangle}
{(m_{Q,0}+m_{q,0})\langle 0|A_0|Qq \rangle}\ ,
\end{equation}
where $Qq$ is a generic heavy-light pseudoscalar meson.
Due to a cancellation of $\sinh (M_{Qq,1})$ (from $\partial_0$)
and the explicit $m_{Q,0}$ in the denominator, one expects $R$ has
a finite limit as $am_{Q,0}\to\infty$.  This is confirmed by our
simulations.
Then
\begin{equation}
\label{eq:NPtad-NOd1}
A_{0}^{{\rm NP}'}=Z^{\rm NP}_A\sqrt{4\kappa_q\kappa_Q}
\sqrt{1 + (b_A+2c_AR)am_{Q,0}}\;
\sqrt{1 + (b_A+2c_AR)am_{q,0}}\;
A_{0}
\end{equation}
gives results
for $\langle 0|A_{0}^{{\rm NP}'}|Qq \rangle$ that are
identical to $\langle 0|A_{0}^{{\rm NP}}|Qq \rangle$  through $\cO(a)$.
However,
because $\kappa_Qam_{Q,0}\to 1/2$ as $\kappa_Q\to0$,
\eq{NPtad-NOd1} has a static limit, unlike (\ref{eq:A-NP}).

The mass dependence of \eq{NPtad-NOd1} is in 
fact very similar to the
Fermilab formalism at tadpole-improved tree level. 
Indeed, from \eqs{tildeQ-I}{tildekappa}, the Fermilab version
of \eq{NPtad-NOd1} is
\begin{equation}
\label{eq:FERMILAB-A-R}
A_{0}^{\rm FNAL}=Z_A\sqrt{4\kappa_q\kappa_Q}
\sqrt{1 + am_{Q,0}/u_0}\;
\sqrt{1 + am_{q,0}/u_0}\;
A_{0} \ ,
\end{equation}
where $Z_A= Z_A^{tad}u_0$ is the renormalization constant
for the axial current without the tadpole factor removed,
and where we have dropped the $d_1$ terms for simplicity.
Hence \eq{NPtad-NOd1} is equivalent to \eq{FERMILAB-A-R}
but with a special
(mass-dependent) value for the tadpole factor: $u_0 = (b_A+2c_AR)^{-1}$.
The similarity to tadpole improvement (within the context
of nonperturbative renormalization) is the reason for the
name ``NP-tad.''   Note that, at tree level, where $b_A=Z_A=u_0=1$
and $c_A=0$, NP-tad is in fact identical to the Fermilab approach.

In practice, we put the $d_1$ terms as well as the corresponding perturbative
subtraction back into the axial current.  Thus
\eq{NPtad-NOd1} becomes
\begin{equation}
\label{eq:A-NPtad}
A^{\rm NP\!-\!tad}_{0}=Z^{\rm NP}_A\sqrt{4\kappa_q\kappa_Q}
\sqrt{1 + (b_A+2c_AR)am_{Q,0}}\;
\sqrt{1 + (b_A+2c_AR)am_{q,0}}\;
A^{\rm d1\!-\!sub}_{0}
\end{equation}
with
\begin{equation}
\label{eq:A-d1-sub}
A^{\rm d1\!-\!sub}_{0} \equiv (1 + 
\alpha_V(q^*)\rho^{({\rm sub})}_A) A_{0} + 
\bar q\gamma_0 \gamma_5\;  a\tilde d_1\vec\gamma\!\cdot\!\! \vec D\,Q 
 -\bar q\;  a\tilde d_1\vec\gamma\!\cdot\!\! \leftvec D\,\gamma_0 \gamma_5 Q \ ,
\end{equation}
where $\rho^{({\rm sub})}_A$
is the IOY perturbative correction
coming from the $d_1$ terms alone, which we extract by comparing the complete
perturbative result computed with and without the terms
\cite{IOY-d1term}.  In \eq{A-d1-sub}, $\tilde d_1=  d_1(a\tilde m_{Q,0})$  
and $\tilde d_1=  d_1(a\tilde m_{q,0})$ in the second and third terms, respectively, with $d_1(am_0)$ given in
\eq{treemasses}.  As before, $d_1(a\tilde m_{q,0})$ is negligible and is included
merely for convenience.

Equations~(\ref{eq:A-NPtad}) and (\ref{eq:A-d1-sub}) define the NP-tad approach.
The errors of NP-tad are formally $\cO(a^2\Lambda_{QCD}^2)$ and $\cO(a^2M_Q^2)$.
Thus, the errors could in principle be large as $M_Q$ increases at
fixed $a$.  The hope is that the requirement that the
decay constants have a limit as $M_q\to\infty$
has forced the $a^2M_Q^2$ (and higher) terms to have small coefficients, 
but this is not proven.
We emphasize that the NP-tad approach is logically neither better
nor worse than the method of Refs.~\cite{APE-fB-BB00,UKQCD-fB00}
for $B$ physics. NP-tad is an attempt to keep 
the nonperturbative $\cO(a)$
information and yet include some higher
effects in a smooth way, 
but there is no {\it a priori} guarantee that 
all or most of such effects are included.  {\it A posteriori}, one can
compare how well the methods scale with lattice spacing.  The scaling of
NP-tad results currently appears to be comparable to that seen with 
standard nonperturbative normalization and extrapolation
in Ref.~\cite{UKQCD-fB00}.\footnote{An earlier version of Ref.~\protect{\cite{UKQCD-fB00}}
showed considerably worse scaling, but that appears to have been associated more with
the use of scale $r_0$ from the potential than with the normalization and extrapolation.}
With only two lattice spacings in the NP-tad data and in \cite{UKQCD-fB00}, 
however, this
comparison is far from definitive.
We therefore use the NP-IOY approach (whose errors are 
better understood), in addition to both the NP-tad method and
standard Wilson fermions, in determining the central 
value and errors.

\section{Computational Details}
\label{sec:details}

\subsection{Lattice Generation and Inversion}
\label{sec:generation}

Table~\ref{tab:lattices} shows the lattice parameters used.  
Quenched lattices are generated using a standard combination
of pseudo heat bath \cite{KENNEDY-PENDLETON} and 
overrelaxed \cite{ORA} updates. 
Successive configurations
are separated by 200 iterations, where each iteration consists of 1 heat
bath and 4 (9 for set H, $\beta=6.52$) overrelaxed sweeps.
The sets J and
CP are new additions to the quenched lattices previously analyzed in
Ref.~\cite{MILC-PRL}. 
Dynamical fermion lattices were
separated
by 10 trajectories (each of unit molecular dynamics time)
of the R algorithm \cite{RALGORITHM}.
(Set G, from HEMCGC \cite{HEMCGC}, is separated
by 10 trajectories of time $1/\sqrt{2}$ in MILC units.)

\begin{table}[tbh!]
\caption{Lattice parameters. The upper group corresponds to quenched
lattices; the lower group, to dynamical lattices with $N_f = 2$
staggered quarks. The set G was generated by HEMCGC \cite{HEMCGC}.
Sets marked by $*$ are new since Ref.~\cite{MILC-PRL}.
Heavy and light Wilson quark propagators were generated on all sets except J
and 5.7-large. On the latter set, which includes lattices of various
sizes,  only light quark propagators were generated.
Normal (``thin-link'') clover propagators were computed on set J and CP1, 
a 199 lattice subset of CP.
(See Table~\protect{\ref{tab:NPparams}}
for thin-link clover parameters.)
Fatlink clover propagators were generated on set CPF (a
99 configuration subset of CP1) and RF (a
98 configuration subset of set R). 
}
\label{tab:lattices}
\begin{tabular}{|c|l|c|c|}
\noalign{\vspace{1cm}}
\hline
name & $\bt\  (a m_q)$ & size & $ \# $ confs. \\
\hline
A & 5.7 & $8^3\times48$ & 200 \\
B & 5.7 & $16^3\times48$ & 100 \\
5.7-large & 5.7 & \ $12^3\times 48$, $16^3\times 48$,  $20^3\times 48$, 
$24^3\times 48$ & \ 403, 390, 200, 184 \\
E & 5.85 & $12^3\times48$ & 100 \\
C & 6.0 & $16^3\times48$ & 100 \\
$*$CP & 6.0 & $16^3\times48$ & 305 \\
$*$J & 6.15 & $20^3\times64$ & 100 \\
D & 6.3 & $24^3\times80$ & 100 \\
H & 6.52 & $32^3\times100$ & 60 \\
\hline
L & 5.445 (0.025) & $16^3\times48$ & 100 \\
N & 5.5 (0.1) & $24^3\times64$ & 100 \\
O & 5.5 (0.05) & $24^3\times64$ & 100 \\
M & 5.5 (0.025) & $20^3\times64$ & 199 \\
P & 5.5 (0.0125) & $20^3\times64$ & 199 \\
$*$U & 5.6 (0.08) & $24^3\times64$ & 202 \\
$*$T & 5.6 (0.04) & $24^3\times64$ & 201 \\
$*$S & 5.6 (0.02) & $24^3\times64$ & 202 \\
G & 5.6 (0.01) & $16^3\times32$ & 200 \\
R & 5.6 (0.01) & $24^3\times64$ & 200 \\
\hline 
\end{tabular}
\end{table}

Unimproved Wilson valence fermion propagators were
generated for all these sets except J.  (On set 5.7-large
only light Wilson propagators were computed.)
Because our calculations were initially limited by slow I/O speeds and lack
of long-term storage, we performed
the calculations of heavy-light meson propagators ``on the fly''
 --- \ie without storing quark propagators.
The hopping parameter expansion of the
heavy quark propagator, as proposed by Henty and Kenway \cite{HENTY},
makes this possible.  Further, the expansion allows us
to study a large number of heavy quark masses, with almost no additional
expense.  For this reason, we continued using the approach of
Ref.~\cite{HENTY} even after faster I/0 and better storage became
available.

In the hopping-parameter approach,
the light quark propagator, for a single
spin-color source, is first computed with standard
methods
(red-black preconditioning;
minimal residual). 
The heavy quark propagator for the same spin-color source
is then computed order by order in the heavy hopping
parameter.  At each order, the contribution
to the meson propagators, summed over space, is stored
to disk.  The full meson propagator for any heavy
hopping parameter, $\kappa_Q$, can then be reconstructed
after the fact by multiplying the stored results
by appropriate powers of $\kappa_Q$ and summing over
iterations as well as
spin and color.
Propagators in the static-light limit, where the
heavy quark mass is taken to infinity, can be
obtained as a by-product of this
procedure.

Our quark sources are Coulomb-gauge Gaussians.
We run the overrelaxed gauge fixer until the
sum of the trace
of all spacelike links (normalized to $1$ when all links
are unit matrices)
changes by less than $7\times10^{-7}$ per pass.
This takes, for example, about 435 passes on set D, and a comparable
number of passes on the other sets.

\begin{figure}[thb!]
\includegraphics[bb = 0 0 4096 4096,
width=4.0truein,height=4.0truein]{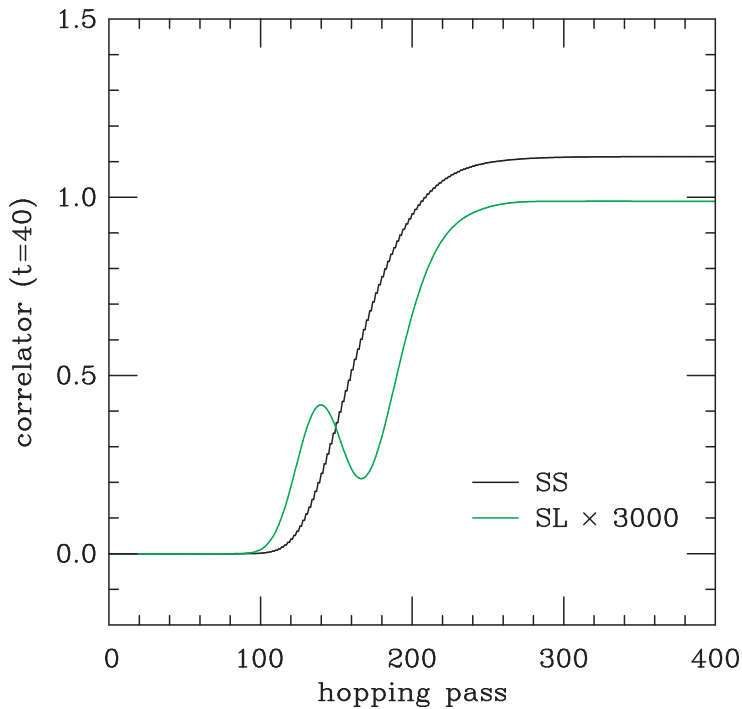}
\caption{Convergence of the hopping parameter expansion for
heavy-light pseudoscalar mesons correlators on set D.
$\kappa_Q=0.1456$; while $\kappa_q=0.1507$, the lightest of the three 
light quarks
analyzed on this set.  The values of the smeared-smeared and smeared-local correlators
at $t=40$ (half-way across the lattice) are shown with the solid and
dashed lines, respectively.
}
\label{fig:hopping}
\end{figure}

Our Wilson light quark propagators are computed
for three values of $\kappa_q$, giving light quark
masses ($m_q$) in the range $0.7 m_s \ltwid m_q \ltwid 2.0 m_s$,
where $m_s$ is the strange quark mass.  
We analyze heavy-light mesons with 
8 to 10 heavy quark masses
per data set, with heavy-light pseudoscalar meson masses ($M_{Qq}$)
in the range $1.25\;\GeV \ltwid M_{Qq} \ltwid 4.0\;\GeV$.
The heavy quark propagators are computed with
400 passes of the hopping parameter expansion.  
Figure~\ref{fig:hopping} shows the convergence on set D of the 
hopping parameter expansion for heavy-light meson correlators at the
maximum time separation (half-way across the lattice).  
The value of $\kappa_Q$ ($0.1456$) used in
Fig.~\ref{fig:hopping} gives a meson mass  of $M_{Qq}\sim\!1.1\; \GeV$ when the light
quark $\kappa_q$ is extrapolated to $\kappa_{u,d}$. Since $1.1\; \GeV$ is slightly
lighter than the lightest value used in our analysis, we 
are confident that the expansion is under control.

Because the heavy-light mesons must be constructed at each of
the 400 orders of the hopping parameter expansion,
it is too expensive to sum the central point of the smeared sinks
over the entire spatial volume, even using FFT's.
Instead, we simply sum
over 16 points in the $L^3$ spatial volume:
the 8 points $(0,0,0)$, $(L/2,0,0)$, $(0,L/2,0)$, $\dots$,
$(L/2,L/2,0)$, $\dots$,
plus the 8 points obtained by adding $(L/4,L/4,L/4)$ to each of the
previous points.
This fixes the lowest non-zero momentum which
contributes to be $(2,2,0)$ (and permutations) in units of $2\pi/L$.
For the heavy-light mesons studied here, these
higher  momentum states are suppressed  sufficiently at asymptotic
Euclidean
time by their higher energy.
However, in the largest physical
volumes, sets  N, O, U, and T, the higher momentum states for the heaviest mesons
are quite close to the ground state,
and we are required to go to large
times ($t_{\rm min}/a\sim 22$--$27$) for the smeared-smeared
propagators in order to make single exponential
fits with good confidence levels.

The static-light mesons have no such suppression,
and the contribution of higher momentum states is limited only
by their overlap with the sources.
Using computed
static-light wavefunctions \cite{WAVEFUNCTIONS},
we find that the contamination in static-light decay constants from
nonzero momentum states is small ($\approx 0.7\%$)
for lattices with spatial size of $\approx 1.5$ fm
(sets A, C, CP, D, E, G, H). However,
on all other (larger) sets the contamination is expected
to be large.  We therefore have performed a dedicated static-light
computation on those lattices, with relative smearing functions taken
from \cite{KENTUCKY} and zero momentum intermediate states enforced
by a complete FFT sum over spatial slices.  In addition, the dedicated
static light computation has been run on sets A and E (because the plateaus from
the hopping method proved to be poor) and sets G and CP (as a check of the
hopping method).  On the latter sets, the two methods give consistent
results,  but we choose the dedicated method 
because the errors are smaller.  Thus, only on sets  C, D, and H is
the hopping approach used for the final analysis of the
static-light mesons.

Standard (``thin-link'') clover improved valence quark propagators
have been generated using stabilized bi-conjugate gradient inversion \cite{BICGS} (for light
quarks) and the hopping parameter expansion (for heavy quarks) 
on quenched sets J and ``CP1,'' a  199-lattice subset
of CP.  
In this case we have 5 light and 5 heavy quark masses, in the same range 
as for the Wilson valence quarks.  For the heavy quarks we
sum, on the fly, the orders in the hopping expansion for a given $\kappa_Q$ 
--- \ie this is a standard inversion, which does not allow
a choice of arbitrary $\kappa_Q$ after the fact.  However, 
our experience with the Wilson
case leads us to believe that a large number of heavy quark
masses is unnecessary: the behavior of $f_{Qq}\sqrt{M_{Qq}}$
with $1/M_{Qq}$ is quite smooth. Further, we are now able
to perform an FFT sum of the meson propagators, so that
zero momentum is enforced and contamination from 
excited states is reduced.   The full Fermilab
formalism allows us to choose heavy quark masses near the
$b$ quark mass in this case; we therefore did not need to compute
static-light mesons here to stabilize an extrapolation.

As explained in Sec.~\ref{sec:clover-formalism},
we take the ALPHA collaboration \cite{ALPHA-CSW-CA-ZA}
values, where available, for the 
normalization and improvement
constants of our clover fermions.  The parameters used are shown in 
Table~\ref{tab:NPparams}.

\begin{table}[tbh!]
\caption{Parameters used for nonperturbative,
thin-link clover fermions (quenched configurations).}
\label{tab:NPparams}
\begin{tabular}{|c|c|c|c|c|c|c|c|}
\hline
set & $\beta$ & $\#$confs. & $c_{SW}$ & $Z_A$ & $ c_A$ & $b_A$  &$b_A$ range \\
\hline
CP1 & 6.0 &  199& 1.769 & 0.7924 & $-0.0828$ & 1.472 & 1.256$\to$1.586   \\
J & 6.15 & 100 & 1.644 & 0.8050 & $-0.0426$ & 1.423 & 1.244$\to$1.510   \\
\hline
\end{tabular} 
\end{table}

\subsection{Covariant Fits}
\label{sec:covariant}

We need to fit correlators in time, 
extrapolate/interpolate in light and heavy quark masses,
and extrapolate in lattice spacing to the continuum. 
In all cases except the last, the data is correlated,
so covariant fits are preferable.
As is well known \cite{SMALL-EIGENVALUES},  however, it requires a large 
statistical sample to determine accurately
the small eigenvalues of the covariance matrix.  With limited
statistics, such eigenvalues will be poorly determined and 
can make the covariant fits unstable. This is a particular problem
in the current analysis because the large time dimension
of our lattices and the fact that we fit two channels
simultaneously means that we often make
fits with 25 or more degrees of freedom.

The technique we use to deal with the problem of small eigenvalues
has many similarities to the methods proposed in \cite{SMALL-EIGENVALUES}
but has some advantages in our analysis.
It is based on a standard approach in factor analysis \cite{FACTOR}.
We first compute the correlation matrix (the covariance matrix, but normalized by
the standard deviations to
have $1$'s along the diagonal) and find its eigenvalues and eigenvectors.
We then reconstruct the correlation matrix from the eigenvectors, but omitting
those corresponding to eigenvalues less than $\lambda_{\rm cut}$, a cutoff.  
The resulting matrix is of course singular.  It is made into an acceptable correlation
matrix by restoring the $1$'s along the diagonal. Finally, the corresponding
covariance matrix is constructed (by putting back the
standard deviations), inverted, and used in the standard
way for making the fits.

The above technique interpolates smoothly between standard covariant fits,
where no eigenvalues are omitted, and noncovariant (uncorrelated) fits, where all
eigenvalues are omitted.  Furthermore, because the correlation
(as opposed to covariance) matrix
is used, the eigenvalues are normalized, with the average eigenvalue
always equal to 1.  This allows us to make a uniform determination
of which eigenvalues to keep, which is very important since we are dealing
with thousands of fits, and it is impossible to examine each fit by
hand.  Our standard procedure is to choose $\lambda_{\rm cut}= 1$, \ie we drop all 
eigenvalues less
than 1. The eigenvectors kept typically account for $90$--$95\%$ of the total covariance.
Indeed, when one changes how the covariance matrix is computed,
(for example, by increasing the number of configurations eliminated
in the jackknife), the eigenvalues smaller than 1 generally change
drastically with our typical sample size of $\sim\!100$.
The approach eliminates unstable, ``pathological'' fits completely.

We have checked that the final results are not significantly affected
when we keep several more (or several fewer) eigenvalues throughout.
Furthermore, on our set with the greatest statistics (set CP, 
305 configurations) we are able to compare with a wide variety
of different cuts on the eigenvalue, as well as standard covariant
fits where all eigenvalues are kept.  We find that central values
almost always agree within one statistical sigma, and usually differ
by much less than that.  In data discussed below (and tabulated
at 
{\tt http://www.physics.wustl.edu/$\!\sim$cb/Nf=2\_tables})
we show for comparison
fits with different eigenvalue cuts for this set.

One disadvantage of the current approach, as well as to the methods
in \cite{SMALL-EIGENVALUES}, is that there is no true quantitative
measure of ``goodness of fit.'' 
When eigenvalues are removed, the truncated chi-squared, $\chi_{\rm cut}^2$,
tends to be (but is not always)
considerably smaller than the ``true'' $\chi^2$ from
a complete covariant fit.  However, our experience has shown that requiring
$\chi_{\rm cut}^2/{\rm d.o.f.}<1$ with  $\lambda_{\rm cut}= 1$ produces fits that
are almost always acceptable by a standard criterion (${\rm CL} > 0.05$, with
CL the confidence level) when the data allow a fully covariant (uncut) fit.
This is not the case for noncovariant fits ($\lambda_{\rm cut}= \infty$).
Such fits may have $\chi_{\rm cut}^2/{\rm d.o.f.}\ll 1$ and yet extend to
a fully covariant fit with extremely small CL.  For example,  exponential
fits to a correlator which include several points clearly outside the plateau
region can still have $\chi_{\rm cut}^2/{\rm d.o.f.}\ll 1$ when $\lambda_{\rm cut}= \infty$,
but not, in our experience, when $\lambda_{\rm cut}= 1$.

At every stage in the analysis, we compute statistical errors by the
jackknife procedure. The covariance matrices are also computed by
jackknife. For the quenched sets, there is no evidence of a non-zero
autocorrelation length.   However, in the dynamical sets, the 
errors typically increase with the number of configurations omitted
in the jackknife until $\sim\! 4$ configurations are omitted.  To be
conservative, we determine our statistical errors and covariance matrices by
omitting 8 configurations at a time (for both dynamical and quenched sets).

\subsection{Correlator Fits and Extraction of Decay Constants}
\label{sec:correlators}

We compute ``smeared-local'' and ``smeared-smeared'' pseudoscalar
meson propagators in each of three cases: heavy-light, static-light,
and light-light (the last with degenerate masses only).
These correlators are defined by
\begin{eqnarray}
\label{eq:GSL}
G_{SL}(t) &=& \sum_{\vec x}\langle 0 | A^R_0(\vec x,t) 
\chi_{5}^\dagger(\vec 0,0) |0\rangle
\\
\label{eq:GSS}
G_{SS}(t) &=& {\sum_{\vec x}}'\langle 0 | \chi_{5}(\vec x,t) 
\chi_5^\dagger(\vec 0,0) |0\rangle\ ,
\end{eqnarray}
where $A^R_0$  is the relevant 
renormalized current, 
namely $A^{\rm KUR}_0$ (\eq{A-KUR}), $A^{\rm STAT}_0$ (\eq{A-STAT}),
$A^{\rm NP\!-\!IOY}_0$ (\eq{A-IOY}), 
$A^{\rm NP\!-\!tad}_0$ (\eq{A-NPtad}),  or $A^{\rm NP}_0$ (\eq{A-NP} --- for light-light
quantities only). 
For $G_{SS}$, the prime on the sum
indicates that only $N_{\rm sink}$ points on a time-slice are included.
As discussed above, for all the Wilson heavy-light and
light-light data, $N_{\rm sink}=16$.
For the clover and the dedicated static computations, the complete
sum is performed (with FFT), so one has $N_{\rm sink}=V\equiv n_xn_yn_z$.
In \eqs{GSL}{GSS},
$\chi_{5}$ is the Gaussian pseudoscalar
source, given by
\begin{equation}
\label{eq:chi5}
\chi_{5}(\vec x,t) = \sum_{\vec y \vec z} e^{-\vec y^2/r_0^2}
e^{-\vec z^2/r_0^2} \bar q(\vec x + \vec y)\gamma_5
Q(\vec x + \vec z) \ .
\end{equation}
The width $r_0$ varies from 2.33 lattice
spacings (set A) to 8 spacings (set H) and is chosen to be
roughly  0.35 fm.
For the Henty-Kenway hopping calculations, the sums in \eq{chi5}
run over even $\vec y$, $\vec z$ only, so that we may exploit
an even-odd decomposition.
In \eq{chi5} and below, we use the notation of the heavy-light case
(quarks $Q$ and $q$) generically:  for light-light formulae, let $Q\to q$; 
for static-light formulae, let $Q\to h$.

\begin{figure}[thb!]
\includegraphics[bb = 0 0 4096 4096,
width=5.0truein]{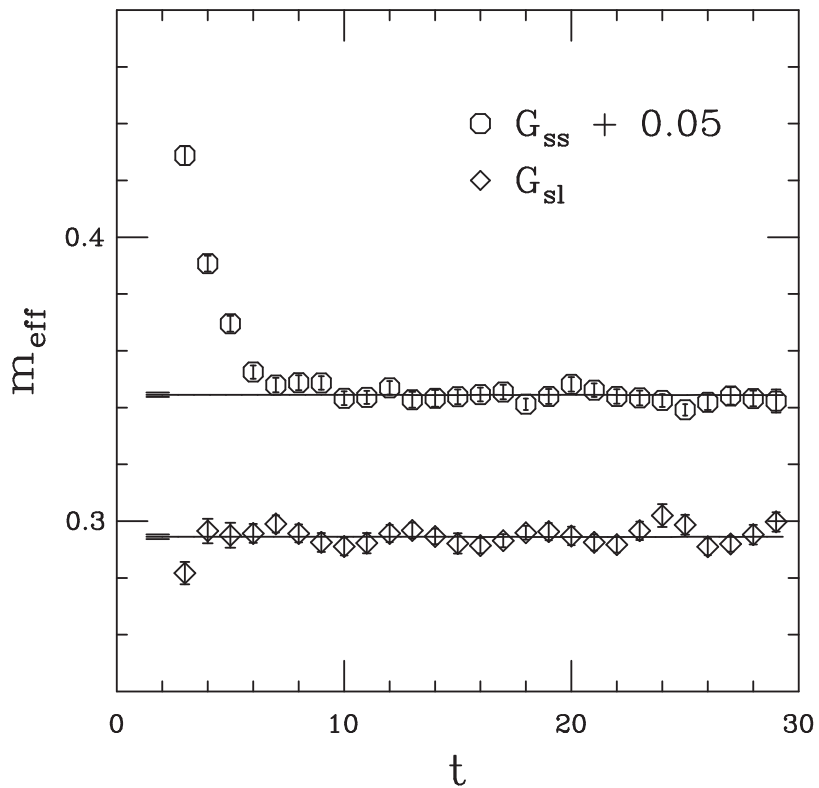}
\caption{Light-light effective masses for set R, $\kappa=0.159$.
The fit ranges are 8 to 31 for $G_{SL}$ and 10 to 31 for
$G_{SS}$.
The smeared-smeared masses are shifted upward for clarity.
The long horizontal lines show the fit value of the 
mass.  The error in the fit value is indicated at the left
end of the fit lines by two short horizontal lines.
}
\label{fig:meff-qq}
\end{figure}

\begin{figure}[thb!]
\includegraphics[bb = 0 0 4096 4096,
width=5.0truein]{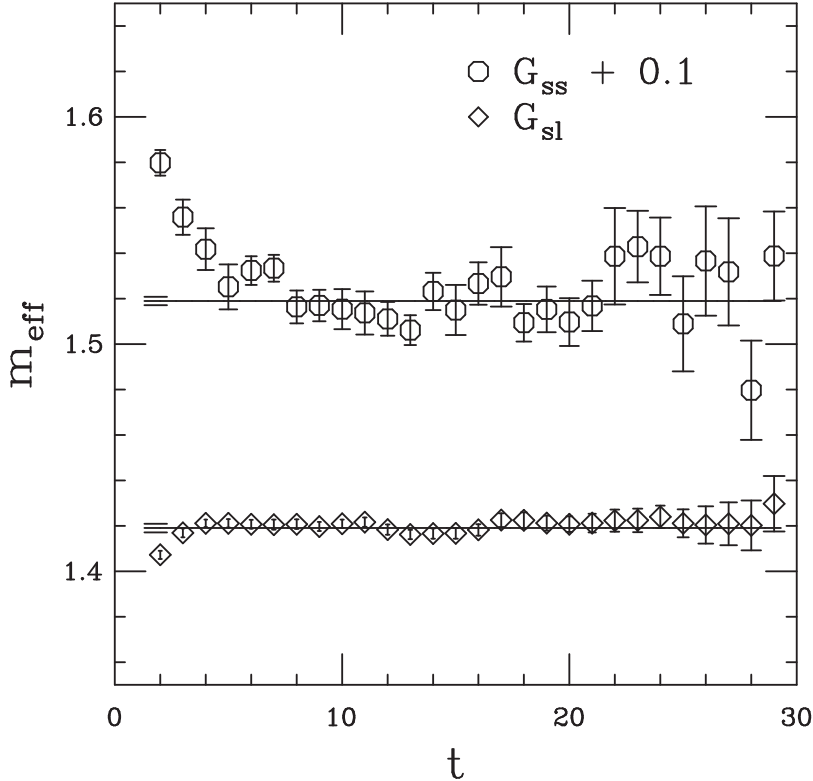}
\caption{Same as Fig.~\ref{fig:meff-qq}, but
for  heavy-light effective masses on set J, 
$\kappa_Q=0.098$, $\kappa_q=0.1347$.
The fit ranges are 12 to 18 for $G_{SL}$ and 8 to 16 for
$G_{SS}$.
}
\label{fig:meff-Qq-clover}
\end{figure}

\begin{figure}[thb!]
\includegraphics[bb = 0 0 4096 4096,
width=5.0truein]{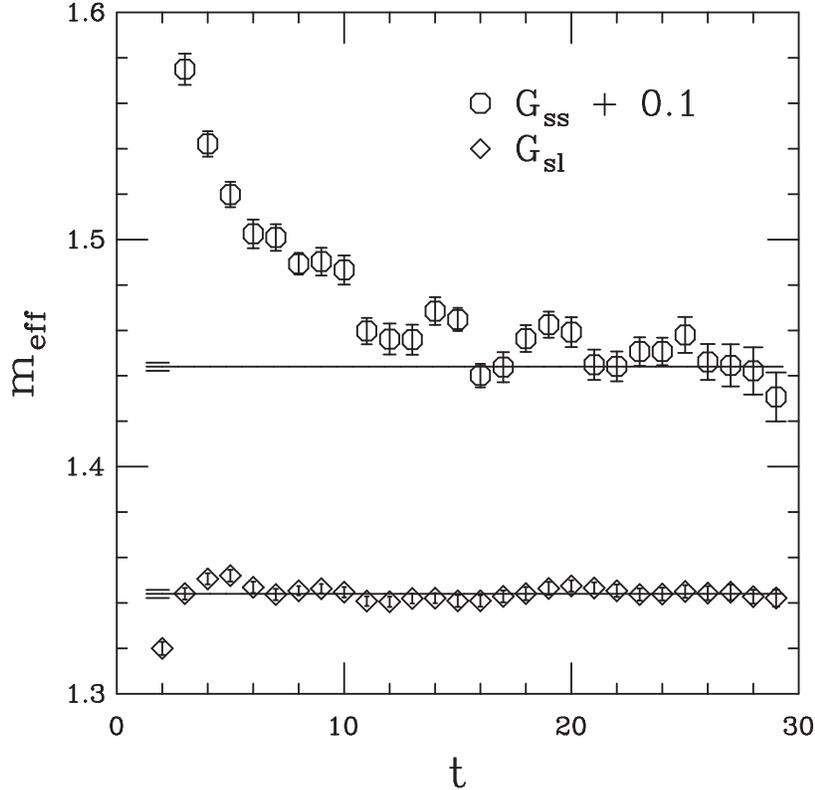}
\caption{Same as Fig.~\ref{fig:meff-qq}, but
for  heavy-light effective masses for set M, $\kappa_Q=0.120$,
$\kappa_q=0.160$.
Fit ranges are 13 to 31 for $G_{SL}$ and 19 to 31 for
$G_{SS}$.
The heavy quark is computed with the Henty-Kenway hopping expansion.
The late plateau in $G_{SS}$ is due to contamination from non-zero
momentum states because the sink is not summed over complete time slices.
}
\label{fig:meff-Qq-hop}
\end{figure}

For large Euclidean time $t$, $G_{SL}$ and $G_{SS}$ are fit 
simultaneously and covariantly 
to single exponential forms,
with the same mass in both channels
\begin{equation}
\label{eq:zetas}
G_{SL} \to \zeta_{SL} e^{-Mt}\; ;
\qquad
G_{SS} \to \zeta_{SS} e^{-Mt}\ .
\end{equation}
In other words, these are fits with three-parameters: $M$,
$\zeta_{SL}$ and $\zeta_{SS}$.
Central values 
use $\lambda_{\rm cut}= 1.0$ throughout, except for
set CP, where $\lambda_{\rm cut}= 0.1$.  Typical effective
mass plots for the light-light and heavy-light cases are
shown in Figs.~\ref{fig:meff-qq}, \ref{fig:meff-Qq-clover} and
\ref{fig:meff-Qq-hop},
respectively.  Here and below, we generally choose $N_f=2$ and clover
data for the plots because the quenched Wilson data has been discussed in more
detail previously \cite{MILC-PRL,PRELIM-MILC}.

We vary the fit range (in $t$) in each channel over
several choices that appear to be in the asymptotic, ``plateau,'' region
for the effective mass.
Combining such choices for the light-light, heavy-light and static-light
cases, we have approximately 25 different
versions of the analysis on each data set.  Our central values are taken
from the version which has the best blend of 
small statistical errors and low, or at least acceptable,
values of $\chi^2_{\rm cut}/{\rm d.o.f.}$.
Here ``acceptable'' is defined, with few exceptions, as 
$\chi^2_{\rm cut}/{\rm d.o.f.}< 1.3$,
with $\lambda_{\rm cut}=1$. (In about 85\% of the central value choices,
$\chi^2_{\rm cut}/{\rm d.o.f.}< 1$.)  The exceptions are: (1) a few fits to heavy-light
mesons outside the ranges of mass that we include in our final determinations
of the decay constants, (2) a few fits to heavy-light mesons on the largest lattices,
where the slow approach to the asymptotic regime for smeared-smeared correlators
computed with the Henty-Kenway approach (see Sec.~\ref{sec:generation}) left us
with somewhat noisy data at large $t$, and (3) the fit to the static-light meson
with heaviest light quark on set G, which had rather noisy data.
In these exceptional cases, we relax our definition of acceptable
$\chi^2_{\rm cut}/{\rm d.o.f.}$ to be less than 1.9, 1.7, and 1.6, respectively.

Of course, with the approximately 900 channels we fit, one should expect that
some fits over truly asymptotic ranges will have poor $\chi^2/{\rm d.o.f.}$
simply because of statistical fluctuations.  However, since our 
$\chi^2_{\rm cut}$ generally underestimates the full $\chi^2$, we have tried to
make choices which are more conservative than a standard criterion, of say,
confidence level $> 0.05$.  

From the fits,
the pseudoscalar decay constants, $f_{Qq}$ for given quark masses
are then obtained {\it via}
\begin{equation}
\label{eq:fQq}
f_{Qq}\sqrt{M_{Qq}} = \frac{\sqrt{2}}{a^{\frac{3}{2}}} 
 \sqrt{\frac{N_{\rm sink}}{V}} \frac{\zeta_{SL}}
{\sqrt{\zeta_{SS}}} \ .
\end{equation}
where we use the definition of the decay constant
that gives $f_\pi=130.7\;\MeV$:
\begin{equation}
\label{eq:fdef}
\left\langle 0\left| A^{\rm cont}_{0}\right|
Qq,\vec{p}=0\right\rangle \equiv -if_{Qq}M_{Qq}\ ,
\end{equation}
with $A^{\rm cont}$ the continuum axial 
current.\footnote{When \protect{\eq{fQq}}
is used for static-light mesons with the Henty-Kenway hopping
approach, an extra factor of $\sqrt{2}$ is required
on the right hand side.  This arises from the fact
that the highest momentum state $(\pi,\pi,\pi)$ 
aliases the zero momentum state with our even-site-only
source, and the higher momenta are not suppressed in Euclidean
time for the static case.}

Data for the masses and decay constants for each of
the sets listed in Table~\ref{tab:lattices} are
posted at
{\tt http://www.physics.wustl.edu/$\!\sim$cb/Nf=2\_tables}\ .
The files ``{\it name}\_qq.dat,'' where ``{\it name}'' is the
set name (A, B, E, \dots) give dimensionless
light-light masses and decay
constants as a function of hopping parameter.  
Similarly ``{\it name}\_Qq.dat'' and ``{\it name}\_Statq.dat''
give dimensionless masses
and values for $a^{3/2}f_{Qq}\sqrt{M_{Qq}}$ for heavy-light
and static-light mesons, respectively. 
For set CP, additional
files, with ``lambda-cut=X'' appended to the name, show the
effect of various truncations of the correlation matrix.
For heavy-light mesons the masses
tabulated are the shifted masses $aM_{Qq,2}$, \eq{mhl2};
for the static-light case they are simply the pole masses.
Time ranges, number of degrees of freedom, and 
$\chi^2_{\rm cut}$ are included for all fits.  For sets CP1 and J, 
the nonperturbative clover lattices, ``{\it name}''
gets a further qualifier in the heavy-light case, which is either ``NP-IOY'' or ``NP-tad''
for the two types of renormalization performed.

To enable the reader to see the effects of various
renormalizations used, as well as to make possible reanalysis
of the data by other groups, we have tabulated
additional raw data. For all sets, we have separately computed
correlators of the bare lattice axial current $A_0$:
\begin{equation}
\label{eq:GSL-bare}
G^{bare}_{SL}(t) = \sum_{\vec x}\langle 0 | A_0(\vec x,t) 
\chi_{5}^\dagger(\vec 0,0) |0\rangle\; ; \qquad A_0 = \bar q\gamma_0\gamma_5 Q .
\end{equation}
We fit $G^{bare}_{SL}$ simultaneously with $G_{SS}$, as in
\eq{zetas}, giving us the three quantities 
$M$,
$\zeta^{\rm bare}_{SL}$ and $\zeta_{SS}$.
We then define 
$\Xi^{\rm bare}$ by
\begin{equation}
\label{eq:Xi-bare}
\Xi^{\rm bare} = \sqrt{\frac{N_{\rm sink}}{V}} \frac{\zeta^{\rm bare}_{SL}}{
\sqrt{\zeta_{SS}}} \ .
\end{equation}
$\Xi^{\rm bare}$ is basically the unrenormalized decay constant.
For example, in the case of our heavy-light Wilson data, for which
the renormalized current is given by \eq{A-KUR}, we have
\begin{equation}
\label{eq:fQq-bare}
f_{Qq}\sqrt{M_{Qq}} = \frac{\sqrt{2}}{a^{\frac{3}{2}}} \; Z_A^{\rm KUR}\;
\sqrt{1-\frac{3\kappa_Q}{4\kappa_c}}\;
\sqrt{1-\frac{3\kappa_q}{4\kappa_c}}\; \Xi^{\rm bare} \ .
\end{equation}
The masses and quantities $\Xi^{\rm bare}$ are posted in the files
``{\it name}\_qq\_bare.dat,'' 
``{\it name}\_Qq\_bare.dat,'' ``{\it name}\_Statq\_bare.dat''.\footnote{Again,
for static-light correlators computed with the hopping approach,
we include an extra factor of $\sqrt{2}$ on the right-hand side
of \eq{Xi-bare}.
In this case the heavy-light masses tabulated are the (pole) 
masses directly from the fits, not the shifted masses.}

For heavy-light mesons with the
nonperturbative clover action, we have posted additional intermediate
data. We define
\begin{eqnarray}
\label{eq:A-dim4}
A^{\rm dim4}_0 &\equiv& \bar q \gamma_0\gamma_5  a\vec\gamma\!\cdot\!\! \vec D\, Q \\
\label{eq:A-d1}
A^{d_1}_0 &\equiv& A_0 + \Big(d_1(a\tilde m_{Q,0}) + 
d_1(a\tilde m_{q,0}) \Big) A^{\rm dim4}_0 \\
\label{eq:A-d1-sub-again}
A^{d_1{\rm \!-\!sub}}_{0} &=& (1 + 
\alpha_V(q^*)\rho^{({\rm sub})}_A) A_{0} + 
\Big(d_1(a\tilde m_{Q,0}) + 
d_1(a\tilde m_{q,0}) \Big) A^{\rm dim4}_0 \\
\label{eq:A-imp}
A^{\rm imp}_0 &\equiv& 
\sqrt{1 + (b_A+2c_AR)am_{Q,0}}\;
\sqrt{1 + (b_A+2c_AR)am_{q,0}}\;
A^{d_1{\rm \!-\!sub}}_{0} \ ,
\end{eqnarray}
where ``imp'' stands for ``improved,''
and where \eq{A-d1-sub-again} is a rewriting of \eq{A-d1-sub},
using the fact that 
the operators 
$\bar{\tilde q}_I \gamma_0 \gamma_5  \vec\gamma\!\cdot\!\! \vec D\,\tilde Q_I$
and $-\bar{\tilde q}_I \vec\gamma\!\cdot\!\! \leftvec D\,  
\gamma_0 \gamma_5  \tilde Q_I$ have equal zero-momentum matrix
elements.
For each current in eqs.~(\ref{eq:A-dim4})--(\ref{eq:A-imp}), we
define a corresponding $\Xi$, as in \eq{Xi-bare}.  Results
for $\Xi^{\rm dim4}$, $\Xi^{d_1}$,
$\Xi^{d_1{\rm \!-\!sub}}$ and 
$\Xi^{\rm imp}$ are tabulated in the files ``{\it name}\_Qq\_intermediate.dat'',
where ``{\it name}'' now is just CP1 or J.
From \eqsfour{A-IOY}{tildeQ-I}{A-dim4}{A-d1}, 
these quantities are related to the decay constants in
the NP-IOY case by
\begin{equation}
\label{eq:Xi-to-f-NPIOY}
f_{Qq}\sqrt{M_{Qq}} = \frac{\sqrt{2}}{a^{\frac{3}{2}}}\; 
\sqrt{1-\frac{3\kappa_Q}{4\kappa_c}}\;
\sqrt{1-\frac{3\kappa_q}{4\kappa_c}}\; \Bigg[
Z_A^{\rm IOY}\; \Xi^{d_1} + Z^{\rm IOY}_{12} \Xi^{\rm dim4} \Bigg].
\end{equation}
Similarly, from \eqsthree{A-NPtad}{A-d1-sub-again}{A-imp}, we have in
the NP-tad case:
\begin{eqnarray}
&&f_{Qq}\sqrt{M_{Qq}} = \frac{\sqrt{2}}{a^{\frac{3}{2}}}  \; Z_A^{\rm NP}
\sqrt{4\kappa_Q\kappa_q}\; \Xi^{\rm imp} 
\label{eq:Xi-to-f-NPtad}
\\
&& \quad = \frac{\sqrt{2}}{a^{\frac{3}{2}}} \; Z_A^{\rm NP}
\sqrt{4\kappa_Q\kappa_q} \sqrt{1 + (b_A+2c_AR)am_{Q,0}}\; 
\sqrt{1 + (b_A+2c_AR)am_{q,0}}\;
\Xi^{d_1{\rm \!-\!sub}}.
\label{eq:Xi-to-f-NPtad-raw}
\end{eqnarray}
In practice, when \eqs{zetas}{fQq} are used to compute $f_{Qq}\sqrt{M_{Qq}}$, 
\eqs{Xi-to-f-NPIOY}{Xi-to-f-NPtad-raw} are obeyed only
up to small corrections.
This is because two separate fits are performed to compute the two terms in
\eq{Xi-to-f-NPIOY}; whereas the quantities are added first and then
fit in \eq{zetas}.  For \eq{Xi-to-f-NPtad-raw}, the discrepancy is due to 
the fact that the factors like $\sqrt{1 + (b_A+2c_AR)am_{Q,0}}$ and their
errors are not included in the fit here, but are factored in later.

Finally, we 
also compute 
smeared-local light-light vector meson propagators.
These are fit 
covariantly to single exponentials (two parameter fits).
Raw data for the vector channel appears in the files
``{\it name}\_qq-vector.dat''.

\subsection{Chiral Extrapolations}
\label{sec:chiral}
Chiral extrapolations/interpolations are needed for the light-light pseudoscalars,
which are used to set the scale (through $f_\pi$) 
and to find $\kappa_c$ and the physical values of $\kappa_{u,d}$ and $\kappa_s$, the
hopping parameters of the up/down and strange quarks.
(We generally determine $\kappa_s$ by adjusting the degenerate pseudoscalar mass to 
$\sqrt{2m_K^2 - m_\pi^2}$,
the tree-level chiral perturbation theory value.)
The light-light vectors provide 
alternative determinations of the scale (through $m_\rho$)
and $\kappa_s$ (through $m_\phi$) and require additional
chiral extrapolations.
The heavy-light and static-light masses and decay 
constants also need to be
extrapolated/interpolated  in light quark mass
to the up/down and strange quark masses. 

We have tried chiral extrapolations using either the
bare light quark mass $m_{q,0}$ 
or the light quark
tadpole-improved kinetic mass $\tilde M_{q,2}$ as the
independent variable.\footnote{As discussed in Sec.~\protect{\ref{sec:clover-formalism}},
we also tried the standard $\cO(a)$ improved quark mass
in the clover case, but the fits were not significantly
different from those with $\tilde M_{q,2}$.}
For both Wilson and clover quarks, the
confidence level of linear fits to 
$M_{Qq}$, $f_{Qq}$, and $f_{qq}$  are 
better with $\tilde M_{q,2}$ than $m_{q,0}$, so we use it from now on in all cases. 

An important question is which functional form one should fit
to.  
Unfortunately, as in other lattice computations to date, we have been forced to
work at fairly large values of light quark mass.  In this region, our data for
decay constants, both $f_{Qq}$, and $f_{qq}$, are quite linear. There is little evidence
for chiral logarithms, which should introduce significant curvature as one approaches
the chiral limit, as emphasized recently by 
Kronfeld and Ryan \cite{KRONFELD} and Yamada \cite{YAMADA}. 
This is presumably not a problem with chiral perturbation theory (\chpt),
but simply an indication that higher order terms (\eg terms quadratic in quark mass) 
are as important as the chiral log terms in
the current mass regime.  
Further, chiral log fits would introduce yet another
parameter in the heavy-light case, the $B$-$B^*$-$\pi$ 
coupling $g$.\footnote{CLEO \cite{CLEO} has 
recently measured the $D^{*+}$ width, which gives, using
lowest order \chpt, a $D$-$D^*$-$\pi$ coupling $g^2\approx0.35$.
However, NLO \chpt\ and the $D^*\to D\pi$ decay
gives $g^2\approx0.07$ \cite{STEWART}; while NLO \chpt\ on
the $D^{*+}$ width (eq.~(21) in \cite{STEWART}) gives $g^2\approx0.22$. 
A recent lattice computation \cite{ABADA} gives $g^2\approx0.45$. There is also
some uncertainty in going from the $D$ to the $B$ system.}
It therefore seems clear to us that fits  of  $f_{Qq}$  to the NLO \chpt\ form would 
require at least four parameters: 
the value in the chiral limit, a linear slope, the coefficient $g^2$, and a higher order
(quadratic?) term introduced to cancel most of the curvature of the logarithms in our
relatively high mass region.
With only three light quark masses on most sets, it is clear that such
an approach is not feasible at present.
In work in progress \cite{MILC-3flavor}, however, we are consistently using 
five light quark masses and expect that we will be able to include chiral logs
and quadratic terms directly in the fits for central values.

For our ``standard'' chiral extrapolations, we thus 
consider only quadratic
and linear fits in $\tilde M_{q,2}$.  For each physical quantity,
we choose one of these fits for the central value, and the other is taken,
where appropriate, to give a standard chiral extrapolation error.
Note that $f_B$ \cite{GOITY} and $f_\pi$ \cite{GASSER-LEUTWYLER}
have similar chiral log effects in full QCD:
\begin{eqnarray}
\label{eq:fb-logs}
f_B & = &  f_B^{0} \Big[ 1 + \frac{1}{16\pi^2f^2} \big(-\frac{3(1+3g^2)}{4} m_\pi^2\ln(m_\pi^2) + \cdots \big)\Big] \\
\label{eq:fpi-logs}
f_\pi & = &  f \Big[ 1 + \frac{1}{16\pi^2f^2} \big(-2m_\pi^2\ln(m_\pi^2) + \cdots \big)\Big]
\end{eqnarray}
Since we fix the lattice scale with $f_\pi$ and
always use the same type of chiral fit for both $f_{Qq}$, and $f_{qq}$,
it is not unreasonable to expect
that much of the systematic effect coming from not including the curvature of \eqs{fb-logs}{fpi-logs}
will cancel.  In Secs.~\ref{sec:chiral-logs} and \ref{sec:chiral-log-errors}, we explain how we 
test this assumption and estimate the chiral logarithm effects in the dynamical case.

\begin{figure}[thb!]
\includegraphics[bb = 0 0 4096 4096,
width=5.0truein]{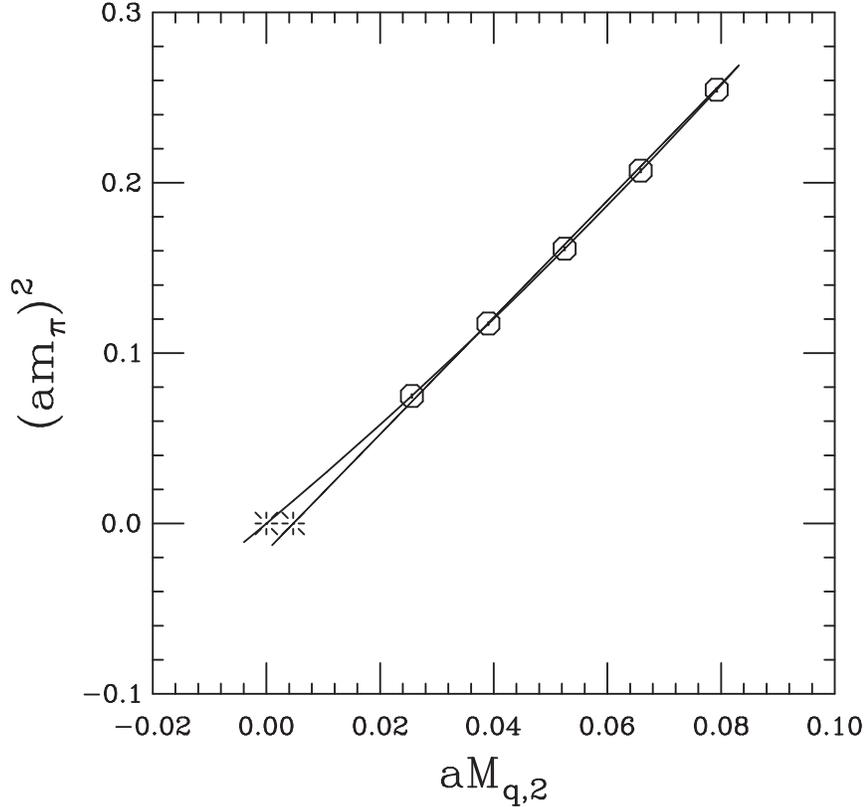}
\caption{$(am_\pi)^2$ \vs $a\tilde M_{q,2}$ for set CP1.
The solid line is a quadratic fit; the dotted line, the alternative
linear fit.  The fits have $\chi^2_{\rm cut}/{\rm d.o.f.}=0.06$
and $36.5$, respectively, with $\lambda_{\rm cut}=1$.
The bursts are the extrapolated points where $(am_\pi)^2=0$.
}
\label{fig:mpi2-CP1}
\end{figure}

Figure \ref{fig:mpi2-CP1} shows the chiral extrapolation of
$m_\pi^2$ to find $\kappa_c$, with  both linear and quadratic
fits.  Since the independent variable,
$\tilde M_{q,2}$, itself depends on $\kappa_c$, such fits
have to be iterated two or three times to find
a self-consistent value of $\kappa_c$ where both
$m_\pi^2$ and $\tilde M_{q,2}$ vanish.  This
has been done only for the quadratic fit in Fig.~\ref{fig:mpi2-CP1} to emphasize
the difference with the linear fit.

Table~\ref{tab:kc} shows the results for $\kappa_c$ and $\chi^2_{\rm cut}$ values
for all the data sets.  Note that the linear fits are uniformly very
poor; while quadratic fits are quite good wherever there are
enough light masses to compute a $\chi^2_{\rm cut}$.  Although using
$am_0$ as the independent variable actually reduces the $\chi^2_{\rm cut}$ values
of the linear fits somewhat, they remain very poor.
These features agree with what was found
in Ref.~\cite{MILC-PRL}, where we performed the light-light
analysis on a large quenched data set at $\beta=5.7$, using
five light quark masses. We therefore use only quadratic
fits/solves from here on for $m_\pi^2$ \vs $a\tilde M_{q,2}$.
Table~\ref{tab:ks} gives the values of $\kappa_s$ resulting
from these fits.

\begin{table}[p]
\caption{Extrapolations of $a^2m_\pi^2$ \vs $a\tilde M_{q,2}$
to find $\kappa_c$. When the number of degrees of freedom is
0 (quadratic fit with 3 $\kappa$ values), a solver is used instead
of a fitter.
For fits, the cutoff on correlation matrix eigenvalues
($\lambda_{\rm cut}$) is 1.  For set 5.7-large, which consists
of several lattice sizes, the central value 
is the average over all sets, the error is a combined weighted
error, and the $\chi^2$ and d.o.f.\ shown are the ones from the
volume $20^3\times48$. 
See Sec.~\protect{\ref{sec:fat}} for a description
of the fat-link computations.
}
\label{tab:kc}
\def\arraystretch{.8}
\begin{tabular}{|c||c|c|c||c|c|c|}
\noalign{\vspace{1cm}}
\hline
name&$\kappa_c$&$\chi_{\rm cut}^2$&d.o.f.&$\kappa_c$&$\chi_{\rm cut}^2$&d.o.f.\\
\hline
\multicolumn{1}{c}{}&\multicolumn{3}{c||}{quadratic fit}&\multicolumn{3}{c}{linear fit} \\
\hline
\hline
\multicolumn{7}{c}{quenched Wilson}\\
\hline
A&0.169433(237)&  -- &0&0.168607(89)& 29.3&1 \\
B&0.169340(100)&  -- &0&0.168383(52)&227.0&1 \\
5.7-large&0.169748(24)&  0.3 &2&0.168862(33)&703.3&3 \\
E&0.161397(124)&  -- &0&0.161046(89)& 20.1&1 \\
C&0.157228(95)&  -- &0&0.156778(46)& 62.5&1 \\
CP&0.157274(74)&  -- &0&0.156906(25)& 44.0&1 \\
D&0.151825(55)&  -- &0&0.151663(35)& 28.0&1 \\
H&0.149368(20)&  -- &0&0.149248(15)& 18.9&1 \\
\hline
\multicolumn{7}{c}{$N_f=2$ Wilson}\\
\hline
L&0.169422(61)&  -- &0&0.168160(50)&299.4&1 \\
N&0.169515(56)&  -- &0&0.168513(25)&496.4&1 \\
O&0.167197(41)&  -- &0&0.166483(30)&306.8&1 \\
M&0.165919(59)&  -- &0&0.165211(33)&192.3&1 \\
P&0.165257(47)&  -- &0&0.164864(28)&144.5&1 \\
U&0.163065(27)&  -- &0&0.162570(16)&426.6&1 \\
T&0.161528(22)&  -- &0&0.161887(27)&155.9&1 \\
S&0.161400(20)&  -- &0&0.160802(12)&447.2&1 \\
G&0.161158(72)&  -- &0&0.160821(45)& 38.7&1 \\
R&0.161167(23)&  -- &0&0.160798(12)&328.9&1 \\
\hline
\multicolumn{7}{c}{quenched clover}\\
\hline
CP1&0.135342(18)&  0.1&2&  0.135168(8) &   109.4 & 3  \\
J&0.135862(20)&  0.0&2&  0.135792(16) &   45.8 & 3  \\
\hline
\multicolumn{7}{c}{fat-link clover ($N=10$, $c=0.045$) }\\
\hline
CPF&0.125558(22)&  0.2&2&  -- &   -- &--  \\
RF&0.125666(25)&  -- &0&  -- &   -- &--  \\
\hline
\end{tabular}
\end{table}

\begin{table}[p]
\caption{Values of $\kappa_s$, the hopping parameter of the
strange quark, from fits to the light-light pseudoscalars and vectors.
In the former case,
we adjust the pseudoscalar mass to 
$\protect{\sqrt{2m_K^2 - m_\pi^2}}$; in the latter, we
adjust the vector mass to $m_\phi$. The values of
$\chi^2$ and degrees of freedom can be found by referring
to the corresponding fits in 
Tables~\protect{\ref{tab:kc}} and \protect{\ref{tab:ainverse}}.
Data from set 5.7-large is combined as in Table~\protect{\ref{tab:kc}}.
See Sec.~\protect{\ref{sec:fat}} for a description
of the fat-link computations.
}
\label{tab:ks}
\def\arraystretch{.8}
\begin{tabular}{|c|@{\hspace{.4in}}c|@{\hspace{.4in}}c|}
\noalign{\medskip}
\cline{1-3}
name&$\kappa_s$ (pseudoscalars) &$\kappa_s$ (vectors)\\
& quadratic fit&linear fit\\
\cline{1-3}
\multicolumn{3}{c}{quenched Wilson}\\
\cline{1-3}
A&0.164331(432)&0.163173(407)\\
B&0.163629(355)&0.163709(215)\\
5.7-large&0.163916(100)&0.163456(78)\\
E&0.158203(170)&  0.157729(351)\\
C&0.154567(97)& 0.154780(229)\\
CP&0.154857(99)&  0.154638(152)\\
D&0.150395(66) & 0.150316(102)\\
H&0.148415(42) & 0.148322(87)\\
\cline{1-3}
\multicolumn{3}{c}{$N_f=2$ Wilson}\\
\cline{1-3}
L&0.164114(201)&  0.164064(184)\\
N&0.164436(213)&  0.164027(202)\\
O&0.162938(121)&  0.162691(120)\\
M&0.161778(205)&  0.161830(206)\\
P&0.161518(141)&  0.161374(106)\\
U&0.159690(81)&  0.159373(103)\\
T&0.158610(121)&  0.157243(148)\\
S&0.158633(68)&  0.158423(50)\\
G&0.158795(94)&  0.158350(146)\\
R&0.158736(55)&  0.158465(85)\\
\cline{1-3}
\multicolumn{3}{c}{quenched clover}\\
\cline{1-3}
CP1&0.133882(41)&  0.133515(63)\\
J&0.134665(45)&  0.134400(56)\\
\cline{1-3}
\multicolumn{3}{c}{fat-link clover ($N=10$, $c=0.045$)}\\
\cline{1-3}
CPF&0.123206(46)&  0.123481(137)\\
RF&0.123440(62)&  0.123236(162)\\
\cline{1-3}
\noalign{\medskip}
\end{tabular}
\end{table}

\begin{figure}[thb!]
\includegraphics[bb = 0 0 4096 4096,
width=5.0truein]{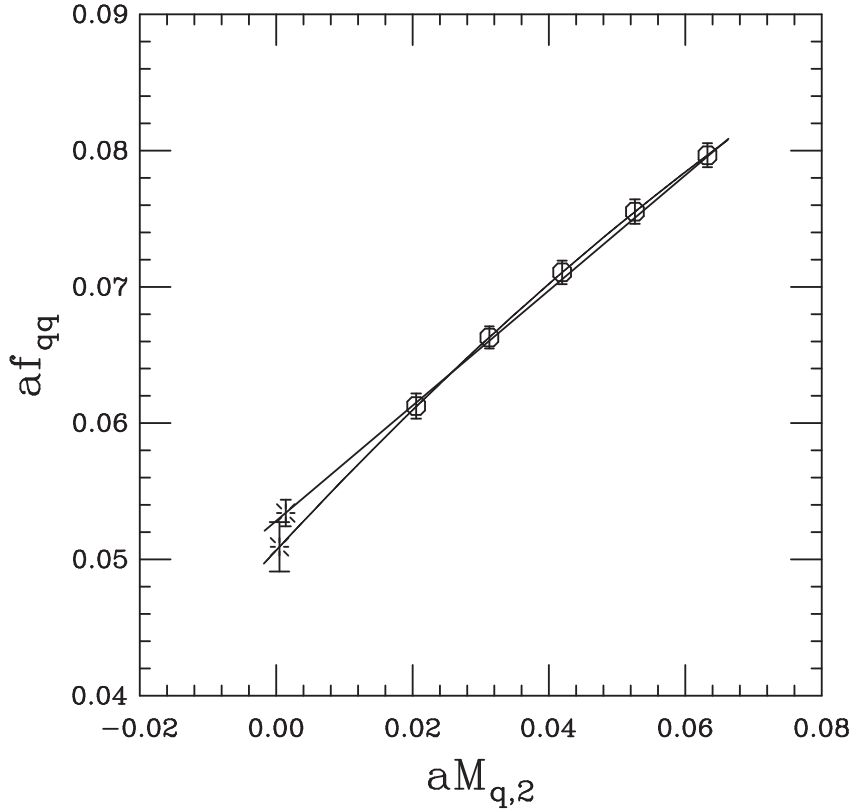}
\caption{The light-light pseudoscalar decay 
constant, $f_{qq}$, \vs $a\tilde M_{q,2}$ for set J.
The solid line is a linear fit; the dotted line, the alternative
quadratic fit.  The fits have $\chi^2_{\rm cut}/{\rm d.o.f.}=1.1$
and $0.005$, respectively, with $\lambda_{\rm cut}=1$ (1 eigenvector of 5 kept).
The bursts are the extrapolated points where $\tilde M_{q,2}$
takes its physical value (\ie $\kappa=\kappa_{u,d}$).  The burst
on the dotted line has been displaced slightly for clarity.
}
\label{fig:fpi-J}
\end{figure}

\begin{figure}[thb!]
\includegraphics[bb = 0 0 4096 4096,
width=5.0truein]{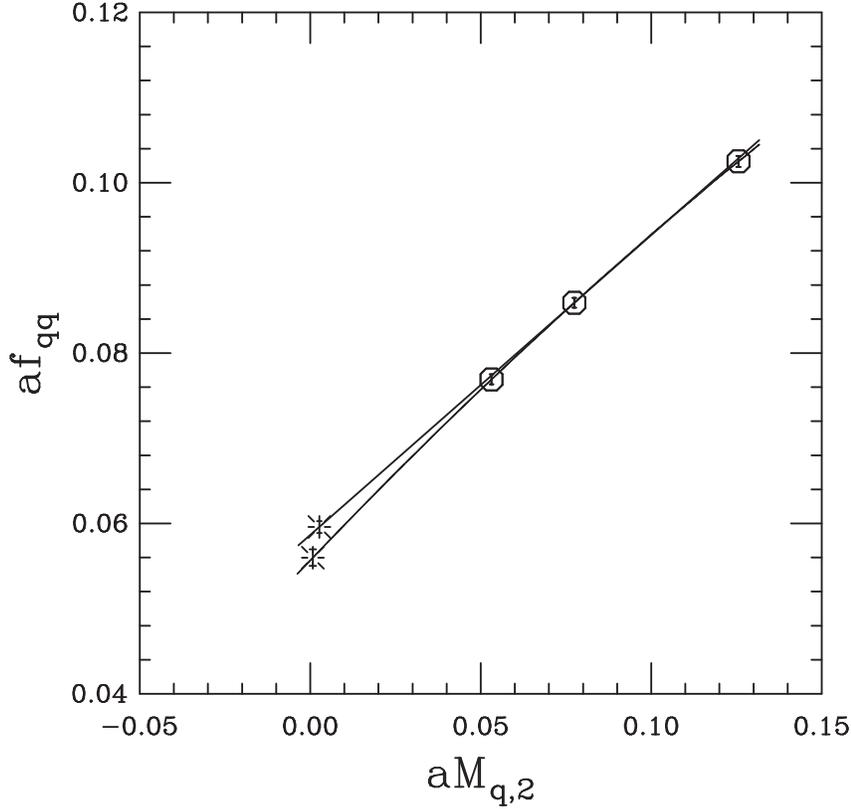}
\caption{Same as Fig.~\protect{\ref{fig:fpi-J}}, but for set R.
The linear fit (solid line) has $\chi^2_{\rm cut}/{\rm d.o.f.}=8.7$,
$\lambda_{\rm cut}=1$ (1 eigenvector of 3 kept);
while the quadratic fit (dotted line) has no degrees of
freedom.
}
\label{fig:fpi-R}
\end{figure}

The case of $f_{qq}$ \vs $a\tilde M_{q,2}$
is a more difficult one. Figures~\ref{fig:fpi-J} and \ref{fig:fpi-R},
and Table~\ref{tab:ainverse} 
show the extrapolations.  Although the linear fits in both figures
appear reasonable to the eye, that in Fig.~\ref{fig:fpi-R}
has a rather high value of $\chi^2_{\rm cut}/{\rm d.o.f}$,
as do many of the other linear fits in Table~\ref{tab:ainverse}.
Where comparisons can be made, the quadratic fits are 
better. This was also true of the large quenched data set at $\beta=5.7$
analyzed in Ref.~\cite{MILC-PRL}. On the other hand, the quadratic
fits often have quite large statistical errors, especially in the Wilson
case where there are only three light quark masses.  Furthermore, on the sets
with the finest lattice spacings (C, CP, D, H, G, R, CP1, J) the linear
fits are generally acceptable (set R is an exception).  For these
reasons, we use the linear fits for the central values
and take the quadratic fits to estimate the ``standard chiral systematic error.''
As mentioned above, the fits in future work \cite{MILC-3flavor} will include chiral logs as well as quadratic terms.

\begin{table}[p]
\caption{Extrapolations of $af_{qq}$ and $m_\rho$ \vs $a\tilde M_{q,2}$
to find the scale, $a^{-1}$, using $f_\pi=0.1307\;\GeV$
and $m_\rho=0.768\;\GeV$. When the number of degrees of freedom is
0 (quadratic fit with 3 $\kappa$ values), a solver is used instead
of a fitter.
For fits, the cutoff on correlation matrix eigenvalues
($\lambda_{\rm cut}$) is 1.
Data from set 5.7-large is combined as in Table~\protect{\ref{tab:kc}}.
See Sec.~\protect{\ref{sec:fat}} for a description
of the fat-link computations.
}
\label{tab:ainverse}
\def\arraystretch{.8}
\begin{tabular}{|c||c|c|c||c|c|c||c|c|c|}
\noalign{\vspace{1cm}}
\hline
name&$a^{-1}\ (\GeV)$&$\chi_{\rm cut}^2$&d.o.f.&$a^{-1}\ (\GeV)$&$\chi_{\rm cut}^2$&d.o.f.&$a^{-1}\ (\GeV)$&$\chi_{\rm cut}^2$&d.o.f.\\
\hline
\multicolumn{1}{|c}{}&\multicolumn{3}{c||}{$f_{qq}$, linear fit}&\multicolumn{3}{c||}{$f_{qq}$, quadratic fit}
&\multicolumn{3}{c|}{$m_\rho$, linear fit} \\
\hline
\hline
\multicolumn{10}{c}{quenched Wilson}\\
\hline
A &1.391(66)&  5.3&1&1.586(122)&--&0&1.413(32)&  1.5&1\\
B &1.311(48)&  6.1&1&1.453(66)&--&0&1.488(25)&  0.0&1\\
5.7-large &1.339(14)&  3.3&3& 1.388(17)&0.3&2&1.450(8)&  9.9&3\\
E &1.780(43)&  3.4&1&1.851(63)&--&0&1.848(73)&  0.1&1\\
C &2.124(54)&  1.8&1&2.266(96)&--&0&2.414(102)&  0.5&1\\
CP &2.211(48)&  1.2&1&2.339(124)&--&0&2.333(51)&  0.2&1\\
D &3.151(91)&  2.8&1&3.331(111)&--&0&3.389(108)&  0.2&1\\
H &4.388(121)&  0.5&1&4.490(146)&--&0&4.489(150)&  0.0&1\\
\hline
\multicolumn{10}{c}{$N_f=2$ Wilson}\\
\hline
L &1.375(30)& 15.7&1&1.585(43)&--&0&1.545(23)&  0.0&1\\
N &1.432(35)&  2.0&1&1.524(53)&--&0&1.524(25)&  0.3&1\\
O &1.568(25)&  2.3&1&1.654(55)&--&0&1.685(20)&  0.9&1\\
M &1.608(47)&  1.0&1&1.680(56)&--&0&1.789(37)&  0.3&1\\
P &1.713(38)&  2.0&1&1.773(50)&--&0&1.853(22)&  2.3&1\\
U &1.800(25)&  3.2&1&1.868(34)&--&0&1.888(25)&  0.3&1\\
T &1.800(37)& 24.3&1&1.939(40)&--&0&1.839(20)&  0.7&1\\
S &2.038(28)&  7.6&1&2.192(31)&--&0&2.157(19)&  0.5&1\\
G &2.243(38)&  4.3&1&2.377(63)&--&0&2.242(52)&  0.0&1\\
R &2.194(26)&  8.7&1&2.306(39)&--&0&2.269(31)&  3.6&1\\
\hline
\multicolumn{10}{c}{quenched clover}\\
\hline
CP1 &1.994(29)&  8.0&3&2.093(65)& 3.9&2&1.908(26)& 0.2&3\\
J &2.447(45)& 3.3&3&2.545(90)& 0.0&2&2.383(39)& 1.4&3\\
\hline
\multicolumn{10}{c}{fat-link clover ($N=10$, $c=0.045$)}\\
\hline
CPF &1.847(18)&14.4&3&1.923(28)& 0.5&2&2.114(61)& 6.1&3\\
RF &1.873(27)& 1.4&1&1.939(69)&--&0&1.920(47)& 0.3&1\\
\hline
\end{tabular}
\end{table}

\begin{figure}[thb!]
\includegraphics[bb = 0 0 4096 4096,
width=5.0truein]{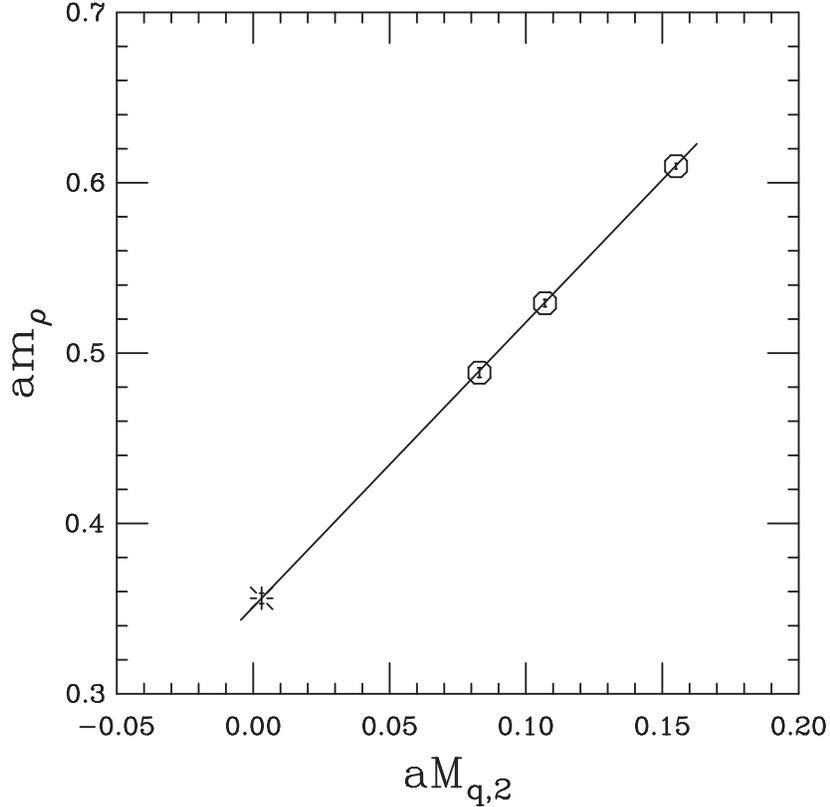}
\caption{Linear extrapolation of the light-light vector meson mass
to the physical point ($\rho$ meson, indicated by the burst). The data
is from set S.
The fit has $\chi^2_{\rm cut}/{\rm d.o.f.}=0.5$ with
$\lambda_{\rm cut}=1$.
}
\label{fig:mrho-S}
\end{figure}

Figure~\ref{fig:mrho-S} shows a typical linear extrapolation of the
light-light vector mass to the physical point for up/down quarks.
As seen in Table~\ref{tab:ainverse}, the linear fits are almost
always quite good, and we use the scale set with $m_\rho$ in this way
as an alternative to that from $f_\pi$.  The same fits
also give an alternative value for $\kappa_s$, using $m_\phi$.  The results are
shown in Table~\ref{tab:ks}. 

Sample chiral fits for heavy-light masses and decay constants
are shown in Figs~\ref{fig:MQq-U-.113} and \ref{fig:fQq-CP1-.09}.
Although the extrapolated data for all the sets are too extensive to tabulate here,
they are available at
{\tt http://www.physics.wustl.edu/$\!\sim$cb/Nf=2\_tables},
in the files 
``{\it name}\_chiral\_mass.dat'' and ``{\it name}\_chiral\_fsqrtM.dat.''
In the latter files we give $f_{Qq}\sqrt{M_{Qq}}$, rather than
$f_{Qq}$, because $f_{Qq}\sqrt{M_{Qq}}$ is what we will need later to
extrapolate/interpolate to the mass of the $B$ and $D$ mesons.
For the decay constants from sets CP1 and J,  ``{\it name}'' includes the 
qualification NP-IOY or NP-tad because $f_{Qq}$ will of course
depend on how the renormalization is done.

\begin{figure}[thb!]
\includegraphics[bb = 0 0 4096 4096,
width=5.0truein]{MQq_U_.113.ps}
\caption{Linear chiral extrapolation of the heavy-light pseudoscalar meson mass,
to the physical point ($\kappa=\kappa_{u,d}$, indicated by the burst).
The data is from set U, $\kappa_Q=0.113$, 
The fit has $\chi^2_{\rm cut}/{\rm d.o.f.}=0.7$ with
$\lambda_{\rm cut}=1$.
}
\label{fig:MQq-U-.113}
\end{figure}

\begin{figure}[thb!]
\includegraphics[bb = 0 0 4096 4096,
width=5.0truein]{fQq_CP1_.09.ps}
\caption{Linear (solid) and quadratic (dotted) chiral extrapolations 
of $f_{Qq}$
to the physical point ($\kappa=\kappa_{u,d}$ indicated by the bursts).
The data is from set CP1, $\kappa_Q=0.09$, with NP-IOY renormalization.
The linear fit has $\chi^2_{\rm cut}/{\rm d.o.f.}=1.0$;
the quadratic, $0.3$.
The parameter $\lambda_{\rm cut}=1$ in both cases.
The burst on the dotted line has been displaced slightly for clarity.
}
\label{fig:fQq-CP1-.09}
\end{figure}

For almost all data sets, the linear chiral fits of the heavy-light masses 
are quite good, at least in the important range
of meson masses between the $D$ and the $B$.
Sets L and T are exceptions, which is perhaps not surprising
since their correlators are quite noisy to begin with, making
it difficult to find good plateaus. Indeed, for set T the data is
noisy enough that covariant chiral fits for heavy-light masses
did not converge, with any choice for
$\lambda_{\rm cut}$ except $\lambda_{\rm cut}=\infty$, \ie noncovariant
fits. However, since the linear fits to heavy-light masses were fine 
on the vast majority of the sets, we believe it is reasonable to 
use them exclusively. Linear fits to the static-light masses are always
acceptable.

The situation for the heavy-light decay constants is similar to
that for the light-light decay constants.  Again, there
is a small amount of curvature in these plots, 
and the direction of curvature is the same as for the light-light case.
(Compare Fig.~\ref{fig:fQq-CP1-.09} with Figs.~\ref{fig:fpi-J} and \ref{fig:fpi-R}.)
Therefore, quadratic chiral fits of $f_{Qq}$
are better than linear ones where the comparison can be made,
but, as before, the quadratic fits/solves lead to significantly larger statistical errors.  
The main difference with the light-light case is that the linear fits typically improve
as the mass of the heavy quark increases, so that by the time the physical $b$ quark
mass is reached they are generally quite reasonable.  The only exceptions
to this rule are the sets CP, for which linear fits are poor over the
whole heavy-quark mass range, and T, which is noisy and again requires
noncovariant fits. In the static-light case, linear
fits are always good. We thus choose linear fits everywhere
for the central values but use quadratic fits for the heavy-lights 
in estimating the systematic error in the standard chiral
extrapolation.  
As discussed above, we expect that extrapolating both light-light and heavy-light
fits in the same way will cancel at least some of the systematic error associate with
curvature and chiral logarithms.  Therefore,
we change from linear to quadratic fits in both
cases at once when we make our estimate of the ``standard'' chiral error.

To summarize: For central values, we use quadratic chiral fits in $a\tilde M_{q,2}$
for $m_\pi^2$ and linear fits for heavy-light and static-light
masses and  all the decay constants.  We call this
combination ``chiral choice I.'' Our standard chiral systematic
errors are found by comparing  the central values with the results of ``chiral choice II:''
quadratic fits  for $m_\pi^2$ and both light-light and heavy-light decay constants, and linear
fits for heavy-light masses  and static-light masses and decay
constants.  For the light-light vector meson masses, which
enter only in various systematic error estimates (alternative
scale determination from $m_\rho$, alternative $\kappa_s$ determination
from $m_\phi$), we always use linear fits.

\subsection{Chiral Logarithm Effects} 
\label{sec:chiral-logs}

The standard chiral systematic error just described does not {\em directly} take into
account the sharp curvature in decay constants at very small quark mass caused by terms of the form
$-m_\pi^2 \ln (m_\pi^2)$ in \eqs{fb-logs}{fpi-logs}.  
Putting aside the issue of scale choice,
an extrapolation in the full theory that ignores the chiral log in $f_{Qq}$ 
is expected to overestimate \fb\ and underestimate \fbsofb\ 
\cite{KRONFELD}  (since $-m_\pi^2 \ln m_\pi^2$ is concave 
down with rapid variation at small mass).  However, because we set the scale with $f_\pi$ and use
the same extrapolations for
$f_{qq}$ and $f_{Qq}$, the effect  on the individual decay constants
is less clear.
Even ratios such as \fbsofb\ are affected indirectly by the scale choice, through
the  fixing of $\kappa_s$, the hopping parameter for the strange quark. (Fixing 
$\kappa_b$, the bottom quark hopping parameter, has little effect on the ratio but does 
represent another scale effect on the individual decay constants.)  It is easy to see
that our scale choice should push \fb\ and \fbsofb\ in the opposite direction of the $f_{Qq}$
extrapolation.  
We can thus hope that such effects largely cancel,  and this is a justification
for taking our central values and errors from the standard linear and quadratic chiral fits
described in the previous section. 

However, to test the above assumption and estimate the errors induced by not
directly fitting with chiral log forms, we need alternative methods of evaluation that
do not involve chiral extrapolations of individual decay constants.
One approach that takes advantage of the fact that the chiral logs in $f_B$ and $f_\pi$
are similar in magnitude is to perform chiral extrapolations only
on the ratio $f_B/f_\pi$ (more precisely, $f_{Qq}/f_{qq}$).
Given the expected range for the parameter $g^2$,
the ratio has a chiral log term of opposite sign from that of $f_B$ alone and either
comparable or greatly reduced magnitude (see \eqs{fb-logs}{fpi-logs} and the footnote
shortly before).  
In practice, since the slope of $f_{qq}$ is greater than $f_{Qq}$,
$f_{qq}/f_{Qq}$ is more linear than $f_{Qq}/f_{qq}$, and we work with the former.

\begin{figure}[thb!]
\includegraphics[bb = 0 0 4096 4096,
width=5.0truein]{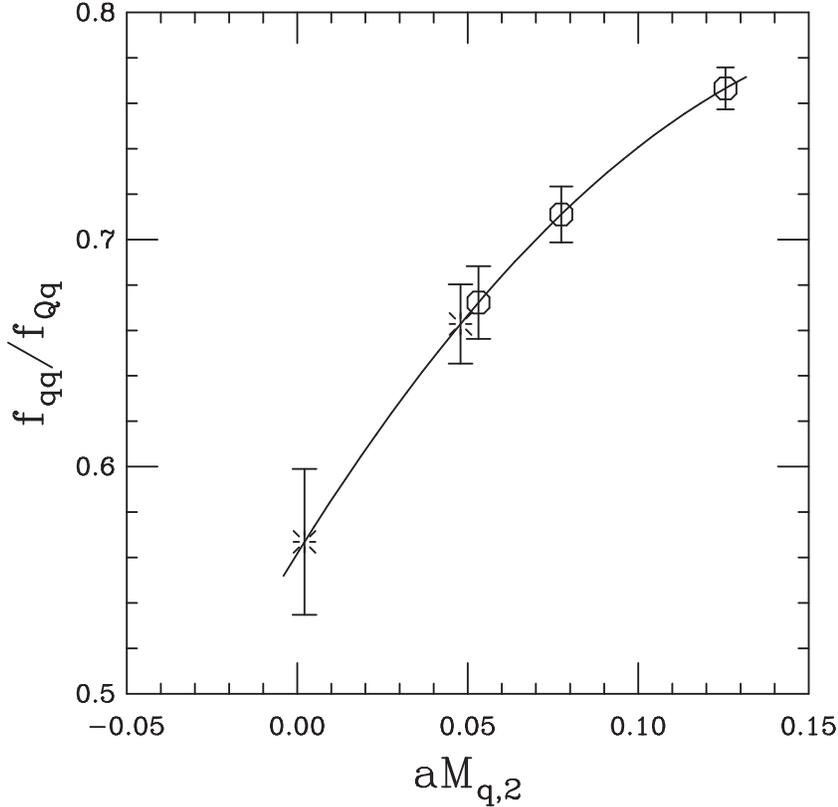}
\caption{Quadratic chiral extrapolation of $f_{qq}/f_{Qq}$ from set
R, with $\kappa_Q=0.125$.  The bursts show the  extrapolation to
$\kappa_s$ and $\kappa_{u,d}$, with these values determined by
method 6 in Table~\ref{tab:decay-logs}.
}
\label{fig:fpiofB-R-kp2-meth6.125}
\end{figure}

Fig.~\ref{fig:fpiofB-R-kp2-meth6.125} shows a chiral extrapolation of $f_{qq}/f_{Qq}$ for set R,
with $\kappa_Q$ chosen so that $M_{Qq}$ is near the $B$  mass.  There is clear curvature, so a
linear fit is not appropriate, and we fit (solve) quadratically.  
To the extent that a residual chiral log remains
in the ratio, the quadratic fit should somewhat overestimate $f_\pi/f_B$ and hence also
\fbsofb.  

However, to take real advantage of the presumed reduction of chiral logs in $f_\pi/f_B$
and known sign of the error in \fbsofb, we must eliminate the dependence of the scale 
and $\kappa_s$ on a chiral extrapolation.  This means that common scale choices
such as $f_\pi$, $m_\rho$ or $m_N$ cannot be used.  Further, we are reluctant to
employ static potential quantities such as the string tension, $r_0$ or $r_1$ because their physical
values are only known phenomenologically, with uncertain errors.
Instead, we look at three more-or-less physical quantities  associated with the $s$ quark:
the vector meson mass $m_\phi$, and the mass and decay constant of a would-be 
$s\bar s$ pseudoscalar, $m_{ss}$ and $f_{ss}$.   To be precise, the $s\bar s$
meson is made of two valence quarks with the physical strange quark mass, but in
a standard sea quark background: either the physical $N_f=3$ case (with $m_u=m_d=\hat m^{\rm phys}$
and $m_s^{\rm phys}$),
or, corresponding more closely to our simulations, the $N_f=2$ case (with $m_u=m_d=\hat m^{\rm phys}$).
Here the superscript ``phys'' stands for the physical mass, and we neglect isospin violations as usual.

The quantity $f_{ss}$ can be related to $f_K$ and $f_\pi$ in 1-loop (NLO) partially
quenched chiral perturbation theory (PQ\chpt), in a manner independent of the
analytic $p^4$ (``Gasser-Leutwyler'') constants.  Using the formulas in Refs.~\cite{SHARPE-PQ2} and
\cite{SHARPE-PQ3} for $N_f=2$ and $N_f=3$, respectively, we find:
\begin{eqnarray}
\label{eq:fss2}
\frac{f_{ss}^{(2)}f_\pi}{f_K^2}\! &=&\! 1+ \frac{1}{16\pi^2f^2}\left(\half m_\pi^2\ln(m_{ss}^2/m_\pi^2) -\half m_{ss}^2 + 
\half m_\pi^2\right) \approx 0.93 \\
\label{eq:fss3}
\frac{f_{ss}^{(3)}f_\pi}{f_K^2}\! &=&\! 1+ \frac{1}{16\pi^2f^2}\left(\threehalves m_\eta^2\ln(m_\eta^2/\Lambda^2)
-\half m_\pi^2\ln(m_\pi^2/\Lambda^2) 
-m_{ss}^2\ln(m_{ss}^2/\Lambda^2)
\right) \approx 0.95 
\end{eqnarray}
where, in the numerical evaluation, we have used 
$f=f_\pi=130.7\;\MeV$ and $m_{ss}^2 = \threehalves m_\eta^2 -\half m_\pi^2$. (The latter
relation is theoretically convenient here because it makes \eq{fss3} explicitly independent of
the chiral scale $\Lambda$.)

For $m_{ss}$, we consider a few choices.  Two tree level relations have already been mentioned:
\begin{eqnarray}
\label{eq:mss-K}
m_{ss}^2 &=& 2m_K^2 -m_\pi^2 \\
\label{eq:mss-eta}
m_{ss}^2 &=& \threehalves m_\eta^2 -\half m_\pi^2
\end{eqnarray}
With Refs.~\cite{SHARPE-PQ2,SHARPE-PQ3}, we can also derive 1-loop relations for $m_{ss}$
similar to \eqs{fss2}{fss3}, although they
do involve analytic terms indirectly through the quark mass ratios.  In particular, we use the
$N_f=2$ result:
\begin{equation}
\frac{m_{ss}^2m_\pi^2}{m_K^4} =  \frac{4\hat m/m_s}{(1+\hat m/m_s)^2}\left[1+ \frac{1}{16\pi^2f^2}\left(
-m_\pi^2\ln(m_{ss}^2/m_\pi^2) +m_{ss}^2 -
m_\pi^2\right)\right] \approx 0.17 \ ,
\label{eq:mss-PQ2}
\end{equation}
where we have taken $m_s/\hat m=24.4$ \cite{LEUTWYLER-MASSES}.

We then perform a series of analyses.  For each, we choose 2 of the quantities $f_{ss}$, $m_{ss}$,
and $m_\phi$, and a method of evaluation for the ``physical'' quantities
$f_{ss}$ (either \eq{fss2} or \eq{fss3}) and/or
$m_{ss}$ (either \eq{mss-K}, \eq{mss-eta} or \eq{mss-PQ2}).  We then fit the ratio
of the 2 chosen quantities as a function of light quark mass ($a\tilde M_{q,2}$) in order to
determine $\kappa_s$.  Generally, only an interpolation or mild extrapolation is required.  
Figures~\ref{fig:mssofss-R-fk2-meth1} and \ref{fig:mssomphi-R-kp22-meth8}
show two examples of such fits, for $m_{ss}/f_{ss}$ and $m_{ss}/m_\phi$,
respectively.  The former uses \eqs{mss-K}{fss2}; while the latter, 
\eq{mss-PQ2}. Note that a slight extrapolation is required to find
$\kappa_s$ in Fig.~\ref{fig:mssofss-R-fk2-meth1}.  
In Fig.~\ref{fig:mssomphi-R-kp22-meth8}, $\kappa_s$ is determined by an
interpolation, which is more similar to the standard analysis.  The difference
between the two situations gives some indication of the errors of the 
procedure.
In Tables~\ref{tab:decay-logs} 
we show the results for $\kappa_s$ for 12 different
versions of such fits, performed on two different dynamical sets of configurations
(the ones with the lightest sea quark masses).

\begin{figure}[thb!]
\includegraphics[bb = 0 0 4096 4096,
width=5.0truein]{mssofss-R-fk2-meth1.ps}
\caption{Quadratic extrapolation of $m_{qq}/f_{qq}$ to $m_{ss}/f_{ss}$ on set R,
with $m_{ss}$ from \eq{mss-K} and $f_{ss}$ from \eq{fss2}. 
The abscissa of the burst gives the value of 
the quark mass, $aM_{q,2}$, at $\kappa_s$.  This particular determination is
called ``method 1'' 
(see text and Table~\ref{tab:decay-logs}).
}
\label{fig:mssofss-R-fk2-meth1}
\end{figure}

\begin{figure}[thb!]
\includegraphics[bb = 0 0 4096 4096,
width=5.0truein]{mssomphi-R-kp22-meth8.ps}
\caption{Quadratic interpolation of $m_{qq}/m^{\rm vec}_{qq}$ to $m_{ss}/m_\phi$ on set R,
with $m_{ss}$ from \eq{mss-PQ2}.
The abscissa of the burst gives the value of 
the quark mass, $aM_{q,2}$, at $\kappa_s$.  This particular determination is
called ``method 8'' 
(see text and Table~\ref{tab:decay-logs}).
}
\label{fig:mssomphi-R-kp22-meth8}
\end{figure}
\begin{table}[p]
\vspace{-0.5truecm}
\caption{Values of $\kappa_s$ and $a^{-1}$ (in GeV) for various
methods of the analysis that do not require (long) chiral extrapolation.
In each box, the upper entry is from
set R; the lower, from set P.  
These should be compared with our central values, which come from
linear chiral extrapolation of $f_{qq}$ (and quadratic interpolation of $m_{qq}^2$):
$\kappa_s= 0.15873(5)$, $a^{-1}=  2.19(3)\;\GeV$ (set R) ,
and $\kappa_s=0.16152(14)$, $a^{-1}=1.71(4)\;\GeV$ (set P).  
We also show the changes (in \MeV) from central values that each method 
produces in decay constants, as well as the average and standard
devation of the mean of those changes.  These quantities are used in
Sec.~\ref{sec:chiral-log-errors} to estimate the systematic effects
of chiral logarithms.
Lines
for which $a^{-1}$ differs by more than 20\%  from the
central value (indicated by `*') are considered unreliable and are eliminated from the averages
}
\label{tab:decay-logs}
\def\arraystretch{.8}
\begin{tabular}{|c|c|c|c|c|c|c|c|}
\noalign{\vspace{0.5cm}}
\hline
{\bf method}& {\bf description} &$\kappa_s$ & $a^{-1}$&\fb &\fbs &\fd &\fds \\
\hline
 \phantom{*} 1& $\kappa_s$: $m_{ss}/f_{ss}$\ ;\qquad $a$:  $ f_{ss}$
& $0.15909(7)$    & $2.38(4)$ & $+11$         & $+17$         & $+ 2$         & $+ 6$ \\
 &$m_{ss}$:   \eq{mss-K}\ ;\qquad $f_{ss}$:   \eq{fss2} 
& $0.16137(15)$    & $1.68(4)$ & $ -3$         & $ -7$         & $+ 7$         & $ -12$ 	 \\
\hline
 \phantom{*} 2&$\kappa_s$:  $ m_{ss}/f_{ss}$\ ;\qquad $a$:  $ f_{ss}$ 
& $0.15926(7)$    & $2.48(4)$ & $+13$         & $+28$         & $+ 0$         & $+13$ \\
 &$m_{ss}$:   \eq{mss-K}\ ;\qquad $f_{ss}:$ \eq{fss3}
& $0.16166(14)$    & $1.76(4)$ & $ -2$         & $+ 3$         & $+ 6$         & $ -4$ 	 \\
\hline
 \phantom{*} 3&$\kappa_s$:  $ m_{ss}/f_{ss}$\ ;\qquad  $a$:  $ f_{ss}$
& $0.15847(8)$    & $2.22(3)$  & $+ 8$         & $+ 5$         & $+ 4$         & $ -3$ \\
 &$m_{ss}$:  \eq{mss-PQ2} \ ; \qquad $f_{ss}$:  \eq{fss2}
& $0.16024(22)$    & $1.55(4)$  & $ -7$         & $ -21$        & $+ 7$         & $ -21$\\
\hline
 \phantom{*} 4&$\kappa_s$:  $ m_{ss}/f_{ss}$\ ;\qquad $a$:  $ f_{ss}$
& $0.15937(7)$    & $2.47(4)$  & $+13$         & $+24$         & $+ 1$         & $+10$\\
 &$m_{ss}$:   \eq{mss-eta} ;\qquad $f_{ss}:$ \eq{fss2}
& $0.16185(14)$    & $1.75(4)$ & $ -2$         & $ -1$         & $+ 6$         & $ -7$          \\
\hline
 \phantom{*} 5&$\kappa_s$:  $ m_{ss}/f_{ss}$\ ;\qquad $a$:  $ f_{ss}$ 
& $0.15953(7)$    & $2.57(4)$ & $+15$         & $+34$         & $ -1$         & $+17$ \\
 &$m_{ss}$:   \eq{mss-eta} ;\qquad $f_{ss}:$ \eq{fss3}
& $0.16211(13)$    & $1.83(5)$ & $ -0$         & $+ 9$         & $+ 5$         & $+ 0$           \\
\hline
 \phantom{*} 6&$\kappa_s$: $m_{ss}/m_\phi$\ ;\qquad $a$:  $ m_\phi$
& $0.15922(6)$    & $2.46(4)$  & $+13$         & $+21$         & $+ 1$         & $+ 7$ \\
 &$m_{ss}$:   \eq{mss-K}\ ;\qquad  $f_{ss}$:   \eq{fss2}
& $0.16232(7)$    & $1.96(2)$ & $+ 3$         & $+11$         & $+ 4$         & $ -2$ \\
\hline
 \phantom{*} 7&$\kappa_s$:  $ m_{ss}/m_\phi$\ ;\qquad $a$:  $ m_\phi$ 
& $0.15922(6)$    & $2.46(4)$ & $+13$         & $+27$         & $+ 1$         & $+13$ \\
 &$m_{ss}$:   \eq{mss-K}\ ;\qquad  $f_{ss}:$ \eq{fss3}
& $0.16232(7)$    & $1.96(2)$ & $+ 3$         & $+16$         & $+ 4$         & $+ 4$ \\
\hline
 \phantom{*} 8&$\kappa_s$:  $ m_{ss}/m_\phi$\ ;\qquad  $a$:  $ m_\phi$ 
& $0.15871(6)$    & $2.33(3)$  & $+10$         & $+10$         & $+ 3$         & $ -0$\\
 &$m_{ss}$:  \eq{mss-PQ2}\ ;\qquad $f_{ss}$:  \eq{fss2} 
& $0.16170(7)$    & $1.89(2)$  & $+ 1$         & $+ 3$         & $+ 5$         & $ -9$\\
\hline
 \phantom{*} 9&$\kappa_s$:  $ m_{ss}/m_\phi$\ ;\qquad $a$:  $ m_\phi$ 
& $0.15946(6)$    & $2.53(4)$ & $+14$         & $+27$         & $ -1$         & $+11$ \\
 &$m_{ss}$:   \eq{mss-eta} ;\qquad  $f_{ss}:$ \eq{fss2}
& $0.16262(7)$    & $2.00(3)$ & $+ 3$         & $+15$         & $+ 3$         & $+ 2$ \\
\hline
 \phantom{*} 10&$\kappa_s$:  $ m_{ss}/m_\phi$\ ;\qquad $a$:  $ m_\phi$
& $0.15946(6)$    & $2.53(4)$  & $+14$         & $+32$         & $ -1$         & $+17$ \\
 &$m_{ss}$:   \eq{mss-eta} ;\qquad  $f_{ss}:$ \eq{fss3}
& $0.16262(7)$    & $2.00(3)$ &  $+ 3$         & $+20$         & $+ 3$         & $+ 8$ \\
\hline
 * 11&$\kappa_s$:  $ f_{ss}/m_\phi$\ ;\qquad $a$:  $ f_{ss}$ 
& $0.16034(107)$    & $2.83(47)$ & $+18$         & $+48$         & $ -5$         & $+27$\\
 * \phantom{11} &$f_{ss}:$ \eq{fss2} 
& $0.16479(71)$    & $2.36(16)$ & $+ 9$         & $+47$         & $ -2$         & $+35$ \\
\hline
 \phantom{*} 12&$\kappa_s$:  $ f_{ss}/m_\phi$\ ;\qquad $a$:  $ f_{ss}$ 
& $0.15881(80)$    & $2.35(21)$ & $+11$         & $+17$         & $+ 3$         & $+ 6$\\
 * \phantom{12} &$f_{ss}:$ \eq{fss3} 
  & $0.16420(67)$    & $2.25(14)$ & $+ 7$         & $+42$         & $ -0$         & $+31$\\
\hline
\hline
&{\bf average} &&& $+6$ &  $+14$ & $+3$ & $+3$ \\
&{\bf standard deviation of mean} &&& $2$ &  $3$ & $1$ & $2$ \\
\hline
\hline
\end{tabular}
\end{table}
Given $\kappa_s$, the next step is to determine the scale, $a^{-1}$.  We consider 
one of the two quantities in the ratio used to determine $\kappa_s$, 
extrapolate or interpolate as needed to reach $\kappa = \kappa_s$, 
and set the result to the ``physical'' value of that quantity.  
The results from either of the two quantities in the ratio
should be consistent; they are.  Figure~\ref{fig:fss-R-fketa2-meth4}
shows a quadratic extrapolation of $f_{qq}$ to $\kappa_s$; $a^{-1}$ is fixed from the result
{\it via}\/ \eq{fss2}.  Although an extrapolation is again required in this particular case,
it is only over a short distance in quark mass.  

\begin{figure}[thb!]
\includegraphics[bb = 0 0 4096 4096,
width=5.0truein]{fss-R-fketa2-meth4.ps}
\caption{Quadratic extrapolation on set R of $f_{qq}$ to $\kappa_s$, which in
turn was found by extrapolation of $m_{ss}/f_{ss}$ using \eqs{mss-eta}{fss2}.
This is method 4 in Table~\ref{tab:decay-logs}.
}
\label{fig:fss-R-fketa2-meth4}
\end{figure}

Once the scale is determined, the standard extrapolation of $m^2_{qq}$ produces
the light quark hopping parameter, $\kappa_{u,d}$.  It is, of course, very close
to $\kappa_c$ in all cases.

Results for various scale determinations
are shown in Table~\ref{tab:decay-logs}.  In most cases, the values of
$a^{-1}$ there are significantly larger than those from the standard linear (or quadratic) extrapolation
of $f_\pi$ (see Table~\ref{tab:ainverse}).  This is not unexpected because extrapolation
from relatively large mass without the chiral log term in \eq{fpi-logs} should overestimate
$af_\pi$. Further, other lattice spacing determinations from light quark physics
(\eg $m_\rho$, $m_N$) also typically involve linear or quadratic chiral extrapolations from rather high masses,
so their generally good agreement with the standard $f_\pi$ scale cannot be used to rule
out the results in Table~\ref{tab:ainverse}.  Indeed, it has been known for some time that
heavy quark physics (charmonium or upsilonium) typically gives scales $\sim 10$ to 20\% larger
than light quark physics.  (For a recent example, see Table I in Ref.~\cite{FERMI-BDlnu}.)
It now appears that at least some of this discrepancy is due to the extrapolation in the light
quark mass.  

On the other hand, the result for $a^{-1}$
is unreasonably large ($\gtwid 30\%$ bigger than the central value)
in three cases in Table~\ref{tab:decay-logs}. (In all other cases the scale
is at most 17\% greater than the central value.) The three cases also have
long extrapolations to find $\kappa_s$ (\ie a value of $\kappa_s$ very close to $\kappa_c$ ---
compare Table~\ref{tab:kc}) 
and very large statistical errors in both $\kappa_s$ and $a^{-1}$. These cases are marked
with asterisks and are omitted from any averages of decay constant effects.

Given $\kappa_s$, $\kappa_{u,d}$, and $a^{-1}$, the ratio $f_{qq}/f_{Qq}$ can now be interpolated/extrapolated
as in Fig.~\ref{fig:fpiofB-R-kp2-meth6.125} to $\kappa_s$ and $\kappa_{u,d}$.
Using the physical $f_\pi$ and the relevant choice for $f_{ss}$, this produces $f_{Qs}$ or $f_{Qu,d}$, which
are then interpolated in heavy quark mass in the same way as in our standard analysis, described 
in the next section.
The differences of the final decay constants from the central values are displayed 
in Table~\ref{tab:decay-logs}
for each of the methods.
These differences will be used  in Sec.~\ref{sec:chiral-log-errors} to estimate the effects
of chiral logs in the dynamical case.  
We also explain there why we think it would be
inappropriate at this stage to {\em correct} the central values by the chiral log effects.
Instead we use the changes shown in Table~\ref{tab:decay-logs}  only to estimate the systematic error.

\subsection{Interpolations in Heavy Quark Mass}
\label{sec:HQET}

We proceed to compute physical decay constants such as $f_B$
and $f_{B_s}$ for each lattice set. Our starting point are the values
of \frootm\ (for $q=u,d$ or $q=s$) as functions of the heavy-quark mass
produced by the chiral fits of the previous two sections.

The static limit is also included where we have it.
According to the heavy quark effective theory (HQET) \cite{HQET},  \frootm\ 
should depend on the heavy meson mass as a polynomial in $1/M_{Qq}$, up to logarithms.
We therefore first divide out the one-loop logarithmic dependence of
the decay constants in the heavy quark limit \cite{LOG}, producing
what we call \frootmp:
\begin{equation}
\label{eq:frootmp}
\efrootmp = \frac{ \efrootm }{ 1 + \alpha_V(q^*) \ln (a M_{Qq})/\pi }\ ,
\end{equation}
where we have ignored the difference between the heavy quark and heavy
meson masses, and where $q^*$ takes the values discussed in
Secs.~\ref{sec:wilson-formalism} and \ref{sec:clover-formalism}.

The data is now expressed in physical units, always using $f_\pi$
to set the scale for the central values.
The quantity \frootmp\ is then plotted \vs $1/M_{Qq}$, where $M_{Qq}$ is 
the kinetic meson mass $M_{Qq,2}$ defined in \eq{mhl2}. 
We fit to a 
polynomial in $1/M_{Qq}$, interpolate to $m_B$, $m_{B_s}$,
$m_D$ or $m_{D_s}$, and then replace the
logarithm in \eq{frootmp}, evaluated at the appropriate meson mass. 
These are always interpolations, not extrapolations, because we have
either the static-light point (all Wilson sets) or 
heavy-light masses above the $B$ (the clover sets CP1 and J, using
the Fermilab formalism).  

For the quenched Wilson data, we do two versions of the polynomial fit: (1) a quadratic fit to
heavy-light mesons in the approximate mass range $2$ 
to $4\;\GeV$ plus the static-light meson (``heavier-heavies'')
and (2) a quadratic fit to
mesons in the approximate mass range $1.25$ to 
$2\;\GeV$ plus the static-light meson (``lighter-heavies'').   
These fits keep just one eigenvector of the correlation
matrix, which corresponds to $\lambda_{\rm cut}=0.9$ to 1.1.
To estimate the effect of leaving out higher powers
in $1/M_{Qq}$ in the fits, we also 
perform,
for the central $q^*$ and chiral fit choices, fit (3):
a cubic fit to all
the mesons in the range $1.25$ to $4\;\GeV$ plus the static-light meson.
The correlation matrix for fit (3) typically has
almost twice the number of eigenvectors as fits (1) and (2), and we 
keep 2 of the them.  This corresponds to $\lambda_{\rm cut}=0.2$ to 1.0.

We make basically the same fits for the
Wilson data on the dynamical lattices.  
The main difference is that we cut off fits
(1) and (3) at approximately $3$, rather than $4$, $\GeV$.  
These lattices are almost all quite large, and, 
as explained in Sec.~\ref{sec:generation}, we have trouble pulling out the lightest
state for very heavy masses on large lattices with our approach to 
the hopping expansion. To make up for some of the points lost by
the reduced upper cutoff on fit (1), we also reduce the lower
cutoff slightly, to $1.8\;\GeV$.

For the quenched clover sets, we make corresponding fits.
However, the mass ranges
are somewhat different because we have only five heavy quark
values, do not have a static point, and, most importantly,
use standard algorithms with FFT, facilitating the  
extraction of the lowest states even for very heavy masses.   
In this case, fit (1) (heavier-heavies)
is a quadratic fit 
over the approximate meson mass range $2.3$
to $6\;\GeV$; while fit (2) (lighter-heavies) is over the mass range
$1.7$ to $3\;\GeV$. 

In central values, we use fit (1) for $f_B$ and $f_{B_s}$
and fit (2) for $f_D$ and $f_{D_s}$. The alternative fits go into
the systematic error estimates, as in Ref.~\cite{MILC-PRL}.
However, for the central values of ratios involving both $B$ and $D$ mesons
(\ie $f_B/f_{D_s}$, $f_{B_s}/f_{D_s}$, and \fbofd),
both numerators and denominators
are taken from fit (2).  As explained 
in Sec.~\ref{sec:quenched}, this tends
to reduce the estimate of the magnetic mass error. 

\begin{figure}[tb!]
\includegraphics[bb = 0 0 4096 4096,
width=5.0truein]{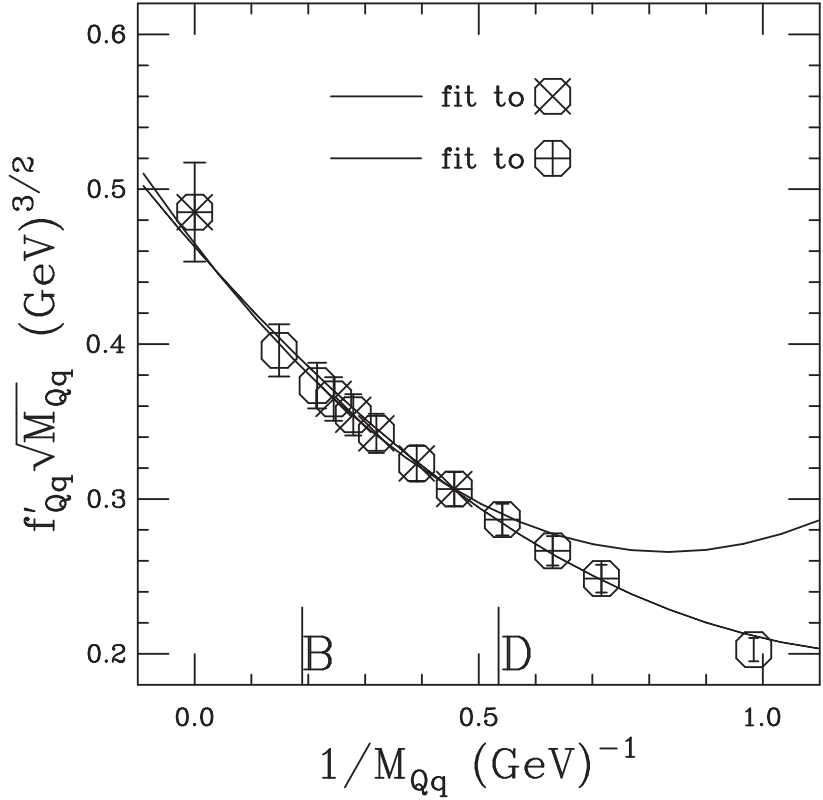}
\caption{\frootmp {\it vs.}\ $1/M_{Qq}$ for set CP (quenched Wilson).
The scale is set by $f_\pi$.
The solid line is fit (1)  (``heavier-heavies")  and includes points
marked with a cross.  The dotted line is fit (2) (``lighter-heavies")
and includes points marked with a plus. The fits have
$\chi^2_{\rm cut}/{\rm d.o.f.}=0.4$ and $0.9$, respectively, with
$\lambda_{\rm cut}=1$ (1 eigenvector kept).
}
\label{fig:frootm-CP}
\end{figure}

\begin{figure}[thb!]
\includegraphics[bb = 0 0 4096 4096,
width=5.0truein]{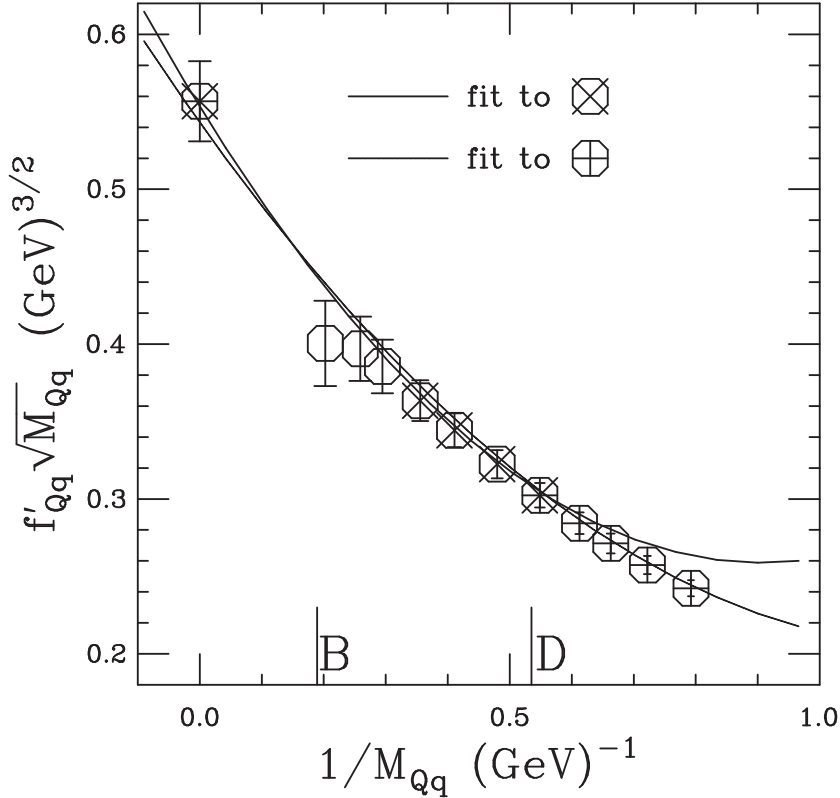}
\caption{Same as Fig.~\ref{fig:frootm-CP}, but for set R (Wilson valence quarks
on dynamical lattices).
The solid line has
$\chi^2_{\rm cut}/{\rm d.o.f.}=0.2$; the dotted line, 0.3.
The tail-off of \frootmp\ in the range $0.2\;\GeV^{-1} < 1/M_{Qq} < 0.3\;\GeV^{-1}$
is attributed to the difficulty in isolating asymptotic states for large masses
and volumes --- see text.
}
\label{fig:frootm-R}
\end{figure}

Figures~\ref{fig:frootm-CP}, \ref{fig:frootm-R}, and
\ref{fig:frootm-J-NP-tad} give examples of the behavior
of \frootmp\ for the quenched Wilson, Wilson on dynamical lattices,
and quenched clover cases, respectively.  Essentially all fits on all sets are acceptable.
In Fig.~\ref{fig:frootm-R},  one can see the tail-off of \frootmp\ for large
heavy quark masses  ($0.2\;\GeV^{-1} < 1/M_{Qq} < 0.3\;\GeV^{-1}$).
As mentioned above, we attribute this to contamination by higher momentum states,
which, for large masses and volumes, are very close in energy to the zero momentum state.
These points
are therefore not included in the fits. 
Note that the term of order $1/M_{Qq}^2$ is not reliably determined in our
data; it changes sign between the Wilson and the clover
cases. This is not surprising since in neither case is the formalism
correct through order  $1/M_{Qq}^2$ for $aM_Q\sim1$.

\begin{figure}[thb!]
\includegraphics[bb = 0 0 4096 4096,
width=5.0truein]{frootm_J_NP-tad.ps}
\caption{Same as Fig.~\ref{fig:frootm-CP}, but for set J (quenched clover),
with NP-tad renormalization.
The solid line has
$\chi^2_{\rm cut}/{\rm d.o.f.}=0.55$; the dotted line has no degrees of freedom.
The solid line is slightly concave down, unlike the case in
Figs.~\ref{fig:frootm-CP} and \ref{fig:frootm-R}.
}
\label{fig:frootm-J-NP-tad}
\end{figure}

Using fits like those in 
Figs.~\ref{fig:frootm-CP}--\ref{fig:frootm-J-NP-tad}, 
we now interpolate the data, 
replace the
perturbative logarithm in \eq{frootmp}, 
and divide by the appropriate $\sqrt{M_{Qq}}$ to find
\fb, \fbs, \fd, and \fds\ for each data set.  The
resulting decay constants and ratios will be extrapolated
to the continuum in Secs~\ref{sec:quenched} and \ref{sec:dynamical}.
Before doing so, however, we repeat the analysis so far for all the
other $\sim\!25$ versions of reasonable plateau choices, as discussed in
Sec.~\ref{sec:correlators}.
We then find the standard deviation of the
results over the other versions and add it in quadrature with the raw
jackknife error of the central value. 
Henceforth, the statistical error of any quantity will be taken
from the result of this procedure.  Typically the procedure increases
the statistical errors by $\sim\!\sqrt{2}$; we believe it 
mitigates any biases introduced from our choice, for the central values,
of the fits with lowest statistical errors and $\chi^2_{\rm cut}/{\rm d.o.f}$.

In Table~\ref{tab:fB-from-sets}, we collect the central
values of \fb, \fbs, \fd, and \fds\ for the various sets.
Similarly, Table~\ref{tab:ratios-from-sets} gives central values
of ratios \fbsofb, \fdsofd, \fbofds, \fbsofds, and \fbofd.

\begin{table}[tbh!]
\caption{Central values of decay constants, in $\MeV$, for each data set.
Statistical errors include the effect of changing the time ranges over
which correlators are fit, as described in the text. 
For sets A and B, the values
reported are those for which the light-light results ($\kappa_c$,
$\kappa_s$ and $a^{-1}$) and their errors are taken from the averages over set 5.7-large.
}
\label{tab:fB-from-sets}
\def\arraystretch{.8}
\begin{tabular}{|c|l|l|l|l|}
\noalign{\vspace{1cm}}
\hline
name&\fb &\fbs &\fd &\fds \\
\hline
\multicolumn{5}{c}{quenched Wilson}\\
\hline
 A & 193.6(8.6) & 234.9(5.8) & 216.6(7.8) & 256.9(6.5) \\
 B & 196.8(11.1) & 236.1(8.2) & 220.7(10.9) & 261.8(7.4) \\
 E & 190.9(7.7) & 219.1(7.8) & 214.7(8.6) & 246.1(8.1) \\
 C & 172.3(8.0) & 206.2(6.8) & 198.8(7.7) & 232.8(6.9) \\
 CP & 177.0(7.8) & 210.8(6.4) & 206.7(6.4) & 238.2(5.7) \\
 D & 174.9(7.5) & 199.8(6.5) & 206.8(7.9) & 232.5(6.3) \\
 H & 180.6(12.1) & 206.7(10.8) & 206.6(10.9) & 232.9(9.0) \\
\hline
\multicolumn{5}{c}{$N_f=2$ Wilson}\\
\hline
 L & 188.6(9.7) & 220.3(10.3) & 214.7(7.4) & 249.5(6.7) \\
 N & 205.7(13.6) & 239.0(10.4) & 222.9(10.6) & 261.5(7.6) \\
 O & 206.8(12.6) & 239.9(10.8) & 230.8(6.3) & 262.4(5.0) \\
 M & 190.6(12.5) & 226.9(10.2) & 215.9(11.5) & 250.2(8.4) \\
 P & 193.1(6.9) & 225.6(6.4) & 212.9(6.7) & 249.5(5.9) \\
 U & 196.5(9.4) & 235.0(7.9) & 224.8(7.2) & 261.2(7.6) \\
 T & 193.3(15.8) & 219.3(12.5) & 209.1(8.3) & 236.3(6.7) \\
 S & 202.6(6.8) & 234.6(5.6) & 223.9(5.2) & 256.6(4.7) \\
 G & 198.5(6.2) & 234.0(6.7) & 220.0(5.0) & 254.7(5.2) \\
 R & 206.2(7.9) & 239.2(8.1) & 223.4(5.4) & 254.6(5.1) \\
\hline
\multicolumn{5}{c}{quenched clover}\\
\hline
 CP1\_NP-IOY & 184.1(5.7) & 212.8(4.4) & -- &  -- \\
 CP1\_NP-tad & 176.1(5.2) & 203.4(3.9) & 196.3(3.7) & 220.0(2.8) \\
 J\_NP-IOY & 176.6(6.3) & 204.4(5.9) & --  & -- \\
 J\_NP-tad & 174.0(6.0) & 201.9(5.7) & 203.5(4.8) & 228.2(4.1) \\
\hline
\end{tabular}
\end{table}

\begin{table}[tbh!]
\caption{Same as Table~\protect{\ref{tab:fB-from-sets}}, but for
ratios of decays constants. 
}
\label{tab:ratios-from-sets}
\def\arraystretch{.8}
\begin{tabular}{|c|l|l|l|l|l|}
\noalign{\vspace{0.5cm}}
\hline
name&\fbsofb &\fdsofd &\fbofds &\fbsofds &\fbofd \\
\hline
\multicolumn{5}{c}{quenched Wilson}\\
\hline
 A  & 1.213(39) &   1.186(21) &   0.768(24) &   0.926(20) &   0.911(27) \\
 B  & 1.200(30) &   1.187(29) &   0.793(31) &   0.928(28) &   0.941(50) \\
 E  & 1.147(34) &   1.146(16) &   0.739(40) &   0.916(33) &   0.846(46) \\
 C  & 1.197(22) &   1.171(17) &   0.740(28) &   0.888(25) &   0.867(32) \\
 CP  &   1.191(21) &   1.152(12) &   0.752(36) &   0.899(26) &   0.867(41) \\
 D  & 1.142(16) &   1.124(16) &   0.755(23) &   0.866(20) &   0.849(26) \\
 H  & 1.145(22) &   1.128(18) &   0.776(28) &   0.888(19) &   0.875(25) \\
\hline
\multicolumn{5}{c}{$N_f=2$ Wilson}\\
\hline
 L  & 1.168(16) &   1.162(16) &   0.777(30) &   0.918(22) &   0.903(38) \\
 N  & 1.162(33) &   1.173(25) &   0.788(37) &   0.927(29) &   0.924(41) \\
 O  & 1.160(25) &   1.137(18) &   0.789(32) &   0.916(23) &   0.897(30) \\
 M  & 1.191(30) &   1.159(26) &   0.793(25) &   0.928(22) &   0.920(25) \\
 P  & 1.168(12) &   1.172(12) &   0.784(16) &   0.914(12) &   0.919(18) \\
 U  & 1.196(28) &   1.162(12) &   0.763(41) &   0.904(37) &   0.886(42) \\
 T  & 1.134(34) &   1.131(16) &   0.840(41) &   0.958(32) &   0.950(41) \\
 S  & 1.158(17) &   1.146(11) &   0.790(19) &   0.927(15) &   0.905(19) \\
 G  & 1.179(14) &   1.158( 8) &   0.788(20) &   0.937(26) &   0.912(22) \\
 R  & 1.160(15) &   1.140( 8) &   0.814(28) &   0.947(20) &   0.927(32) \\
\hline
\multicolumn{5}{c}{quenched clover}\\
\hline
 CP1\_NP-IOY & 1.156(16) &   1.120( 8) &   --  &   --  &   -- \\
 CP1\_NP-tad &1.155(16) &  1.121( 9) &  0.800(19) &  0.923(12) &  0.896(18) \\
 J\_NP-IOY &  1.157(14) &  1.117( 8) &  -- &  -- &  -- \\
 J\_NP-tad &  1.160(14) &  1.121( 8) &  0.772(21) &  0.892(18) &  0.866(24) \\
\hline
\end{tabular}
\end{table}

\section{Quenched Approximation Results}
\label{sec:quenched}

Final results and errors in the quenched approximation are determined
much as in Ref.~\cite{MILC-PRL}.  However, there are some significant differences,
especially for the continuum extrapolation and the estimate of the associated
errors.  We discuss our methods in detail where they differ from \cite{MILC-PRL};
where the methods are the same, we include only a very brief description for
completeness.

We begin with the continuum extrapolation of
various quantities.  We focus on \fb, \fbs, \fbsofb, and \fds,
which are probably the most important,  phenomenologically.
Figures \ref{fig:fb-quenched}--\ref{fig:fds-quenched}
show the data for these quantities as a function of lattice spacing.
The behavior of the other decay constants and ratios is similar.

\begin{figure}[thb!]
\includegraphics[bb = 0 0 4096 4096,
width=5.0truein]{fb_quenched.ps}
\caption{$f_B$ {\it vs.}\ $a$\/ for quenched lattices; the scale is
set by $f_\pi$.  Diamonds are results
with Wilson light quarks and Wilson or static heavy quarks. Octagons
and crosses are results with nonperturbative clover heavy and light quarks;
``NP-IOY'' (octagons) and ``NP-tad'' (crosses) differ  in how
the renormalization of the heavy quarks is performed (see text).
For clarity, the octagons have been moved slightly to
the right, and the fit to the crosses has been 
slightly lowered.
}
\label{fig:fb-quenched}
\end{figure}

\begin{figure}[thb!]
\includegraphics[bb = 0 0 4096 4096,
width=5.0truein]{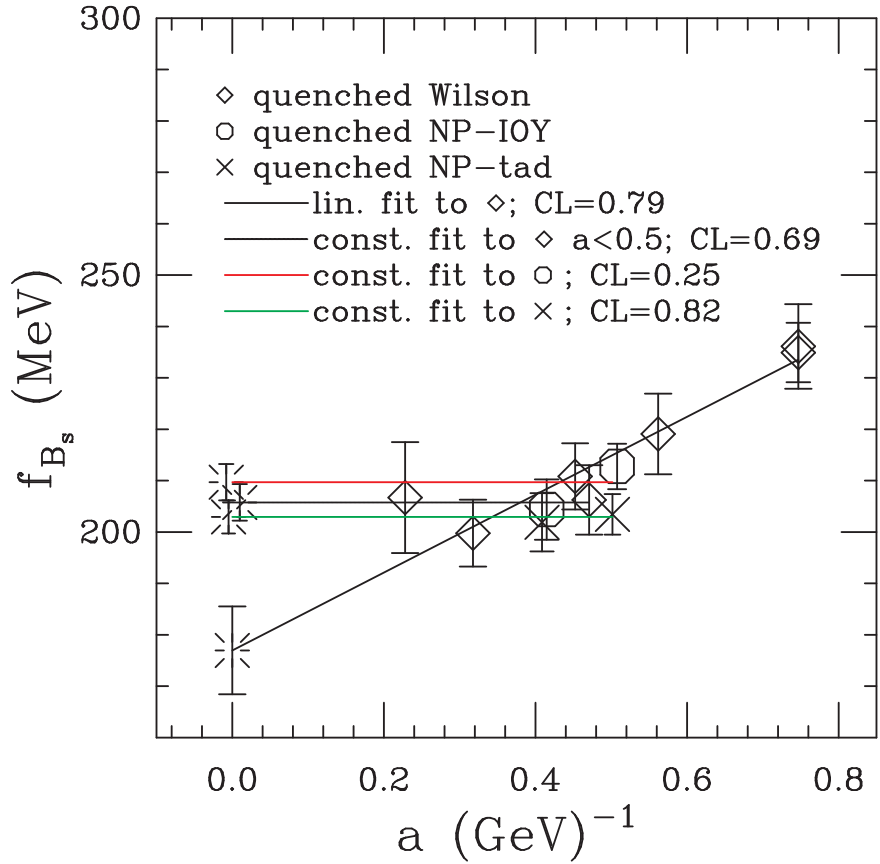}
\caption{Same as Fig.~\ref{fig:fb-quenched}, but for \fbs.
For clarity, the octagons have been moved slightly to
the right.
}
\label{fig:fbs-quenched}
\end{figure}

\begin{figure}[thb!]
\includegraphics[bb = 0 0 4096 4096,
width=5.0truein]{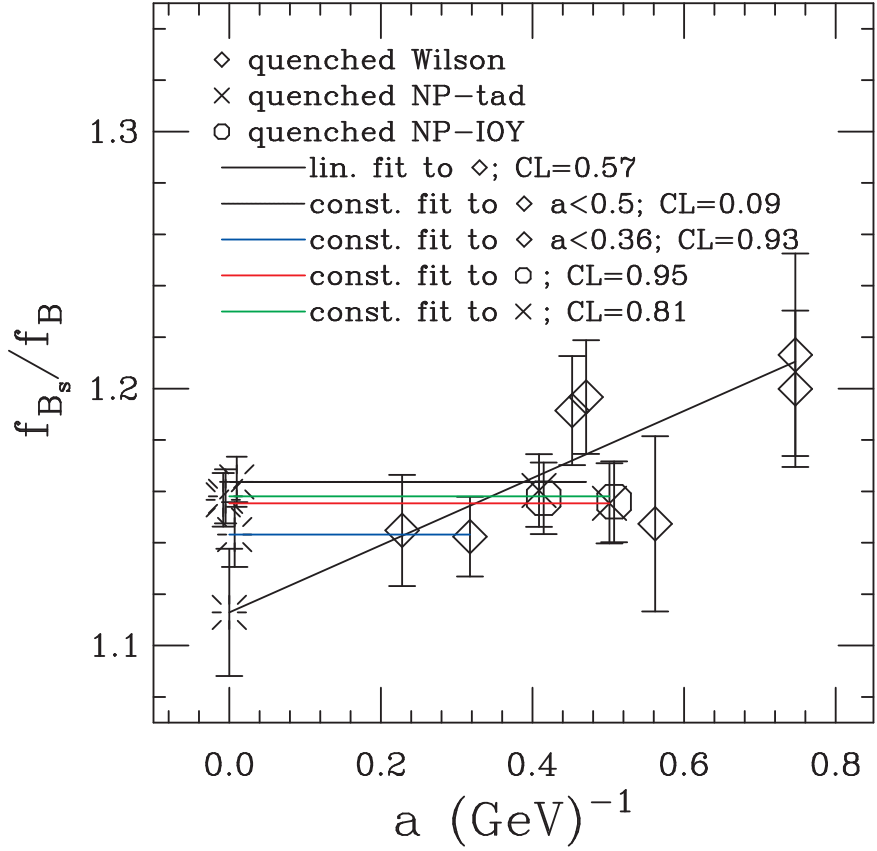}
\caption{\fbsofb\ {\it vs.}\ $a$ for quenched lattices. Labels are
the same as in Fig.~\ref{fig:fb-quenched}, but one additional fit is
shown: a constant fit to the two diamonds (Wilson quark results)
with smallest $a$ ($a<0.36 \GeV^{-1}$).
For clarity, the octagons have been moved slightly to
the right and the fit to the octagons has been
slightly lowered.
}
\label{fig:fbsofb-quenched}
\end{figure}

\begin{figure}[thb!]
\includegraphics[bb = 0 0 4096 4096,
width=5.0truein]{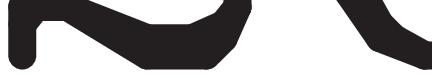}
\caption{\fds\  {\it vs.}\ $a$ for quenched lattices. 
Labels are the same as in Fig.~\ref{fig:fb-quenched}, but the
NP-IOY points have been omitted because the perturbative calculation
is not available at the relevant lattice masses.
}
\label{fig:fds-quenched}
\end{figure}

It is not {\it a priori} obvious how to extrapolate decay constants and
ratios to the continuum. As discussed in Sec.~\ref{sec:wilson-formalism},
our Wilson valence results have $\alpha_V^2$ errors as well as
errors of the form $a\Lambda_{QCD}\times h(aM_Q)$.
Here $h(aM_Q)$ is a calculable (in perturbation theory) function that
is expected to be $\cO(1)$ everywhere.\footnote{Despite the fact that
static quarks are trivially $\cO(a)$ improved, the function
$h(aM_Q)$ does not vanish even as $M_Q\to\infty$ for fixed $a$, because
the Wilson light quarks still have $\cO(a)$ errors.} Since $aM_Q\gtwid 1$
for our entire range of $a$ values, the assumption of
a dominantly linear dependence on
$a$ is only one possibility.  A practical alternative is the assumption
that, for $a$ smaller than some value, the errors are small enough that
the difference with continuum values is negligible --- so that extrapolation
with a constant function is warranted.

We confront these assumptions with the data in
Figs.~\ref{fig:fb-quenched}--\ref{fig:fds-quenched}. For the Wilson
valence data,  we show linear fits over all $a$
and constant fits for $a<0.5\;\GeV^{-1}$ 
($\beta\ge6.0$).  Both types of fits are generally quite good.  The exception
is the constant fit for the ratio \fbsofb.  In \cite{MILC-PRL}, the relatively
poor confidence level of
the constant fit for \fbsofb\ (or \fdsofd)
relative to that of the linear fit led us to
choose a linear extrapolation for the central value of the ratios.  That, 
in turn,
required choosing the linear extrapolation for the central values of the decay
constants themselves, since it would be inconsistent to assume linear
behavior for \fbsofb\ but constant behavior for \fb\ and \fbs\ separately.
Note, however, that if we just look at the two finest lattices
($a<0.36\;(\GeV)^{-1}$, $\beta\ge6.3$), the behavior of \fbsofb\ is
quite consistent with a constant; such a fit is also shown in
Fig.~\ref{fig:fbsofb-quenched}.
 
The new quenched clover data, shown in 
Figs.~\ref{fig:fb-quenched}--\ref{fig:fds-quenched} for both the NP-IOY and 
NP-tad schemes, has clarified the 
situation somewhat. The discretization errors here should be considerably 
smaller than for Wilson valence quarks.  As discussed in Sec.~\ref{sec:clover-formalism}
the errors are formally $\cO(a^2\Lambda_{QCD}^2)$ and either $\cO(\alpha_V^2)$ (NP-IOY),
or $\cO(a^2M_Q^2)$ (NP-tad).  Because there will be also be a function
like $h(aM_Q)$ in this case, the actual behavior with $a$ when $aM_Q\sim1$
is likely to be complicated.  The best we can do with just two clover data
points is to assume that the errors are small enough that a constant
extrapolation is warranted; such fits are shown in
Figs.~\ref{fig:fb-quenched}--\ref{fig:fds-quenched}.
Comparable extrapolation of clover data with a constant was
performed in Refs.~\cite{FERMILAB,JLQCD98}. 

For \fbsofb\ the clover data show very little $a$ dependence and give
a result compatible with the various constant fits to the  small-$a$
Wilson data. The clover results are not compatible with
the linear extrapolation of the Wilson
data, which is now seen to give a rather low result.  
Recent preliminary quenched
results \cite{MILC-3flavor} with clover valence quarks on 
Symanzik improved glue are also incompatible with the Wilson linear 
extrapolation.
For our central quenched value of \fbsofb\ or \fdsofd\ we therefore drop the
linear Wilson extrapolation and average the four
constant extrapolations: 
two for Wilson ($a<0.5\;\GeV^{-1}$ and $a<0.36\;\GeV^{-1}$)
and two for clover (NP-IOY and NP-tad).\footnote{Although the calculation
in Ref.~\cite{IOY} is not well controlled at the $D$ mass,
the NP-IOY procedure may be used for \fdsofd\ because the renormalizations
cancel. Note that NP-tad involves the light quark mass in
the renormalizations (see eqs.~\protect{\ref{eq:Rdef}} and 
\protect{\ref{eq:A-NPtad}}) so does not
give identical results to NP-IOY even for \fbsofb\ or \fdsofd.}
The systematic error of the
continuum extrapolation is then taken as the standard deviation of the
four individual
extrapolations. 
The other decay constant ratios (\fbofds, \fbsofds, and \fbofd) are treated
similarly, although there is one fewer result to average,
since NP-IOY is not applicable.

Though we have dropped the linear extrapolation from the analysis of the
ratios, it is not inconsistent to include it in
the analysis of the decay constants themselves. Indeed, for \fb\ and
\fbs, the downward trend of the clover data as $a$ decreases makes
it difficult to rule out the linear extrapolation of the Wilson data.  
On the other hand, constant clover extrapolations do give results closer
to the constant Wilson extrapolations than to the linear Wilson extrapolations.
For \fds\ (and \fd, not shown), the situation is reversed:
The clover data has an upward trend as $a$ decreases; yet constant
clover extrapolations give results (slightly) 
closer to the linear Wilson extrapolations
than to the constant Wilson extrapolations. To obtain the central
values of the decay constants, we therefore average the results of
all the extrapolations and take the standard deviation of the results as the
continuum extrapolation error.  For \fb\ and \fbs\
a total of four fits are included: linear Wilson, constant Wilson,
constant NP-IOY and constant NP-tad.  For \fd\ and \fds\, there
are three fits, since NP-IOY is omitted.

As described in Sec.~\ref{sec:chiral}, we estimate the chiral extrapolation
errors
by comparing (after continuum extrapolation) the central values (which use
``chiral choice I'')
with those obtained by changing the chiral fits of the heavy-light
and light-light decay constants from linear to quadratic 
(``chiral choice II'').\footnote{In the dynamical case, we 
attempt to estimate an additional chiral error coming from
chiral logarithms by performing a separate chiral extrapolation of $f_{qq}/f_{Qq}$.
(See Sec.~\ref{sec:chiral-logs} and \ref{sec:chiral-log-errors}.) This is not feasible in the 
quenched case:  the quenched chiral logs in $f_{Qq}$ have coefficients  with unknown
magnitude and sign \cite{BOOTH}; while $f_{qq}$ has no quenched logs at all at one loop \cite{QChPT}.}
The errors of the chiral extrapolation and other systematic errors within the quenched approximation
are collected in Tables~\ref{tab:decay-syst-quench} and
\ref{tab:ratio-syst-quench}.  Note that the quoted chiral errors
are all positive. This can be traced to the effect of the quadratic extrapolation
of $f_{qq}$ (used to set the scale through $f_\pi$), which is clearly, though slightly,
concave down (see Figs.~\ref{fig:fpi-J} and \ref{fig:fpi-R}).  The concavity
in $f_{Qq}$ in the region of the $B$ is less pronounced.
 
\begin{table}[tbh!]
\caption{Central values ($f_\pi$ scale) and errors  in $\MeV$
for the quenched decay constants. 
The statistical errors and the effects of excited states
are combined, as described at the 
end of Sec.~\ref{sec:HQET}.  
Errors marked with explicit $+$ or $-$ signs are treated
as signed;  all others are treated as symmetric.
The scale  and $\kappa_s$ errors are not included in the
total error {\it within} the quenched approximation
but are shown for completeness.
}
\label{tab:decay-syst-quench}
\def\arraystretch{.8}
\begin{tabular}{|c|c|c|c|c|}
\noalign{\vspace{0.5cm}}
\hline
&\fb &\fbs &\fd &\fds \\
\hline
 central value & 	173.0 & 198.8 & 199.5  & 223.2 \\
\hline
\multicolumn{5}{c}{errors}\\
\hline
 statistics \& excited states &	5.7 & 4.7 & 5.6 & 4.6 \\
 continuum extrapolation &      8.7 & 14.8 & 4.9 & 11.1 \\
 chiral extrapolation &       $+8.6$& $+9.4$ & $+4.3$ & $+6.5$ \\
 perturbative & 	        9.6 & 14.1 & 6.6 & 10.9 \\
 magnetic mass &         	1.7 & 1.9 & 5.1 & 5.7 \\
 1/M fit & 	                2.0 & 1.7  & 0.5 & 0.2 \\
 finite volume &  $+2.8$ $-8.5$ & $+1.0$ $-7.8$ & $+3.8$ $-5.1$ & $+4.3$ $-4.1$ \\
\hline
 scale (change to $m_\rho$) & $-3.5$ 	 & $-5.9$ & $+4.2$ & $+4.0$ \\
 $\kappa_s$ (change to $\phi$) & --  & $+3.7$ & -- & $+2.3$ \\
\hline
\end{tabular}
\end{table}

\begin{table}[tbh!]
\caption{Same as Table~\protect{\ref{tab:decay-syst-quench}}
but for decay constant ratios. 
}
\label{tab:ratio-syst-quench}
\def\arraystretch{.8}
\begin{tabular}{|c|c|c|c|c|c|}
\noalign{\vspace{0.5cm}}
\hline
&\fbsofb &\fdsofd &\fbofds &\fbsofds  &\fbofd \\
\hline
 central value & 	1.155 & 1.128 & 0.769  & 0.891 & 0.871 \\
\hline
\multicolumn{5}{c}{errors}\\
\hline
 statistics \& excited states &	0.011 & 0.008 & 0.015 & 0.012 & 0.016\\
 continuum extrapolation &      0.009 & 0.012 & 0.017 & 0.019 & 0.013 \\
 chiral extrapolation &       $+0.003$& $+0.014$ & $+0.009$ & $+0.009$ & $+0.020$ \\
 perturbative & 	        0.008 & 0.011 & 0.015 & 0.020 & 0.017 \\
 magnetic mass &                0.000 & 0.000 & 0.015 & 0.018 & 0.017 \\
 1/M fit & 	                0.001 & 0.000  & 0.006 & 0.008 & 0.009 \\
 finite volume &  $+0.012$ $-0.013$ & $+0.008$ $-0.000$ & $+0.025$ $-0.000$ & $+0.003$ $-0.009$ & $+0.028$ $-0.000$ \\
\hline
 scale (change to $m_\rho$) & $+0.001$ 	 & $-0.001$ & $+0.009$ & $+0.008$ & $+0.007$ \\
 $\kappa_s$ (change to $\phi$) & $+0.025$ & $+0.018$ & $-0.009$ & $+0.004$ & -- \\
\hline
\end{tabular}
\end{table}

The perturbative error is estimated by varying, over a ``reasonable range,'' the
values of $q^*$ used in the one-loop renormalization constants.  For Wilson
fermions, we take the range for the heavy-light currents to be $1/a \le q^* \le 2.86/a$, 
with $1.43/a$ the central
value, as described in Sec.~\ref{sec:wilson-formalism}. Similarly,
for the light-light Wilson currents $q^*$ ranges between $1/a$ to $4.63/a$, with
$2.32/a$ the central value. In the clover case, perturbation theory for
the heavy-light currents is only relevant for NP-IOY. For central values, we
take $q^*=3.34/a$ (set CP1) and
$q^*= 2.85/a$ (set J), which come from the static-light calculation of \cite{CB-TD} 
with the corresponding clover coefficients. The scale $q^*$ is then allowed to range between
$1/a$ and twice the central value.
For light-light clover currents,
only $b_A$ is treated perturbatively; the central value for $q^*$ is taken to be
$1/a$ (see Sec.~\ref{sec:clover-formalism}). This gives the central values
for $b_A$ shown in Table~\ref{tab:NPparams}.  The upper end of the range of $b_A$ shown
comes from taking $q^*=0.7/a$; the lower end, from using ``boosted
perturbation theory'' with $g^2= 6/(\beta <P>)$. ($ <P>$ is the mean plaquette, normalized
to have maximum 1). This is equivalent to taking a $q^*$ of
roughly $5.25/a$, so we are using a rather conservative range.

As mentioned in Sec.~\ref{sec:wilson-formalism},
there is a systematic error associated with the fact that $c_{mag}\equiv M_2/M_3$
is not equal to 1 with Wilson fermions.
Because $c_{mag}$ has complicated dependence on $a$, this error is not
removed by any of the simple extrapolations available to us.  One may argue
that the residual effect is just one particular discretization 
error and therefore has already been included.    However, if one models this error
for both linear and constant extrapolations
using \eq{treemasses} for $M_2$ and $M_3$
(along the lines of what was done in Refs.~\cite{JLQCD98} and \cite{MILC-PRL}),
one finds that the error is larger
with a constant extrapolation but has the same (but unknown) sign in both cases.
Therefore we believe it reasonable to include as an additional error 
the linear extrapolation estimate of the Wilson magnetic mass error.
From the tadpole improved tree-level model, one estimates these errors as
$\sim\!2\%$ for $f_B$ and $\sim\!3\%$ for $f_D$
(see \cite{MILC-PRL}).  An alternative estimate comes from the comparison of
the results of interpolations to the physical heavy meson masses using
the ``heavier-heavies'' (fit (1) -- see Sec.~\ref{sec:HQET}) with those using the
``lighter-heavies'' (fit (2)):  the lighter masses are affected
much less by the magnetic mass error, and the static point is not affected at all.
We take the larger of the two estimates as our magnetic mass error
for Wilson fermions.  

The magnetic mass error is absent for clover fermions.
Therefore, in our final error budgets 
(Tables~\ref{tab:decay-syst-quench}, \ref{tab:ratio-syst-quench}) we multiply the
Wilson magnetic mass error by $1/2$ or $2/3$, depending on the relative
number of Wilson and clover estimates that go into the central value.

Note that the magnetic mass errors in the tables
are considerably smaller for $B$ mesons than for $D$ mesons,
despite the fact that the difference between $M_2$ and $M_3$ increases with the
lattice mass.   The point is that the magnetic mass errors
are  systematic effects on the $1/M_Q$ corrections, and such
corrections are inherently bigger for $D$ mesons than for $B$'s.  Further, 
especially large errors can be introduced if $1/M_Q$ fits in the range of the $B$ are
extrapolated back to the $D$ region.  For that reason we always use
fit (2) (``lighter-heavies'') for central values
of ratios that involve both $D$'s and $B$'s: \fbofds, \fbsofds, and \fbofd.

The remaining two systematic errors, the effect of the interpolation in $1/M_{Qq}$ and the finite 
size errors, are estimated just as in Refs.~\cite{MILC-PRL}.
For the central values, we truncate the fit of $f_{Qq}\sqrt{M_{Qq}}$ {\it vs}. $1/M_{Qq}$ 
at quadratic order.  We estimate the error thereby introduced by
changing to cubic fits (with mass range $1.25$
to $4$ GeV, plus the static point when available). The errors found are $\sim\!1\%$;
this is what one would expect if the mass scale of the cubic term is 
$\sim\!0.75\; \GeV$, roughly the scale size found in 
the linear and quadratic terms.

We estimate the finite volume effects by finding the fractional difference
between results on set A (spatial size $\sim\!1.2$~fm) and set B ($\sim\!2.5$~fm).
Since set A is smaller than the other quenched lattices ($\sim\!1.3$--$1.5$~fm)
and B is much larger,  this should bound the finite volume error.  
To be conservative, we consider both: (a) the difference when all 
quantities are computed individually on sets A and B and (b) the difference 
when the light-light quantities are held fixed to their values from set 5.7-large.
Since $f_\pi$ generally suffers larger finite size effects than $f_{Qq}$,
these two estimates typically have opposite signs; in that case we include both
estimates as signed errors. When the estimates have the same sign, however,
we simply choose the larger.

Tables~\ref{tab:decay-syst-quench} and \ref{tab:ratio-syst-quench} also show errors
associated with fixing the scale (changing from $f_\pi$ to $m_\rho$) and fixing
$\kappa_s$ (changing from using the pseudoscalars to using the $\phi$ meson).
Logically, these should be considered errors {\it of} the quenched approximation, not
{\it within} the quenched approximation, and are not included in this
section.  Indeed, the question ``what is $f_B$ in the
quenched approximation?''\ is only well defined when one specifies how the scale
is fixed.  Even in the continuum limit, different scale choices 
(and different ways of fixing $\kappa_s$ for strange-quark quantities)
must give different results in the quenched approximation.  The differences 
should of course  go away in the continuum limit of the full theory.
In Sec.~\ref{sec:dynamical}, where we attempt to quote results that
can be directly compared with experiment, such errors are taken into account.

Our final results for heavy-light decay constants {\it within} the quenched
approximation (fixing the scale by $f_\pi$) are 
\begin{eqnarray}
\label{eq:fBquench}
f_B & = & 173 (6) (16) \;\MeV; \quad
f_{B_s} =  199 (5) ({}^{+23}_{ -22}) \;\MeV\nonumber\\
f_D  & = &  200 (6) (11) \;\MeV; \quad
f_{D_s}  =  223 (5)({}^{+18}_{-17}) \;\MeV\nonumber\\
\frac{f_{B_s}}{f_B} & = & 1.16 (1) (2) ; \quad
\frac{f_{D_s} }{f_D}  =  1.13 (1)(2) \\
\frac{f_{B}}{f_{D_s}} & = & 0.77 (2) ({}^{+4}_{-3}) ; \quad
\frac{f_{B_s} }{f_{D_s}}  =  0.89 (1) ({}^{+4}_{-3}) \nonumber\\
\frac{f_B}{f_D} & = & 0.87 (2) ({}^{+5}_{-3}) \ . \nonumber
\end{eqnarray}
The errors are statistical and systematic (within the quenched
approximation), respectively. Relevant systematic errors 
in Tables~\ref{tab:decay-syst-quench} and \ref{tab:ratio-syst-quench} have been
combined in quadrature.  Errors whose signs are not likely to be reliably
determined by our procedures (continuum extrapolation,  perturbation theory,
magnetic mass, $1/M$ fits) have been treated as symmetric errors. The others
(chiral extrapolation and finite volume)
have been treated as signed errors.
The results in \eq{fBquench} differ from those in
Ref.~\cite{MILC-PRL} due to:
\begin{itemize}
\item[]{1.\ } Inclusion of new data from sets CP, CP1 and J. 

\item[]{2.\ }  Setting the central value of
the heavy-light scale from the static-light calculation of Ref.~\cite{CB-TD},
rather than that of Ref.~\cite{HERNANDEZandHILL}.

\item[]{3.\ } Other changes in analysis, motivated by the new runs.
The most important of these is
the way we find the central value of the continuum extrapolation (as discussed
above, we now average our four possible versions rather than taking only the 
linear Wilson fit).
In addition, the details of the error estimate for the chiral extrapolation have changed. 
Some alternative chiral fits used 
previously --- \eg linear fits of $m_\pi^2$ \vs quark mass --- are 
convincingly excluded by the new data.
\end{itemize}

\section{Results with Dynamical Quarks}
\label{sec:dynamical}
\subsection{Continuum Extrapolation}
\label{sec:continuum}
\begin{figure}[thb!]
\includegraphics[bb = 0 0 4096 4096,
width=5.0truein]{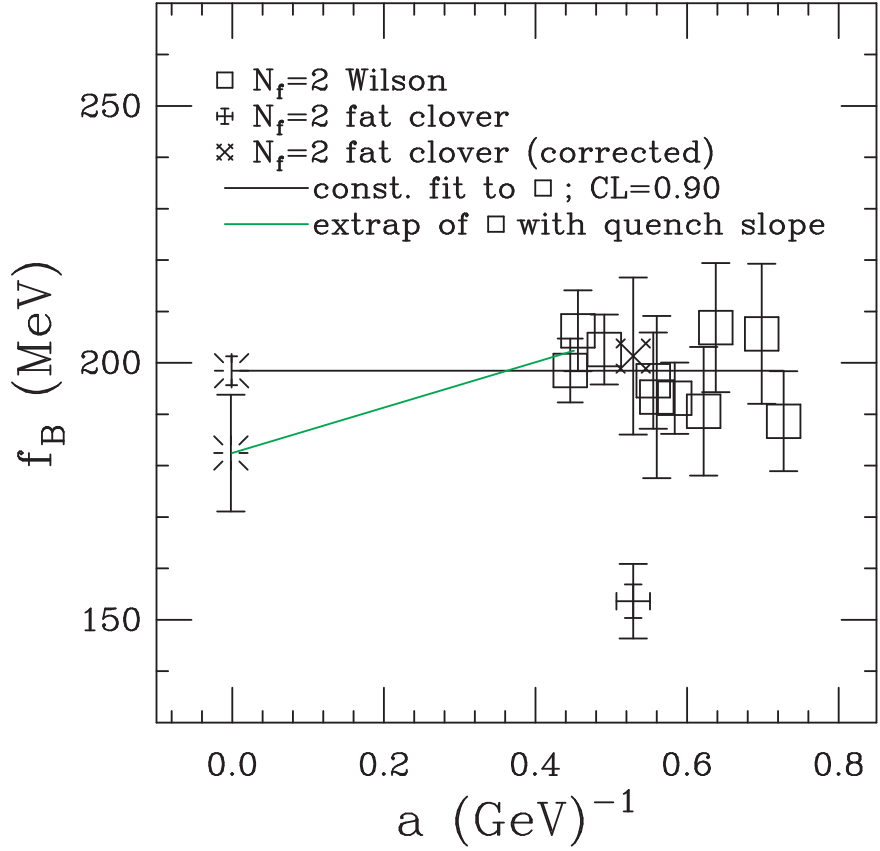}
\caption{$f_B$ {\it vs.}\ $a$ for dynamical $N_f=2$ lattices; 
a few points have been moved a slight distance horizontally for clarity.
Squares are results
with light Wilson valence quarks and Wilson or static heavy valence quarks. 
From left to right, the squares come from $\beta=5.6$ (sets G, R, S, T, U),
$\beta=5.5$ (sets P, M, O, N), and  $\beta=5.445$ (set L).
The solid line is a fit of all the Wilson results to a constant.
The dashed line shows what would happen if the dynamical results
decreased for smaller lattice spacing with the same slope as the
linear fit to the corresponding quenched data.
The fancy plus is the result with fat-link clover valence quarks
(light and heavy) on set RF. The fancy cross shows the  ``corrected'' value (see
text).  The fat clover data (corrected or uncorrected) is not included 
in our final results. 
}
\label{fig:fb-dynamical}
\end{figure}

\begin{figure}[thb!]
\includegraphics[bb = 0 0 4096 4096,
width=5.0truein]{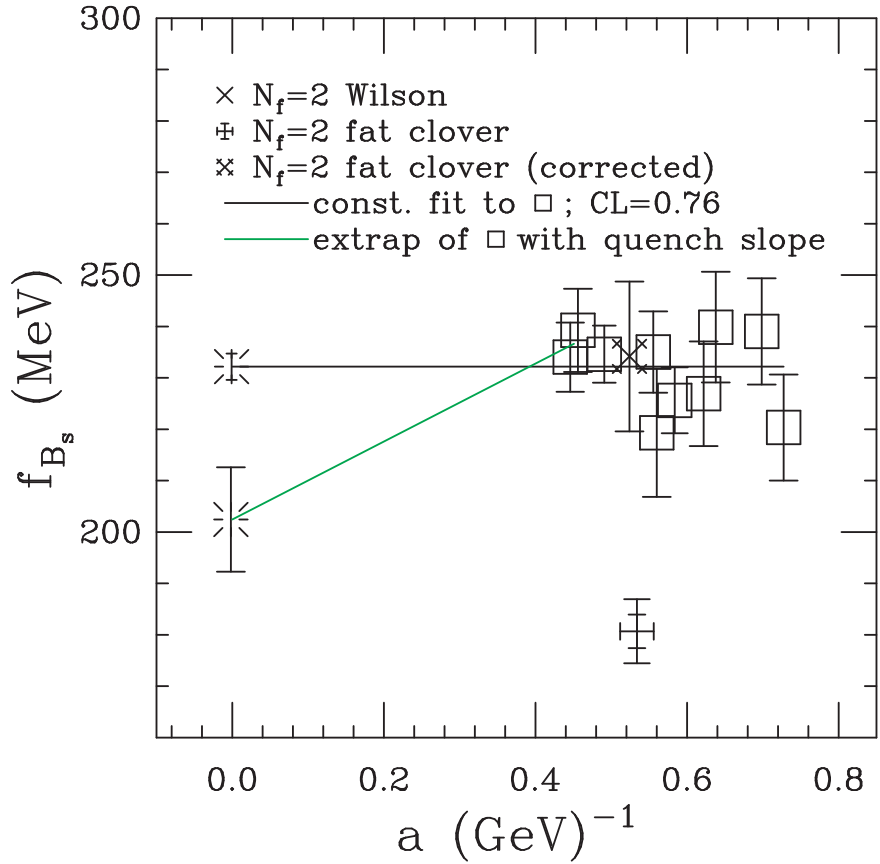}
\caption{Same as Fig.~\ref{fig:fb-dynamical}, but for \fbs.
}
\label{fig:fbs-dynamical}
\end{figure}

\begin{figure}[thb!]
\includegraphics[bb = 0 0 4096 4096,
width=5.0truein]{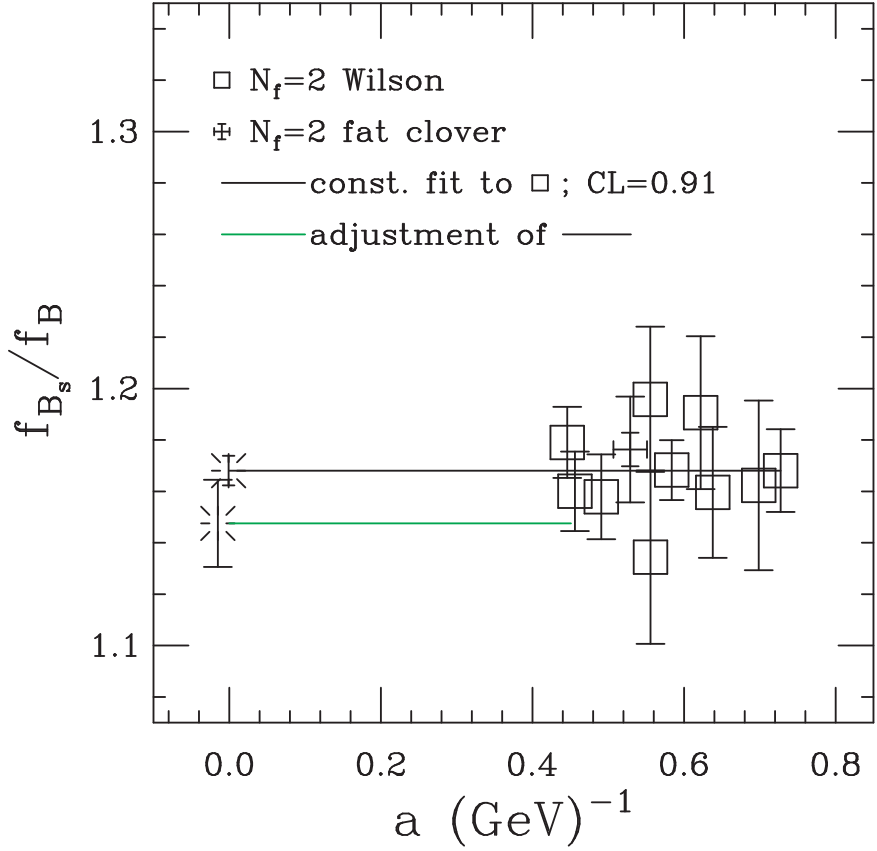}
\caption{\fbsofb\ {\it vs.}\ $a$ for dynamical $N_f=2$ lattices. Labels are
the same as in Fig.~\ref{fig:fb-dynamical}, but no correction to
the fatlink clover result is needed for the ratio of decay constants. 
The alternative dashed line assumes a drop when $a\to0$ that
is the same as the difference in the quenched
case between a constant fit to the results from the highest three $\beta$ values
and a  constant fit to those from the highest two $\beta$ values.
}
\label{fig:fbsofb-dynamical}
\end{figure}

Dynamical $N_f=2$ results for
\fb, \fbs, \fbsofb, and \fds\
as a function of lattice spacing are shown in Figs.~\ref{fig:fb-dynamical}, 
\ref{fig:fbs-dynamical}, \ref{fig:fbsofb-dynamical}, 
and \ref{fig:fds-dynamical}, respectively.  Leaving aside the ``fat clover''
results for now, the data in all cases seem to favor constant fits; indeed,
the best linear fits have very small slopes.
Note however that the smallest lattice spacing here is $\sim\!0.45\;(\GeV)^{-1}\approx
0.09\;$fm; whereas in the quenched case we have data down 
to $\sim\!0.23\;(\GeV)^{-1}\approx 0.045\;$fm.  It is thus possible that
the apparent independence of lattice spacing is due to the cancellation of two
effects: (1) an overall decrease as lattice spacing decreases, which was one of the 
alternatives considered in the quenched case, and
(2) the turning on of short distance dynamical fermion effects 
as one moves away from the quite coarse spacings  of sets L and N.  The latter
effect could be exacerbated by staggered flavor violations, which would be
especially large on the coarsest lattices and which would reduce the 
effective number of dynamical flavors.

If the above possibility is realized, then the $N_f=2$ results could well begin to
decrease for still smaller lattice spacings as the quenched-like behavior sets in. 
For the decay constants, we therefore consider two alternative extrapolations:
the constant extrapolation of all $N_f=2$ data, 
and a linear extrapolation that begins at the average
value of the results on the two finest lattices (sets R and G) and then continues
to the continuum limit with the quenched slope (see Figs~\ref{fig:fb-dynamical}, 
\ref{fig:fbs-dynamical}, and \ref{fig:fds-dynamical}.)  
For ratios of decay constants, we
ruled out the linear extrapolation in the quenched case. Yet the two finest quenched
lattices (D and H) have in general lower values for the ratios 
than the averages that include the quenched sets (C and CP) that
are comparable to the finest $N_f=2$ lattices.
The two alternatives for ratios are therefore taken to be (1) the constant
extrapolation of all $N_f=2$ data, and (2) the first extrapolation reduced by 
the quenched difference: (average of C, CP, D, and H) - (average of D and H).
Figure~\ref{fig:fbsofb-dynamical} shows these alternatives.
In all cases we then take the central value to be the average of the two
alternatives, and the error of the continuum extrapolation to be the ``sample
standard deviation'' of the two (dividing by $n-1=1$, not $n=2$).
Central values and errors for the $N_f=2$ data are shown in 
Tables~\ref{tab:decay-syst-dynam} and \ref{tab:ratio-syst-dynam}.

\begin{table}[tbh!]
\caption{Central values ($f_\pi$ scale) and errors  in $\MeV$
for the dynamical ($N_f=2$) decay constants. 
As in Tables~\ref{tab:decay-syst-quench} and \ref{tab:ratio-syst-quench},
the statistical errors and the effects of excited states
are combined.  The errors above the line (\ie up to and including
finite volume errors) are treated as errors within the 
$N_f=2$ partially quenched approximation.  Errors below the line are 
treated as errors of
that approximation. 
In general, errors marked with explicit $+$ or $-$ signs are treated
as signed, and  other errors are treated as symmetric.  The exception is
partial quenching, where we do not take the sign seriously but show it
nevertheless in parentheses.
}
\label{tab:decay-syst-dynam}
\def\arraystretch{.8}
\begin{tabular}{|c|c|c|c|c|}
\noalign{\vspace{0.5cm}}
\hline
&\fb &\fbs &\fd &\fds \\
\hline
 central value & 	190.5 & 217.3 & 214.9  & 241.0 \\
\hline
\multicolumn{5}{c}{errors}\\
\hline
 statistics \& excited states &	7.1 & 6.4 & 6.1 & 5.2 \\
 continuum extrapolation &      11.3 & 21.0 & 8.5 & 18.7 \\
 valence chiral extrapolation &  $+16.6$& $+14.7$ & $+7.5$ & $+8.3$ \\
 perturbative & 	        12.0 & 18.5 & 8.2 & 15.1 \\
 magnetic mass &         	3.8 & 4.4 & 8.4 & 9.7 \\
 1/M fit & 	                2.6 & 2.7  & 1.0 & 0.9 \\
 finite volume &  $+7.7\to0.0$ & $+5.2\to0.0$ & $+3.4\to0.0$ & $-0.1\to0.0$ \\
\hline
 partial quenching &      $(+)2.4$& $(-)3.0$ & $(+)3.4$ & $(-)3.8$ \\
 scale (change to $m_\rho$) & $+10.6$ 	 & $+8.7$ & $+5.4$ & $+3.9$ \\
 $\kappa_s$ (change to $\phi$) & --  & $+3.9$ & -- & $+2.3$ \\
 missing dynamical $s$ quark & $+8.7$  & $+9.2$ & $+7.7$ & $+8.9$ \\
\hline
\end{tabular}
\end{table}

\begin{table}[tbh!]
\caption{Same as Table~\protect{\ref{tab:decay-syst-dynam}}
but for decay constant ratios. 
}
\label{tab:ratio-syst-dynam}
\def\arraystretch{.8}
\begin{tabular}{|c|c|c|c|c|c|}
\noalign{\vspace{0.5cm}}
\hline
&\fbsofb &\fdsofd &\fbofds &\fbsofds &\fbofd \\
\hline
 central value & 	1.158 & 1.142 & 0.793  & 0.922 & 0.913 \\
\hline
\multicolumn{5}{c}{errors}\\
\hline
 statistics \& excited states &	0.011 & 0.009 & 0.016 & 0.013 & 0.016\\
 continuum extrapolation &      0.015 & 0.014 & 0.005 & 0.004 & 0.001 \\
 valence chiral extrapolation &$-0.016$& $+0.005$ & $+0.032$ & $+0.019$ & $+0.037$ \\
 perturbative & 	        0.012 & 0.011 & 0.034 & 0.043 & 0.042\\
 magnetic mass &                0.003 & 0.002 & 0.024 & 0.028  & 0.027\\
 1/M fit & 	                0.001 & 0.000  & 0.014 & 0.020 & 0.015 \\
 finite volume &  $-0.019\to0.000$ & $-0.018$  & $+0.026\to0.000$ & $+0.010\to0.000$ &  $+0.016\to0.000$ \\
\hline
 partial quenching  &       $(-)0.023$& $(-)0.026$ & $(+)0.027$ & $(-)0.021$ & $(+)0.008$ \\
 scale (change to $m_\rho$) & $-0.010$ 	 & $-0.005$ & $+0.015$ & $+0.017$ &  $+0.015$ \\
 $\kappa_s$ (change to $\phi$) & $+0.014$ & $+0.017$ & $-0.008$ & $+0.004$ & -- \\
 missing dynamical $s$ quark & $+0.001$ & $+0.007$ & $+0.012$ & $+0.015$  & $+0.021$ \\
\hline
\end{tabular}
\end{table}

\begin{figure}[thb!]
\includegraphics[bb = 0 0 4096 4096,
width=5.0truein]{fds_dynamical.ps}
\caption{Same as Fig.~\ref{fig:fb-dynamical}, but for \fds.
}
\label{fig:fds-dynamical}
\end{figure}

\subsection{Fat Links}
\label{sec:fat}
In the above discussion of the continuum extrapolation, we ignored the fat-link clover
results.  If taken at face value, these results would imply the existence of 
extremely large discretization errors. We therefore need to examine 
the fat-link computations in detail.
These computations use valence quarks --- both heavy and light ---  with the
standard clover action, but with gauge links that have first been
``fattened'' by $N$ iterations of APE smearing \cite{APE-SMEAR}.
The coefficient of the sum of the staples is $c/6$ and that
of the forward link is $1-c$; a projection back into $SU(3)$
is included after each smearing step.  The fat-link 
results (set RF) displayed in Figs.~\ref{fig:fb-dynamical}--\ref{fig:fds-dynamical}
have $N=10$, $c=0.45$. The clover coefficient  $c_{\rm SW}$ is taken
to have its tree-level value (1.0); this is also approximately the tadpole-improved value,
since the fattening strongly suppresses tadpole contributions.
Physically, APE smearing corresponds roughly to a Gaussian smearing
of the fermion-gauge field interaction over a range
 $\langle x^2 \rangle \simeq cN/6$ \cite{CBTDPisa}.

Various kinds of fat links have come to play a major role in lattice
simulations in the last few years. The motivation for introducing them in
the context of Wilson-like fermions \cite{FAT1}  was that they
improve the chiral properties of the fermions. This happens in several
(related) ways:  First, fat links reduce additive mass renormalization.
They also suppress exceptional configurations,
 which present a severe challenge to clover
computations on our dynamical lattices \cite{SEMILEP-LAT99,MILC-LAT99}.
(This occurs because they shrink the range of the real eigenmodes
of the Dirac operator.)
Finally, in perturbation theory, fat links bring the vector and axial
vector renormalization constants $Z_V$ and $Z_A$ (as well as the scalar
and pseudoscalar renormalization constants) closer together.

Simulations of light quark systems with a variety of fat link actions
at lattice spacings in the range 0.1--0.2 fm show little dependence of
 physical observables on the amount of fattening, even for the very
aggressive amount of fattening of the simulations we report here.
For many quantities, this amount of fattening 
also gives quite small discretization errors \cite{FAT-SCALING}.

We take the light-light
renormalization coefficients for fat-link clover fermions
from the perturbative calculations of Ref.~\cite{CB-TD}.
The heavy-lights (for which perturbative calculations do not exist)
are normalized using the {\it static-light} results of \cite{CB-TD}.  
Although, one expects that this should be roughly correct for 
the large values of $aM$ at the B meson, it introduces a possibly
serious source of systematic error into the fat-link results.

As first reported in Ref.~\cite{MILC-LAT99}, the fat-link clover results for 
decay constants are seen to be much smaller than the apparent
 continuum-limit results of 
the Wilson quarks.   
Simulations of $\bar Q Q$ systems with fat link quarks also
show that fattening
suppresses the magnitude of vector-pseudoscalar mass splitting.
A measurement of  the heavy quark potential gives some qualitative
understanding of both effects:
the attractive short distance piece of the potential is
washed away by the fattening. This is shown
in Fig. \ref{fig:smeared-potential}, where we compute the
static potential using $c=0.45$, $N=10$ APE-smeared fat links
at quenched $\beta=5.85$.
The loss of this part of the potential leads to a suppression
of the heavy quark wave function at the origin.
Although this is an effect that would vanish in the continuum limit (for fixed
$N$, $c$), it could introduce large scaling violations for
short-distance-sensitive quantities.

\begin{figure}[thb!]
\includegraphics[bb = 0 0 4096 4096,
width=5.0truein]{smeared-potential.ps}
\caption{Static potential at quenched $\beta=5.85$ with and without $c=0.45$, $N=10$
APE-smeared  fattening.
}
\label{fig:smeared-potential}
\end{figure}

To study more directly the effect of fattening on  heavy-light decay constants,
we have computed the decay constants with clover fermions
 on a 99 lattice subset of
quenched set CP1, which we call CPF.
We have tried  four different levels of fattening: $c=0.45$
 with $N=2$, 6, and 10,
and $c=0.25$ with $N=7$.
In these cases, $c_{\rm SW}$ is
set equal to the tadpole-improved tree level value $1/u_0^3$, with
$u_0$ determined by the plaquette computed with the smeared links. 
The renormalization constants are determined in the same way as for the
dynamical case (set RF).
A comparison of two of the smearing levels 
with the thin-link clover computations 
is shown in Fig.~\ref{fig:fat-link}. 
The fat-link $f_B$ values are considerably suppressed compared to those from 
the thin links, which in turn are consistent with the results
of continuum-extrapolated quenched Wilson fermions 
(see Sec.~\ref{sec:quenched}).

\begin{figure}[thb!]
\includegraphics[bb = 0 0 4096 4096,
width=5.0truein]{fat-link.ps}
\caption{Effect of smearing on quenched $f_B$.  The thin clover points
are at $\beta=6.0$ and $6.15$ (sets CP1 and J); the fat, at  $\beta=6.0$ (set CPF).
The extrapolation of the thin clover results to the continuum is also shown.
}
\label{fig:fat-link}
\end{figure}

Figure \ref{fig:fat-link} shows that the suppression produced by the
lowest and highest levels of fattening are consistent.  
In fact, there is
not much difference in the values of the heavy-light decay constants among
the four different levels of fattening we studied,
even though the amount of smoothing introduced into the short-distance potential is quite
different for the four cases.
Furthermore, the
light-light decay constants with fat clover and thin clover links differ by only
$\sim 7\%$: Compare the $f_\pi$-determined lattice
spacings of sets CP1 and CPF in Fig.~\ref{fig:fat-link}, or see 
Table~\ref{tab:ainverse}.
Note that in the light-light
case we are using the correct renormalization factors from Ref.~\cite{CB-TD}.
This suggests that the $\sim\!25\%$
suppression of  heavy-light decay constants for our fat links may be due more 
to the use of the incorrect renormalizations (static-light instead of heavy-light) than
to scaling violations from the smoothing of the short-distance potential.  
Be that as it may,  these quenched studies show 
that the fat clover $N_f=2$ results may be ignored, at least until fat-link heavy-light
renormalization constants are available.

An alternative approach would be to try to correct the fat-link clover dynamical
results by the factor (thin link quenched)/(fat link quenched) at a comparable
lattice spacing. We can do this since the lattice spacings for sets CPF (quenched)
and RF ($N_f=2$) are quite close. (See Table~\ref{tab:ainverse}.)
The corrected fat-link results shown in Figs.~\ref{fig:fb-dynamical},
\ref{fig:fbs-dynamical}, and \ref{fig:fds-dynamical} are consistent with
the Wilson $N_f=2$ results.
However, we judge that the reasons for the fat-link suppression are not
well enough understood to be confident that the correction factor is
the same in the quenched and dynamical cases. We therefore
drop  the fat clover $N_f=2$ results and use the Wilson results only.

We emphasize that fat-link actions are formally neither better nor worse than actions with
thin links --- the differences lie in only in the composition and strength
of higher dimensional (irrelevant) operators. However, from a
practical point of view one is interested in actions for which particular quantities 
scale well with lattice spacing. 
Fat links are intended to improve chirality, but
chirality is a property of light quarks, not heavy ones. In hindsight,
there is no physical motivation to construct or use fat link actions for
heavy quarks. Some recent developments \cite{Hasenfratz:2001hp}
for fat-link actions for light quarks have been influenced by our negative experience 
--- one of the design criteria is to minimize effects such as are shown in 
Fig.~\ref{fig:smeared-potential}. 
We are currently studying the behavior of decay constants simulated with 
thin-link heavy quarks and fat-link light quarks.

\subsection{Partial Quenching and Chiral Extrapolation}
\label{sec:PQ}

Our central values with $N_f=2$ are computed in the ``partially
quenched'' approximation: dynamical quark configurations are treated
as fixed backgrounds and chiral extrapolation is performed in the valence quark
mass only.  
The main justification for using the partially
quenched approximation can be seen qualitatively in 
Figs.~\ref{fig:fb-dynamical}--\ref{fig:fds-dynamical}: For our range of
dynamical quark masses and with our statistical and systematic errors, 
there is no obvious trend
in the decay constants when the dynamical quark mass is varied at fixed $\beta$.   
(This statement is examined in more detail below.)

The standard systematic error associated with the valence-mass chiral
extrapolation is then estimated in exactly the same way as in the quenched approximation
(comparison of ``chiral choice I'' with ``chiral choice II'' --- see Secs.~\ref{sec:chiral}
and \ref{sec:quenched}).  Effects of chiral logarithms at very low quark mass
are considered separately in Sec.~\ref{sec:chiral-log-errors}.

To estimate the systematic error due to 
partial quenching, we perform a complete additional analysis in 
the ``fully unquenched'' theory, where the
light ($u,d$) valence quark mass on a given lattice set 
is interpolated or extrapolated to the value of the dynamical mass on that set.
Since the valence and dynamical quarks are simulated with different lattice
actions, the equality must be defined by some physical quantity. We demand
that the pseudoscalar (``pion'') have the same mass with either action.
We then perform chiral extrapolations of $f_{Qq}$ with 
$m_{q,\rm valence}=m_{q,\rm dynamical}$
using data from sets at fixed $\beta$: either 
$\beta=5.6$ (sets G, R, S, T, U) or $\beta=5.5$ (sets P, M, O, N).
Such extrapolations must be performed in physical units because they involve
different sets with different lattice spacings.  To set the scale, we use as usual
$f_{qq}$, extrapolated in valence quark mass to the physical $u,d$ 
point, \ie $f_\pi$.  Note that the scale is set in a partially-quenched manner. 
However the fully unquenched theory is recaptured once the dynamical mass is itself
extrapolated to the physical $u,d$ point.\footnote{This approach could be dangerous
if the dependence of $f_{qq}$ on the dynamical quark mass at fixed valence mass were
so violent that the chiral extrapolation of $f_{Qq}$ in
physical units became uncontrolled.  This does not appear be the case, as seen in
Figs.~\protect{\ref{fig:fb-chiral-dyn-5.6}}--\protect{\ref{fig:fdsofd-chiral-dyn-5.5}}
below.  However,  in work in progress \protect{\cite{MILC-3flavor}}, 
we employ a safer approach, in which 
the dynamical lattices have matched scales 
set independently of the valence quarks using the static quark potential.}

In Fig.~\ref{fig:fb-chiral-dyn-5.6} we show the chiral extrapolation of
$f_B$ with $m_{q,\rm valence}=m_{q,\rm dynamical}$ at $\beta=5.6$.
We call the dependent variable ``$f_B$'' because the heavy quark has already
been interpolated to the $b$ quark mass, as in Sec.~\ref{sec:HQET}.
As independent variable, we use the pseudoscalar mass squared, $m_{qq}^2$, and
extrapolate to $m^2_{qq}= m^2_\pi$.  Note that the linear fit is
excellent, even though it includes very heavy $m_{qq}$ values.  
However, if we restricted
$m_{qq}$ to a safer range for a chiral extrapolation ($m^2_{qq}< 0.6 (\GeV)^2$),
the results would be essentially unchanged. The behavior of \fd\ is very similar to that
of \fb.

\begin{figure}[thb!]
\includegraphics[bb = 0 0 4096 4096,
width=5.0truein]{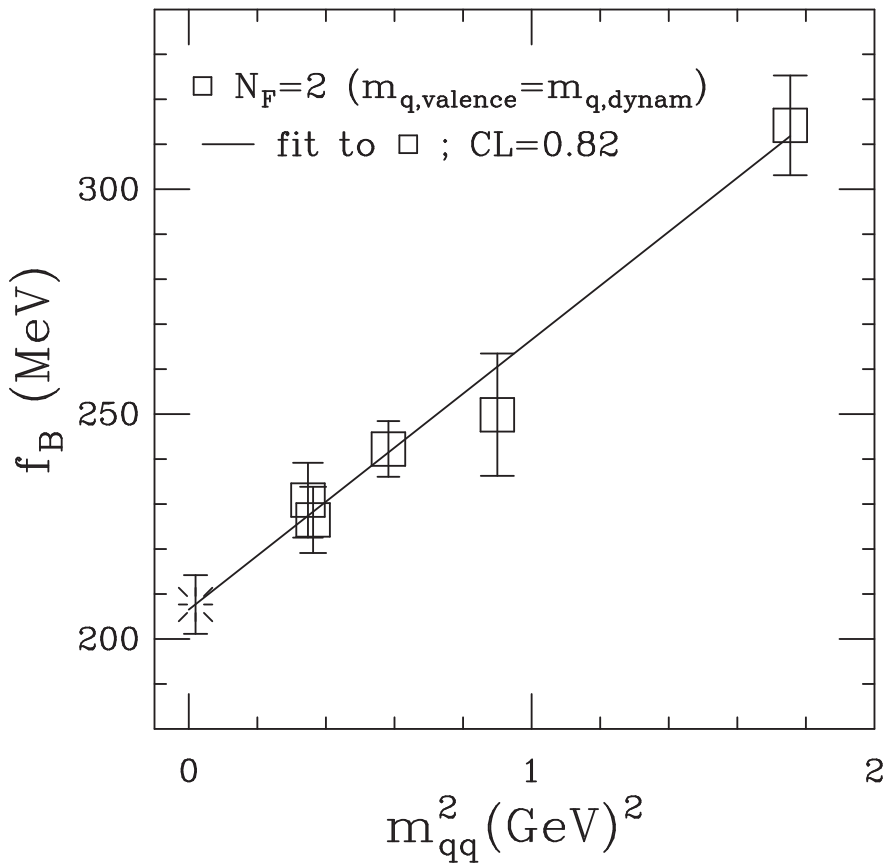}
\caption{``Fully unquenched'' chiral extrapolation of $f_B$ at $\beta=5.6$
(sets G, R, S, T, U).  The burst shows the extrapolated value when
$m^2_{qq}= m^2_\pi$.
}
\label{fig:fb-chiral-dyn-5.6}
\end{figure}

\begin{figure}[thb!]
\includegraphics[bb = 0 0 4096 4096,
width=5.0truein]{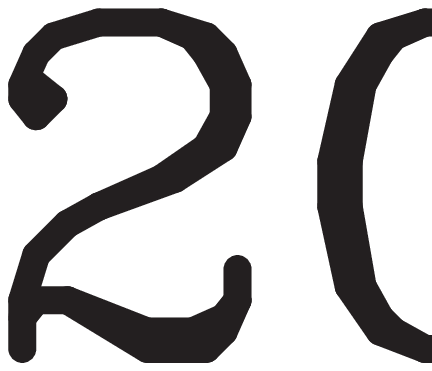}
\caption{Same as Fig.~\protect{\ref{fig:fb-chiral-dyn-5.6}} but for \fbs.
The valence quark masses do not vary but are held fixed at the masses
of the $b$ and $s$.
The fit is linear (not a constant), but has quite small slope.
}
\label{fig:fbs-chiral-dyn-5.6}
\end{figure}

Figure~\ref{fig:fbs-chiral-dyn-5.6} shows \fbs\ as a function of the dynamical
quark mass at $\beta=5.6$.  
The light valence quark mass has already been interpolated to the strange quark
mass, and only the dynamical $u,d$ quark mass is varied. With 
the current statistical and discretization errors,
there is little evidence here for dynamical quark mass dependence (using
an $f_\pi$ scale.) This may be due, at least partially, to staggered flavor 
violations, which reduce the effective range over which the dynamical
mass varies. Note, however, that there {\it is}\/ a significant difference when 
one compares these dynamical mass points to the 
infinite mass case (the quenched approximation):  compare Figs.~\ref{fig:fbs-quenched} and
\ref{fig:fbs-dynamical}.   The behavior of \fds\ is nearly
identical to that seen in Fig.~\ref{fig:fbs-chiral-dyn-5.6}; 
the other decay constants, such as \fb, have  similar behavior when they are plotted as a function
of the dynamical mass for fixed valence mass.

The chiral extrapolation of \fdsofd\ as a function of dynamical quark mass
(represented by the dynamical $m_{qq}^2$) is shown in Fig.~\ref{fig:fdsofd-chiral-dyn-5.5}.
 For \fd, the light valence quark mass is put equal to the
dynamical mass; while for \fds, it is kept equal to the physical strange mass.
Since \fds\ has fixed valence quark mass, it, like \fbs, changes 
little with dynamical quark mass;
while \fd\ varies more or less linearly, like \fb.  We therefore fit \fdsofd\
to the inverse of a linear function in $m_{qq}^2$, \ie to $1/(c+dm_{qq}^2)$,
with $c$ and $d$ allowed to vary.  The ratio \fbsofb\ is fit in the same way;
while the ratios \fbofds, \fbsofds, and \fbofd\ are fit to linear functions.
(The latter two ratios are, like \fbs,  almost independent of the dynamical
quark mass, and so the fitting form makes little difference as long as constant
behavior is allowed.)

\begin{figure}[thb!]
\includegraphics[bb = 0 0 4096 4096,
width=5.0truein]{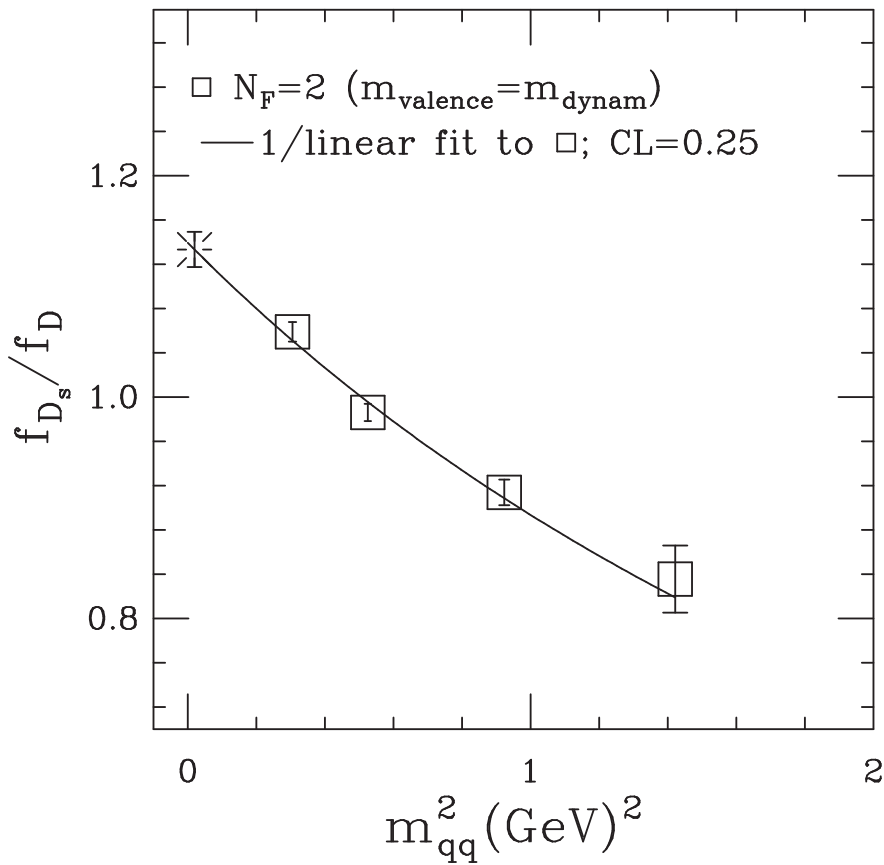}
\caption{
Chiral extrapolation of \fdsofd\  at $\beta=5.5$
(sets P, M, O, N) with the light valence quark mass in \fd\ equal to
the dynamical quark mass.  
The quantity $1/(\efdsofd)$  is fit to a linear function.
}
\label{fig:fdsofd-chiral-dyn-5.5}
\end{figure}

We can now examine the dependence of the fully unquenched quantities
on lattice spacing.   
Unfortunately, we can perform the fully unquenched analysis only at
two $\beta$ values, 5.5 and 5.6, for each of which lattice sets exist with four 
different dynamical quark masses.   
At the third  $\beta$ value of our dynamical simulations ($\beta=5.445$),
we have only a single dynamical mass ($am=.025$, set L). We attempt a chiral
extrapolation there by using the average of the (physical) parameters describing the
$m_{qq}^2$ dependence at $\beta=5.6$ and $5.5$ (as determined above).
Each parameter has a statistical error estimated by propagating 
the statistical errors of the $\beta=5.6$ and $5.5$ data, and a systematic error
taken to be the difference between the average
value and the  $\beta=5.5$ value.  The overall error at $\beta=5.445$ is then
determined by adding in quadrature the intrinsic statistical error from set L and
the statistical and systematic errors coming from the chiral extrapolation.
The amount of chiral extrapolation required for set L is actually quite small
because the physical dynamical quark mass there is close to the smallest masses
available at $\beta=5.6$ and $5.5$.
Therefore the errors introduced by our ``synthetic'' chiral extrapolation at
$\beta=5.445$ do not appear to be large.  
However, the fact that the third data point in the fully unquenched
analysis must be obtained in this way is another reason why we 
prefer the partially quenched analysis for the central values.

\begin{figure}[thb!]
\includegraphics[bb = 0 0 4096 4096,
width=5.0truein]{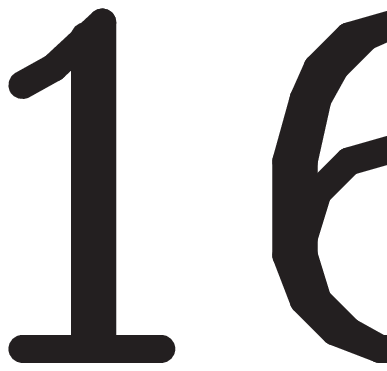}
\caption{
Lattice spacing dependence of \fb\ after fully unquenched chiral extrapolation.
From left to right, the points represent: $\beta=5.6$, $5.5$, and $5.445$.
When there is more than one lattice spacing at a given $\beta$,
the points are plotted at the lattice spacing of the finest lattice
(lowest dynamical mass). Thus $\beta=5.6$ and $5.5$ are represented by
the lattice spacing of sets R and P, respectively. The fit is to a constant.
}
\label{fig:fb-dyn-notPQ}
\end{figure}

\begin{figure}[thb!]
\includegraphics[bb = 0 0 4096 4096,
width=5.0truein]{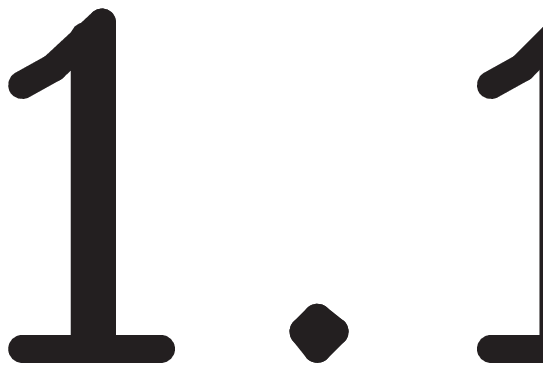}
\caption{Same as Fig.~\protect{\ref{fig:fb-dyn-notPQ}}, but for \fbsofb.
}
\label{fig:fbsofb-dyn-notPQ}
\end{figure}

Figs.~\ref{fig:fb-dyn-notPQ} and \ref{fig:fbsofb-dyn-notPQ} show the lattice
spacing dependence of \fb\ and \fbsofb\ after the fully unquenched chiral
extrapolations.  Like the partially quenched data of Figs.~\ref{fig:fb-dynamical}
and \ref{fig:fbsofb-dynamical}, the fully unquenched data is quite consistent with constant
behavior in $a$.  The other decay constants and ratios behave similarly.
The difference between the result of the constant fits in the
fully unquenched and partially quenched cases is defined to be the
systematic error of partial quenching, and is listed for the various
quantities in Tables~\ref{tab:decay-syst-dynam} and \ref{tab:ratio-syst-dynam}.  
Given the issues in the fully
unquenched analysis, we believe that this error determination is merely
a rough estimate of the magnitude of the effect and do not take the sign
of the difference seriously.  We therefore symmetrize this error in the final
error analysis.

\subsection{Rough Estimate of Chiral Logarithm Effects}
\label{sec:chiral-log-errors}
As discussed in Sec.~\ref{sec:chiral-logs}, our rather heavy light mass values
preclude a detailed study of chiral logarithms.  However, an extrapolation
of $f_{qq}/f_{Qq}$ (rather than individual decay constants), coupled with methods
of determining $\kappa_s$ and $a^{-1}$ without significant chiral extrapolation,
should provide an indication of the effect of the logarithms at light quark mass. 
Recall that, in the full theory, the coefficient of the chiral logs in $f_{qq}$
is probably larger than in $f_{Qq}$.  This means that any errors in coming from a
quadratic  extrapolation of  $f_{qq}/f_{Qq}$ should be opposite to those in
our standard extrapolations of $f_{Qq}$ itself  --- especially for heavy-light
decay constant ratios, which
are less sensitive to the scale determination.  In particular, the $f_{qq}/f_{Qq}$ approach
should overestimate \fbsofb, just as our standard approach may underestimate it.   
Indeed, the most significant change from the central value occurs in \fbsofb\ and is positive.

Table~\ref{tab:decay-logs} of Sec.~\ref{sec:chiral-logs} shows the changes in the decay
constants with various methods for fixing $\kappa_s$ and $a^{-1}$.  Changes in the ratios
are given in Table~\ref{tab:ratio-logs}.  After eliminating the three lines in each table
marked with asterisks (see Sec.~\ref{sec:chiral-logs}), we average the changes
in decay constants and ratios and find the standard deviations
of the means.  With the exception of the quantity \fdsofd, the averages in all cases
are positive and larger than the standard deviations
of the means.  We define the ``error due to chiral logarithm effects'' in these cases as
the signed (positive) number which is the sum of the average and the standard deviation
of the mean.  This is slightly more conservative than just taking the straight average.
For \fdsofd, where the average is consistent with 0.00, we take the error as the (unsigned)
standard deviation of the mean.
 
\begin{table}[p]
\caption{Estimates of the effects (in MeV) of chiral logarithms
on the extrapolation of decay constant ratios. 
For descriptions of the methods, as well as
$\kappa_s$  and $a^{-1}$, see  Table \protect{\ref{tab:decay-logs}} in Sec.~\ref{sec:chiral-logs}. 
Lines
indicated by a `*' are eliminated from the averages.
}
\label{tab:ratio-logs}
\def\arraystretch{.8}
\begin{tabular}{|c|c|c|c|c|c|}
\noalign{\vspace{0.5cm}}
\hline
{\bf method}&\fbsofb &\fdsofd &\fbofds &\fbsofds & \fbofd \\
\hline
 \phantom{*} 1 & $+0.02$       & $+0.02$       & $+0.02$       & $+0.05$       & $+0.04$ \\
 & $ -0.02$      & $ -0.09$      & $+0.07$       & $+0.01$       & $+0.00$\\
\hline
 \phantom{*} 2  & $+0.06$       & $+0.06$       & $+0.01$       & $+0.06$       & $+0.06$ \\
   & $+0.02$       & $ -0.05$      & $+0.05$       & $+0.03$       & $+0.02$ \\
\hline
 \phantom{*} 3  & $ -0.02$      & $ -0.03$      & $+0.04$       & $+0.03$       & $+0.01$\\
  & $ -0.07$      & $ -0.13$      & $+0.08$       & $ -0.00$      & $ -0.02$\\
\hline
 \phantom{*} 4 & $+0.04$       & $+0.04$       & $+0.02$       & $+0.06$       & $+0.05$ \\
  & $+0.01$       & $ -0.06$      & $+0.06$       & $+0.02$       & $+0.01$ \\
\hline
 \phantom{*} 5  & $+0.08$       & $+0.08$       & $+0.00$       & $+0.07$       & $+0.07$ \\
   & $+0.05$       & $ -0.03$      & $+0.04$       & $+0.04$       & $+0.03$ \\
\hline
 \phantom{*} 6  & $+0.03$       & $+0.03$       & $+0.03$       & $+0.06$       & $+0.05$ \\
  & $+0.04$       & $ -0.03$      & $+0.06$       & $+0.06$       & $+0.05$ \\
\hline
 \phantom{*} 7  & $+0.05$       & $+0.05$       & $+0.01$       & $+0.06$       & $+0.05$ \\
   & $+0.07$       & $ -0.00$      & $+0.05$       & $+0.06$       & $+0.05$ \\
\hline
 \phantom{*} 8  & $ -0.01$      & $ -0.02$      & $+0.04$       & $+0.04$       & $+0.03$\\
   & $+0.01$       & $ -0.07$      & $+0.08$       & $+0.05$       & $+0.04$\\
\hline
 \phantom{*} 9  & $+0.05$       & $+0.05$       & $+0.02$       & $+0.06$       & $+0.06$ \\
   & $+0.06$       & $ -0.01$      & $+0.05$       & $+0.06$       & $+0.06$ \\
\hline
 \phantom{*} 10  & $+0.07$       & $+0.08$       & $+0.00$       & $+0.06$       & $+0.06$ \\
   & $+0.08$       & $+0.02$       & $+0.04$       & $+0.06$       & $+0.06$ \\
\hline
 * 11  & $+0.12$       & $+0.15$       & $ -0.01$      & $+0.10$       & $+0.11$ \\
 * \phantom{11}  & $+0.18$       & $+0.18$       & $ -0.02$      & $+0.11$       & $+0.11$ \\
\hline
 \phantom{*} 12 & $+0.02$       & $+0.01$       & $+0.02$       & $+0.06$       & $+0.03$ \\
 * \phantom{12} & $+0.17$       & $+0.15$       & $ -0.01$      & $+0.09$       & $+0.10$ \\
\hline
\hline
{\bf average} &  $+0.03$ & $-0.00$ & $+0.04$ & $+0.05$ & $+0.04$ \\
{\bf stand.\ dev.\ of mean} &  $0.01$ & $0.01$ & $0.01$ & $0.00$ & $0.01$ \\
\hline
\hline
\end{tabular}
\end{table}

The chiral logarithm effects, while quite significant 
in the case of \fbsofb\ and some of the other ratios,
appear to be considerably smaller than has been anticipated in
Refs.~\cite{KRONFELD,YAMADA}.  We believe this due to the fact that
we set the scale in our central values using $f_\pi$ and extrapolate
the light-light and heavy-light decay constants in the same manner.
Thus, much of the chiral logarithm effects, which are similar in
$f_\pi$ and $f_B$, cancel.

On the other hand, we emphasize that our estimate of the chiral logarithm effects is, for a variety
of reasons, rather rough.
First of all, the changes in the decay constants and ratios vary a great deal among the 
different methods and configurations shown in the tables.  Indeed, the standard 
deviation (as opposed to the standard deviation of the mean) of a change is typically 
the same size as the average change and is sometimes larger.  Secondly, our approach
relies on \chpt\ to find the quantities $m_{ss}$ and $f_{ss}$, and \chpt\ is not necessarily 
rapidly convergent for $s$ quarks.  We have also performed only a partially
quenched analysis of this issue.  Because of the size of the errors,
we have not attempted to extrapolate the dynamical quarks 
to their physical masses.  Finally, we note that there is an inherent (though presumably
small) inconsistency in our determinations of $\kappa_s$ and $a^{-1}$, which indirectly
use the physical values of  $f_\pi$, $f_K$, $m_\pi$ and $m_K$ (or $m_\eta$ or $m_\phi$).  
We cannot force all these quantities to have their physical values at once in
a theory without a dynamical strange quark.  For this reason, it is unclear for example whether it is
better to use $N_f=2$ or $N_f=3$ PQ\chpt\ in finding $f_{ss}$; we hope that our range of methods
gives a reasonable range of results.

Given the crude nature of the chiral log error, we believe that it would be inappropriate at this
stage to use the computations described in Sec.~\ref{sec:chiral-logs} to correct our central values.
Instead, we use them only for error estimates.

\subsection{Final Error Estimates and Results}
\label{sec:other-errors}

The magnetic mass error in
Tables~\ref{tab:decay-syst-dynam} and \ref{tab:ratio-syst-dynam}
is estimated with almost the same method as we used
for the quenched calculation.  The only difference is that here all the
valence quarks are of Wilson type,
so that there is no reduction of the  magnetic mass error in the final error budget
for the relative number of Wilson and clover estimates.
The perturbative and $1/M$ fit errors in the tables
are determined in exactly the same manner as in the quenched approximation.

The errors due to finite volume are studied by comparing results on
sets R and G, both of which have $\beta=5.6$ and $am=0.01$, but which
have spatial volumes $24^3$ and $16^3$,
respectively.  Note that all but one of
our $N_f=2$ sets are large (spatial size $\sim\!2.1$--$3.3$~fm);
only set G is comparable in size ($\sim\!1.4$~fm) to the quenched lattices.
The difference between sets R and G is therefore likely to be a considerable
overestimate of the actual finite volume error. Despite this, the differences
are almost never statistically significant.  Here a ``significant difference'' is defined
as one that is larger than the sum, in quadrature, of the statistical errors of
the two sets.  When the difference is insignificant, we set the finite volume
error to zero, as indicated in 
Tables~\ref{tab:decay-syst-dynam} and \ref{tab:ratio-syst-dynam} by the
notation ``$\to 0.0$.''  The only case where we find a significant 
($\sim \! 1.6\sigma$) effect  is in \fdsofd.

The total systematic error {\it within}\/ the current approximation 
(partially quenched $N_f=2$ theory) is then taken to be the sum of all
the systematic errors above the line 
in Tables~\ref{tab:decay-syst-dynam} and \ref{tab:ratio-syst-dynam}:
continuum extrapolation, valence chiral extrapolation, perturbative,
magnetic mass,  $1/M$ fit, and finite volume errors.  Since these errors show
no evidence of correlations, we perform the sum in quadrature. We do, however,
treat positive and negative errors separately, since the valence
chiral extrapolation error represents a binary choice and has a well
determined sign.

We still need to estimate the error {\it of}\/ the partially quenched $N_f=2$
approximation. One measure of this error has already been discussed:
the partial quenching error.
The effect of the missing third
light virtual quark (the $s$ quark) is estimated in a direct way by assuming
a simple linear dependence of the decay constants on the number of
dynamical flavors.   The error is thus chosen to be one half the difference
of the $N_f=2$ and quenched calculations. This estimate is labeled
``missing dynamical $s$ quark'' in
Tables~\ref{tab:decay-syst-dynam} and \ref{tab:ratio-syst-dynam}.
We also estimate the effect in two indirect ways: by determining the change
in the results when (1) the scale is fixed by $m_\rho$ (instead of $f_\pi$), and
(2) for strange quark quantities, when $\kappa_s$ is fixed by the vector meson
sector ($m_\phi$) instead of the pseudoscalars.  In full QCD (and with
no other systematic errors!), these differences should vanish, so their size is
an estimate of the distance we are from the full theory.  

The total error of the partially quenched
$N_f=2$ approximation is then defined to be the maximum of the four estimates
below the line in
Tables~\ref{tab:decay-syst-dynam} and \ref{tab:ratio-syst-dynam}:
partial quenching, scale, $\kappa_s$, and missing dynamical $s$ quark.
The latter three estimates have a well-determined sign, and we therefore
find the maximum positive and maximum negative error separately.
(As discussed above the partial quenching error is treated symmetrically.)
For the
individual decay constants, the scale and  missing dynamical $s$ quark estimates
are always largest; while the error in the ratios are almost always dominated
by the partially quenched error.

Finally we include an additional error
 due to the fact that our extrapolations
from rather large light quark masses cannot see the chiral logarithms directly.
This error is estimated in Sec.~\ref{sec:chiral-log-errors}. We emphasize that
it is necessarily crude.

Our final results for heavy-light decay constants,
including the effects of dynamical quarks,
are: 
\begin{eqnarray}
\label{eq:fBdynam}
f_B & = & 190 (7) ({}^{+24}_{ -17}) ({}^{+11}_{ -2})  ({}^{+8}_{-0})\;\MeV; \quad
f_{B_s} =  217 (6) ({}^{+32}_{ -28})  ({}^{+9}_{ -3}) ({}^{+17}_{-0}) \;\MeV\nonumber\\
f_D  & = &  215 (6) ({}^{+16}_{ -15}) ({}^{+8}_{ -3})({}^{+4}_{-0})  \;\MeV; \quad
f_{D_s}  =  241 (5)  ({}^{+27}_{ -26}) ({}^{+9}_{ -4}) ({}^{+5}_{-0}) \;\MeV\nonumber\\
\frac{f_{B_s}}{f_B} & = & 1.16 (1) (2) (2) ({}^{+4}_{-0}) ; \quad
\frac{f_{D_s} }{f_D}  =  1.14 (1)({}^{+2}_{ -3})(3) (1) \\
\frac{f_{B}}{f_{D_s}} & = & 0.79 (2) ({}^{+5}_{-4})  (3) ({}^{+5}_{-0}) ; \quad
\frac{f_{B_s} }{f_{D_s}}  =  0.92 (1) (6) (2) ({}^{+5}_{-0}) \nonumber\\
\frac{f_B}{f_D} & = & 0.91 (2) ({}^{+6}_{-5}) ({}^{+2}_{-1}) ({}^{+5}_{-0})\nonumber \ .
\end{eqnarray}
Here the errors are, respectively,  statistical, systematic within the $N_f = 2$
partially quenched approximation, 
the systematic errors of that approximation (due to partial quenching and 
the missing virtual strange quark), and an estimate of the effect of chiral
logarithms.

The result for $f_{D_s}$ is consistent with experimental results;
Ref.\ \cite{FDS-EXPT} obtains
$f_{D^+_s}=280(19)(28)(34) \;\MeV$, which the  {\it Review of Particle Properties}
cites as ``the best and most recent value'' \cite{PDG}.
Our $N_f=2$ values are consistent with recent results of
CP-PACS \cite{CPPACS} and preliminary results of JLQCD \cite{JLQCD-LAT01}, 
though our central values 
of the decay constants
and ratios \fbsofb\ and \fdsofd\ are somewhat lower than those of CP-PACS.

\section{Conclusions and Future Directions}
\label{sec:conclusions}

Equation \ref{eq:fBdynam} and Tables~\ref{tab:decay-syst-dynam},
\ref{tab:ratio-syst-dynam}, \ref{tab:decay-logs} and \ref{tab:ratio-logs} summarize our results.  
Chiral extrapolation, continuum extrapolation and perturbation theory
are generally the  biggest sources of  errors for the decay constants,
while partially quenching, the missing $s$ quark, and the magnetic mass 
are also important for many of the ratios.  Because the lattice $u,d$
have necessarily been rather heavy, as they have in other lattice calculations
to date, the effects of chiral logarithms
at low quark mass have only been investigated crudely and indirectly.
We believe that is the error on which we have least control at present.

Work in progress \cite{MILC-3flavor} addresses many of the above issues.  
Improved actions have decreased the continuum extrapolation errors significantly, as well
as eliminated the separate magnetic mass error.
A dynamical $s$ quark is now explicitly included. 
Further, since the new computations use a wide range of both dynamical and
light valence quark masses  we hope to treat the chiral logarithms
explicitly within a partially quenched framework \cite{CB-MG-PQQCD,BOOTH}.  This
should provide more direct evidence about the issue about the size of chiral logarithm effects 
\cite{KRONFELD,YAMADA}, as well as eliminate the explicit
partial quenching error.

Future calculations will use staggered light quarks, as has already been investigated
in conjunction with NRQCD heavy quarks \cite{WINGATE}.  This will allow for
very light valence masses and therefore make possible a detailed study of chiral
logarithms.
To improve the chiral extrapolations
still more, one-loop chiral perturbation theory calculations that take into account
staggered taste\footnote{We prefer the term ``taste'' for the quantum number introduced by doubling in order
to distinguish from physical flavor.} violation will be needed.  Such calculations for pseudoscalar meson
masses already exist; those for
heavy-light decay constants are in progress \cite{CB-CHIRAL}.

The next step after that is likely to involve perturbation theory.  Once the other
errors have been reduced, the errors of one-loop perturbative calculations will no
longer be acceptable.  Higher order calculations using automated
methods \cite{TROTTIER} or non-perturbative computations
will be required.

We thank the HEMCGC collaboration for
use of lattice set G, and  K.-I.\ Ishikawa for sharing
of unpublished results. 
This work was supported by the U.S. Department of Energy under contracts
DOE -- DE-FG02-91ER-40628,      
DOE -- DE-FG03-95ER-40894,      
DOE -- DE-FG02-91ER-40661,      
DOE -- DE-FG02-97ER-41022       
and
DOE -- DE-FG03-95ER-40906       
and National Science Foundation grants
NSF -- PHY99-70701              
and
NSF -- PHY00--98395.            
Calculations for this project were performed 
at Oak Ridge National Laboratory Center for Computational Sciences,
San Diego Supercomputer Center,
Indiana University,  National Center for Supercomputing Applications,
Pittsburgh Supercomputer Center, 
Maui High Performance Computing Center, Cornell Theory Center,
CHPC (University of Utah), 
and Sandia National Laboratory.

\vfill\eject

\end{document}